\documentclass[fleqn,usenatbib]{mnras}

\usepackage{newtxtext,newtxmath}

\usepackage[T1]{fontenc}
\usepackage{ae,aecompl}

\usepackage{graphicx}	
\newsavebox\CBox

\usepackage{amsmath}	
\usepackage{amsfonts}
\usepackage{bm}
\usepackage{hyperref}
\usepackage{threeparttable}
\usepackage{multirow}
\usepackage{subcaption}
\usepackage{framed}
\usepackage{rotating}

\title[Optimal ALMA image of the HUDF in the era of JWST]{An optimal ALMA image of the Hubble Ultra Deep Field in the era of JWST: obscured star formation and the cosmic far-infrared background}

\author[Hill et al.]
{Ryley Hill,$^{1}$\thanks{E-mail: ryleyhill@phas.ubc.ca}
Douglas Scott,$^{1}$
Derek J. McLeod,$^{2}$
Ross J. McLure,$^{2}$
Scott C. Chapman$^{3,4,1}$
\newauthor
and James S. Dunlop$^{2}$
\\
$^{1}$Department of Physics and Astronomy, University of British Columbia, 6225 Agricultural Road, Vancouver, BC, V6T 1Z1, Canada\\
$^{2}$Institute for Astronomy, University of Edinburgh, Royal Observatory, Edinburgh, EH9 3HJ, UK\\
$^{3}$Department of Physics and Atmospheric Science, Dalhousie University, Halifax, NS, B3H 4R2, Canada\\
$^{4}$NRC -- Herzberg Astronomy and Astrophysics, 5071 W Saanich Rd, Victoria, BC, V9E2E7, Canada
}

\date{November 2023}

\pubyear{2023}

\begin{document}
\label{firstpage}
\pagerange{\pageref{firstpage}--\pageref{lastpage}}
\maketitle

\begin{abstract}
\noindent We combine archival ALMA data targeting the Hubble Ultra Deep Field (HUDF) to produce the deepest currently attainable 1-mm maps of this key region. Our deepest map covers 4.2\,arcmin$^2$, with a beamsize of 1.49\,arcsec$\,{\times}\,1.07\,$arcsec at an effective frequency of 243\,GHz (1.23\,mm). It reaches an rms of 4.6\,$\mu$Jy\,beam$^{-1}$, with 1.5\,arcmin$^2$ below 9.0\,$\mu$Jy\,beam$^{-1}$, an improvement of ${>}\,$5\,per cent (and up to 50\,per cent in some regions) over the best previous map. We also make a wider, shallower map, covering 25.4\,arcmin$^2$. We detect 45 galaxies in the deep map down to 3.6$\sigma$, 10 more than previously detected, and 39 of these galaxies have JWST counterparts. A stacking analysis on the positions of ALMA-undetected JWST galaxies with z$\,{<}\,$4 and stellar masses from 10$^{8.4}$ to 10$^{10.4}$\,M$_{\odot}$ yields 10\,per cent more signal compared to previous stacking analyses, and we find that detected sources plus stacking contribute (10.0${\pm}$0.5)\,Jy\,deg$^{-2}$ to the cosmic infrared background (CIB) at 1.23\,mm. Although this is short of the (uncertain) background level of about 20\,Jy\,deg$^{-2}$, we show that our measurement is consistent with the background if the HUDF is a mild (${\sim}\,2\sigma$) negative CIB fluctuation, and that the contribution from faint undetected objects is small and converging. In particular, we predict that the field contains about 60 additional 15\,$\mu$Jy galaxies, and over 300 galaxies at the few $\mu$Jy level. This suggests that JWST has detected essentially all of the galaxies that contribute to the CIB, as anticipated from the strong correlation between galaxy stellar mass and obscured star formation.

\end{abstract}


\section{Introduction}

The Hubble Ultra Deep Field (HUDF) is probably the most well-studied extragalactic region of the sky, containing some of the deepest optical and near-IR exposures obtained to-date \citep[e.g.][]{Beckwith2006,illingworth2013,koekemoer2013,eisenstein2023}. Studies of the HUDF can tell us about how galaxies have evolved from the earliest times to the present day. We now know that a significant fraction of the cosmic star formation occurred within distant galaxies full of dust \citep[e.g.][]{Casey2015,Koprowski2017}, making them effectively invisible at optical and near-IR wavelengths, for all but the deepest images. Yet these galaxies are bright at millimetre (mm) and submillimetre (submm) wavelengths, and so in order to complete our understanding of galaxy evolution, we must also survey the HUDF in this waveband.

The best telescope at these wavelengths available today is the Atacama Large Millimeter/submillimeter Array (ALMA). ALMA has been used several times to survey the HUDF at wavelengths around 1\,mm (\citealt{Dunlop2017}; ASAGAO, \citealt{Hatsukade2018}; ASPECS, \citealt{Gonzalez-lopez2020}; GOODS-ALMA, \citealt{Gomez-guijarro2022}), and it has also been used to follow up interesting individual targets within the HUDF \citep[e.g.][]{Fujimoto2017,Cowie2018}. Yet in contrast to the tens of thousands of galaxies detected in the optical, only a few dozen mm-bright galaxies have been found in these follow-up observations so far.  Although often referred to as `SMGs' (for `submillimetre galaxies'), since we are working here at wavelengths slightly longer than 1\,mm, we will refer to them with the more generic acronym DSFG (for `dusty star-forming galaxy').

In order to detect more galaxies around 1\,mm with ALMA, we can combine all of the data into a single image. In a similar way, two decades ago, a `super-map' of the GOODS-N field was constructed by combining SCUBA data sets at 850\,$\umu$m \citep{Borys2003}, and multiwavelength counterparts from the {\it Hubble Space Telescope\/} ({\it HST\/}) were identified and used to study the properties of the detected submm-luminous sources \citep{Pope2005}. With this same motivation we have undertaken an archival project to combine all of the ALMA observations of the HUDF taken around 1\,mm, with the goal of finding new DSFGs, identifying their counterparts, assessing how much of the background light has been resolved and providing an image that can be used by others for stacking analyses.

In Section~\ref{sec:data} we describe how we retrieve the data from the ALMA archive and combine it into single continuum images in $uv$ space. In Section~\ref{sec:results} we provide our new galaxy catalogue and compare it to previous catalogues. In Section~\ref{sec:cib} we describe how our results lead to a new estimate of the resolved fraction of the cosmic infrared background (CIB) at 1\,mm.  We discuss in Section~\ref{sec:improvements} improvements in our data products and results compared to what was previously available at 1\,mm in the HUDF and we conclude in Section~\ref{sec:conclusion}.  Appendix~\ref{sec:appendix} presents a wider (and shallower) map and gives a supplementary list of sources, while Appendix~\ref{sec:appendixB} provides cutouts of our ALMA sources overlaid over {\it JWST\/} F356W imaging. In Appendix~\ref{sec:alternative} we outline an alternative approach to combining data subimages in real space, which provides a test of the reliability of our $uv$-space combination.

\section{Data retrieval and processing}
\label{sec:data}

\subsection{Obtaining archival ALMA data}

We focus on ALMA Band~6, which spans a wavelength (frequency) range of 1.1--1.4\,mm (210--270\,GHz). This band currently contains the most extensive ALMA observations of the HUDF, and so this is where we expect to produce the deepest archival map. To begin, we queried the ALMA archive\footnote{\url{https://almascience.nrao.edu/asax}} for all Band-6 observations centred at 03$^{\rm h}$32$^{\rm m}$39.0$^{\rm s}$ $-$27$^{\circ}$47$^{\prime}$29.1$^{\prime\prime}$ and overlapping within a radius of 1.5\,arcmin. There are a total of 12 unique programmes that satisfy this criterion, and we selected all of these for our combined map. For some programmes, only a fraction of the total time was spent on the HUDF (e.g.\ for individual pointings of targeted galaxies or for large surveys extending beyond the HUDF), and for these cases we only selected the data that overlap with our region of interest. For more details see Table~\ref{table:obs}, where we summarize all of the data used here to produce the combined map.
We also downloaded the raw data from the ALMA archive centred at the same position, but extending out to 3.5\,arcmin, which we used to make a shallower but larger map; there are a total of seven additional programmes satisfying this criterion, and these are listed in Table~\ref{table:obs_outer}.

For each of the observations, we downloaded the raw $uv$ data and calibrated it using the provided \textsc{ScriptForPI} and the \textsc{CASA} \citep{Mcmullin2007} version used by the observatory at the time of the observations.  We split the science targets from the calibration targets, then time-averaged the data by 30\,s and averaged the frequency channels by a factor of 4 in order to reduce the volume of data. These tasks were carried out using the Canadian Advanced Network for Astronomy Research (CANFAR) platform \citep{Major2019}, which provides easy access to all versions of \textsc{CASA}, as well as ample storage space and memory.

\subsection{Obtaining archival ALMA images}

In addition to downloading the $uv$ data for each programme, we also obtained the imaging products made available by the observatory. These data are used for understanding the transmission function of the combined data (see Section~\ref{sec:cib}) and for testing an alternative data combination method (see Appendix~\ref{sec:alternative}). For every tuning of every target given in Table~\ref{table:obs}, the observatory provides a single image made using the multi-frequency synthesis (MFS) mode, where visibilities in each channel are mapped to a single $uv$-plane, and therefore represent the mean value of the sky at a characteristic frequency (usually the central frequency of the channels) weighted by a spectral function. For programmes carried out in earlier ALMA cycles, we use the Additional Representative Images for Legacy \citep[ARI-L;][]{Massardi2021}, which are reduced in a way more similar to later cycles. These images are primary-beam-corrected, but the primary-beam image is also available for download. The final number of HUDF images downloaded from the ALMA archive is 61, along with their corresponding 61 primary-beam images.

\begin{table*}
\centering
\caption{ALMA projects downloaded from the ALMA archive and used to create the final combined 1-mm image.}
\label{table:obs}
\begin{threeparttable}
\begin{tabular}{lccccc}
\hline
Project ID & Target name(s) & Frequency range & \phantom{0}Map rms$^{\rm a}$ & Synthesized beam$^{\rm b}$ & \phantom{0}Sky coverage$^{\rm c}$\\
 & & [GHz] & \phantom{0}[$\umu$Jy\,beam$^{-1}$] & [arcsec$\,{\times}\,$arcsec] & \phantom{0}[arcmin$^2$]\\
\hline
2012.1.00173.S$^{\rm d}$ & HUDF & 211.2--231.2 & \phantom{0}24 & 0.62$\,{\times}\,$0.52 & \phantom{0}7.05\\
2013.1.00718.S$^{\rm e}$ & UDF1 & 212.2--272.0 & \phantom{0}15 & 1.77$\,{\times}\,$0.90 & \phantom{0}1.15\\
2013.1.01271.S & UDF6462 & 223.3--244.1 & \phantom{00}9 & 0.33$\,{\times}\,$0.26 & \phantom{0}0.31\\
2015.1.00098.S$^{\rm d}$ & HUDF-JVLA-ALMA & 244.3--271.8 & \phantom{0}57 & 0.20$\,{\times}\,$0.16 & 34.7\phantom{0}\\
2015.1.00543.S$^{\rm d}$ & GOODS-S & 255.1--274.7 & 130 & 0.26$\,{\times}\,$0.22 & 54.2\phantom{0}\\
2015.1.00664.S & KMOS3DGS4-24110 & 273.2--274.7 & \phantom{0}91 & 0.15$\,{\times}\,$0.14 & \phantom{0}0.22\\
2015.1.01096.S & UDF-640-1417 & 222.0--255.2 & \phantom{0}11 & 0.99$\,{\times}\,$0.74 & \phantom{0}0.29\\
2015.1.01447.S & UDF0 & 212.0--272.0 & \phantom{00}6 & 1.24$\,{\times}\,$0.91 & \phantom{0}0.29\\
2015.A.00009.S & Bo15Hz27 & 226.4--245.0 & \phantom{0}13 & 0.73$\,{\times}\,$0.46 & \phantom{0}0.30\\
2016.1.00324.L$^{\rm d}$ & UDF\_mosaic\_1mm & 212.1--272.0 & \phantom{0}12 & 1.37$\,{\times}\,$0.93 & \phantom{0}3.72\\
2017.1.00755.S$^{\rm d}$ & GOODS-S & 255.1--274.9 & 120 & 1.31$\,{\times}\,$0.82 & 54.1\phantom{0}\\
2018.1.00567.S & ASAGAO27, 35, 40, 45 & 244.3--262.9 & \phantom{0}24 & 0.65$\,{\times}\,$0.47 & \phantom{0}1.03\\
\hline
\end{tabular}
\begin{tablenotes}
\item $^{\rm a}$ Approximate map rms estimated using the observatory MFS data products. For programmes with a single tuning, this is the rms of the primary-beam-uncorrected map after masking all known sources. For programmes with multiple tunings, we follow the same procedure and then estimate the rms of the weighted mean of the images as $1/\big( \Sigma_i 1/\sigma_i^2 \big)$. The actual rms from combining images at different tunings in $uv$ space using the MFS mode will generally be less than this estimate.
\item $^{\rm b}$ Maximum synthesized beam across all frequencies of a given data set. The actual synthesized beam from combining data taken in different tunings using the MFS mode may differ slightly from this estimate.
\item $^{\rm c}$ The sky coverage overlapping with our data selection criteria, which may be less than the total sky coverage of the given programme.
\item $^{\rm d}$ Programmes used to make a mosaic with uniform noise.
\item $^{\rm e}$ The imaging products from this data set were not used to create the combined image in the image plane (see Appendix~\ref{sec:alternative}) because the beamsize is much larger than for the other data sets.
\end{tablenotes}
\end{threeparttable}
\end{table*}

\subsection{Data combination in the \textit{uv} plane}
\label{sec:data_uv}

Our main goal is to produce the deepest possible map of the central HUDF region, which essentially corresponds to the footprint of ASPECS \citep[see][]{Gonzalez-lopez2020}.  We do so by combining all of the available data within this region.  Our secondary goal is to produce the deepest possible map of the wider HUDF region, corresponding essentially to the ASAGAO footprint \citep[see][]{Hatsukade2018}. To do this, we carried out the following procedure.

All of the time-averaged and frequency-averaged $uv$ data were imaged using the standard \textsc{CASA} task \textsc{tclean}. For the data presented in Table~\ref{table:obs}, the pixel size was set to 0.2\,arcsec. We ran \textsc{tclean} in MFS mode, which, as discussed above, scales $uv$ points observed at different frequencies to a single characteristic frequency, which in our case is the average frequency of the data being combined, 243\,GHz (or 1230\,$\umu$m). In principle spectral variations in the observed sources can lead to imaging artefacts; however, for frequency ranges of about 30\,per cent these artefacts are expected to be small \citep{sault1999}. The frequencies that are combined to produce our map range from about 211 to 275\,GHz, or roughly 30\,per cent, thus we do not attempt to correct for potential artefacts. We also chose natural weighting (each $uv$ point is weighted by its instrumental noise), but with a 250\,k$\lambda$ $uv$ taper. We set the cutoff of the map to 0.2 times the primary beam, which is where the footprint of our map roughly matches the footprint of the ASPECS map. We also produced a map with no $uv$ taper, setting the pixel size to 0.18\,arcsec (because the synthesized beam is somewhat smaller). Lastly, we cleaned the image down to 20\,$\umu$Jy\,beam$^{-1}$, similar to the cleaning level chosen by ASPECS. The final synthesized beamsize of the $uv$-tapered map is $1.49\,{\rm arcsec}\,{\times}\,1.07\,{\rm arcsec}$, while for the map with no $uv$ taper the synthesized beamsize is $1.32\,{\rm arcsec}\,{\times}\,0.92\,{\rm arcsec}$. The pixel rms in the tapered image (which we use for further analysis in this paper) has a minimum value of 4.6\,$\umu$Jy\,beam$^{-1}$ in the deepest region, while the rms is less than 9.0\,$\umu$Jy\,beam$^{-1}$ over an area of 1.5\,arcmin$^2$. Lastly, we tested tapers of 150\,k$\lambda$ and 50\,k$\lambda$, but we found that the trade-off in sensitivity for the larger synthesized beams led to fewer detected sources. We note that we found similar source flux densities across the different tapers, meaning that at 250\,k$\lambda$ we are not resolving out much flux. On the other hand, the map with no $uv$ taper resolves a larger number of sources, making photometry more challenging.

For the shallower, larger map, we exclude all of the observations targeting the deep central region (this includes all of the ASPECS data, the data from \citealt{Dunlop2017}, and two deep pointings towards the eastern corner of the HUDF); this is to ensure that the synthesized beam of the larger map is not dominated by data from the central region (see Table~\ref{table:obs_outer} for details). This map is therefore mostly a combination of the ASAGAO survey data and the GOODS-ALMA survey data (both high and low resolution). However, we include all of the ASAGAO follow-up observations even if they lie within the deep central region (programme ID 2018.1.00567.S), since they roughly match the array configurations and frequencies of the ASAGAO survey and the GOODS-ALMA survey. We set the pixel size to 0.06\,arcsec for the map with no $uv$ tapering, and 0.12\,arcsec for the map with 250\,k$\lambda$ $uv$ tapering, and set the cutoff of the map to 0.4 times the primary beam, which produced a map with a footprint roughly similar to the ASAGAO footprint. All other \textsc{tclean} parameters were kept the same as described above, except for the cleaning threshold, which was increased to 100\,$\umu$Jy\,beam$^{-1}$ (similar to the level chosen by ASAGAO). The final synthesized beamsize of the $uv$-tapered map is $0.87\,{\rm arcsec}\,{\times}\,0.64\,{\rm arcsec}$, while for the map with no $uv$ taper the synthesized beamsize is $0.36\,{\rm arcsec}\,{\times}\,0.27\,{\rm arcsec}$. The pixel rms in the tapered image (which we also use for further analysis in this paper) has a typical value of 60\,$\umu$Jy\,beam$^{-1}$ outside of individual pointings, and 20\,$\umu$Jy\,beam$^{-1}$ within individual pointings. Similar to the deep map, we tested tapers of 150\,k$\lambda$ and 50\,k$\lambda$, but we found that the 250\,k$\lambda$ taper recovers the most sources without resolving out too much flux.

Lastly, we produced maps with uniform noise properties by excluding programmes with individual pointings targeting known galaxies, and instead focusing on large survey programmes. Specifically, for our deep central mosaic we combined data from the programmes 2012.1.00173.S, 2015.1.00098.S, 2015.1.00543.S, 2016.1.00324.L and 2017.1.00755.S, while for the shallower, larger map we only used data from the programmes 2015.1.00098.S, 2015.1.00543.S and 2017.1.00755.S (see Tables~\ref{table:obs} and \ref{table:obs_outer}, respectively). We used the same \textsc{tclean} parameters as above, generating versions with no $uv$ tapering and with 250\,k$\lambda$ $uv$ tapering. The final synthesized beamsizes were effectively unchanged compared to the maps that included individual pointings.

For the remaining analysis in this paper we focus on the tapered maps with all of the available data combined. The additional maps, with no tapering and uniform noise properties, were used to check the reliability of our flux densities, and we found no systematic differences between the measurements.

To calculate the corresponding noise maps for the combined images, we first created a mask using catalogues of previously-detected galaxies. In particular we took the catalogues from \citet{Dunlop2017}, ASAGAO \citep{Hatsukade2018}, ASPECS \citep{Gonzalez-lopez2020} and GOODS-ALMA \citep{Gomez-guijarro2022}. For each galaxy we masked a region 3 times larger than the beamsize, and then multiplied this by the signal map. We next made cutouts of 5 times the beamsize around each pixel and calculated the rms. The resulting noise map was then smoothed with a Gaussian kernel with the same size as the cutout regions to remove artefacts. 

All of these data products are made publicly available, including the primary-beam-corrected maps, the noise maps, the primary beam maps and the synthesized beam maps.\footnote{\url{https://doi.org/10.5683/SP3/YWBVWH}}  The final signal map, noise map and signal-to-noise ratio (S/N) map for the deep central mosaic with a 250\,k$\lambda$ taper, including individual pointings, is shown in Fig.~\ref{fig:snr}, while the same maps for the larger, shallower mosaic are shown in Fig.~\ref{fig:snr_outer}. The deep central mosaic covers an area of 4.2\,arcmin$^2$ out to our primary beam threshold of 0.2, while the large shallow mosaic covers an area of 25.4\,arcmin$^2$ out to our primary beam threshold of 0.4.

\subsection{Optical and infrared galaxy catalogues}

The HUDF has been the target of extensive multiwavelength observations that we can use to complement our 1.23-mm image. The Cosmic Assembly Near-IR Deep Extragalactic Legacy Survey (CANDELS) catalogue of galaxies in the GOODS-S field\footnote{\url{https://archive.stsci.edu/prepds/candels}} \citep{Guo2013} contains a summary of much of the deep optical-to-near-IR imaging in the HUDF, and it is expected that most of the ALMA-detected galaxies should have a counterpart in this catalogue. Detections in the CANDELS catalogue were obtained from {\it HST\/}'s WFC3 instrument in the F160W band, so it is a 1.5-$\umu$m-selected catalogue. The entire catalogue covers a much larger region than our deep map, but the deepest region of the catalogue covers roughly the same area. 

The first public data release from the the {\it JWST\/} Advanced Deep Extragalactic Survey (JADES) NIRCam survey of the HUDF is also now available\footnote{\url{https://archive.stsci.edu/hlsp/jades}} \citep{Rieke2023}, in addition to the JWST Extragalactic Medium-band Survey \citep[JEMS,][]{Williams2023}.  While the JADES survey is still ongoing and will eventually be deeper, it already contains considerably more galaxies per unit area than the CANDELS catalogue. JADES images range from 0.9\,$\mu$m to 4.4\,$\mu$m, and we chose to select our catalogue at $3.5\,\umu$m, as a compromise between the longest possible wavelength and the depth. Our entire deep map is covered by the JADES survey at approximately uniform sensitivity. Throughout this paper we primarily make use of the JADES imaging, with the CANDELS catalogue used to demonstrate new improvements thanks to {\it JWST}.

\subsubsection{JADES photometric redshifts and stellar masses}
\label{M_star-z_data}

The source catalogues, photometry measurements and spectral-energy-distribution (SED) fitting carried out using the public JADES imaging will be described in detail in McLeod et al.~(in prep.) Here, we provide a brief summary of the relevant results used in this paper.

In order to construct photometry catalogues, we ran \textsc{SourceExtractor} \citep{BertinArnouts1996} in dual-image mode, with NIRCam F356W used as the detection image. All imaging was homogenized to match the point-source function (PSF) of the F444W imaging in order to minimise any colour systematics arising from differences in the PSF. We performed isophotal photometry on all available JADES and JEMS NIRCam bands, as well as ancillary \textit{HST} ACS imaging from the Hubble Legacy Fields GOODS-South data release \cite[see][]{illingworth2013,whitaker2019} and ground-based VIMOS $U$-band imaging over GOODS-South \citep{Nonino2009}. To extract robust photometry from this lower-resolution imaging, we utilised \textsc{TPHOT} \citep{Merlin2015}, using positional and surface-brightness information from the higher-resolution NIRCam imaging as input.

To calculate photometric redshifts and hence stellar masses, we performed SED fitting using \textsc{LePhare} \citep{ArnoutsIlbert2011}, with a Bruzual-Charlot \citep{BruzualCharlot2003} template set, incorporating a Chabrier \citep{Chabrier2003} initial mass function. The template set includes a Calzetti \citep{Calzetti2000} dust-attenuation law, with range of reddening $A_{V}=[0.0,6.0,0.2]$. We used two metallicities, 0.2\,Z$_{\odot}$ and Z$_{\odot}$, and included exponentially declining star-formation histories with $\tau$ ranging between 0.1 and 15\,Gyr. Where the object has a known spectroscopic redshift, we fixed to this redshift in obtaining the stellar mass.

The median uncertainty in our photometric redshifts for all of our HUDF galaxies is $\Delta z / (1\,{+}\,z)\,{\simeq}\,0.06$. For stellar masses, in this paper we only care about relative uncertainties, since we only use this parameter to sort galaxies into different bins. The absolute uncertainties in stellar masses are expected to be large, but the relative uncertainties should be small, so ignore them.

\begin{figure*}
\begin{center}
\includegraphics[width=0.33\textwidth]{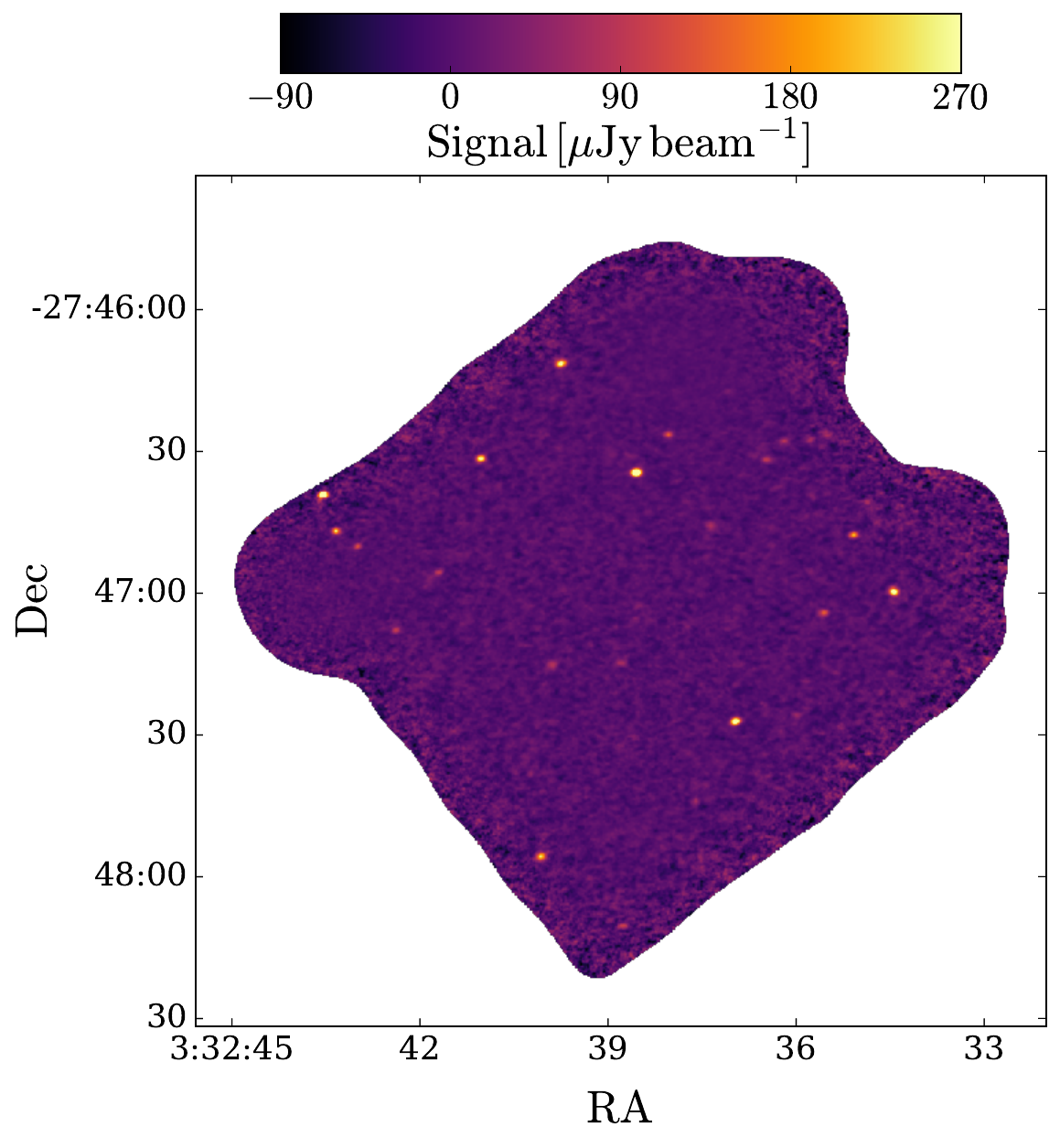}
\includegraphics[width=0.33\textwidth]{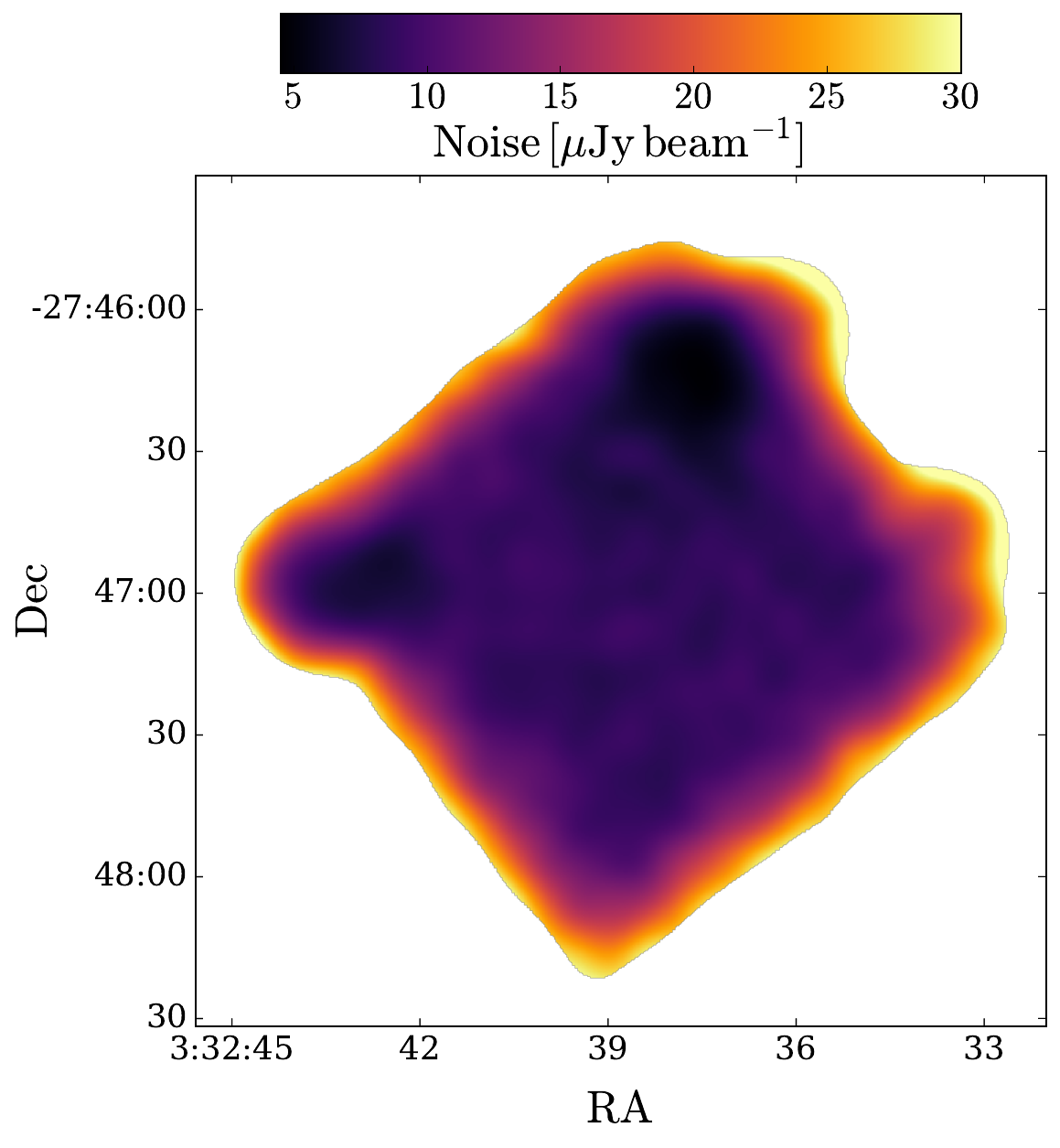}
\includegraphics[width=0.33\textwidth]{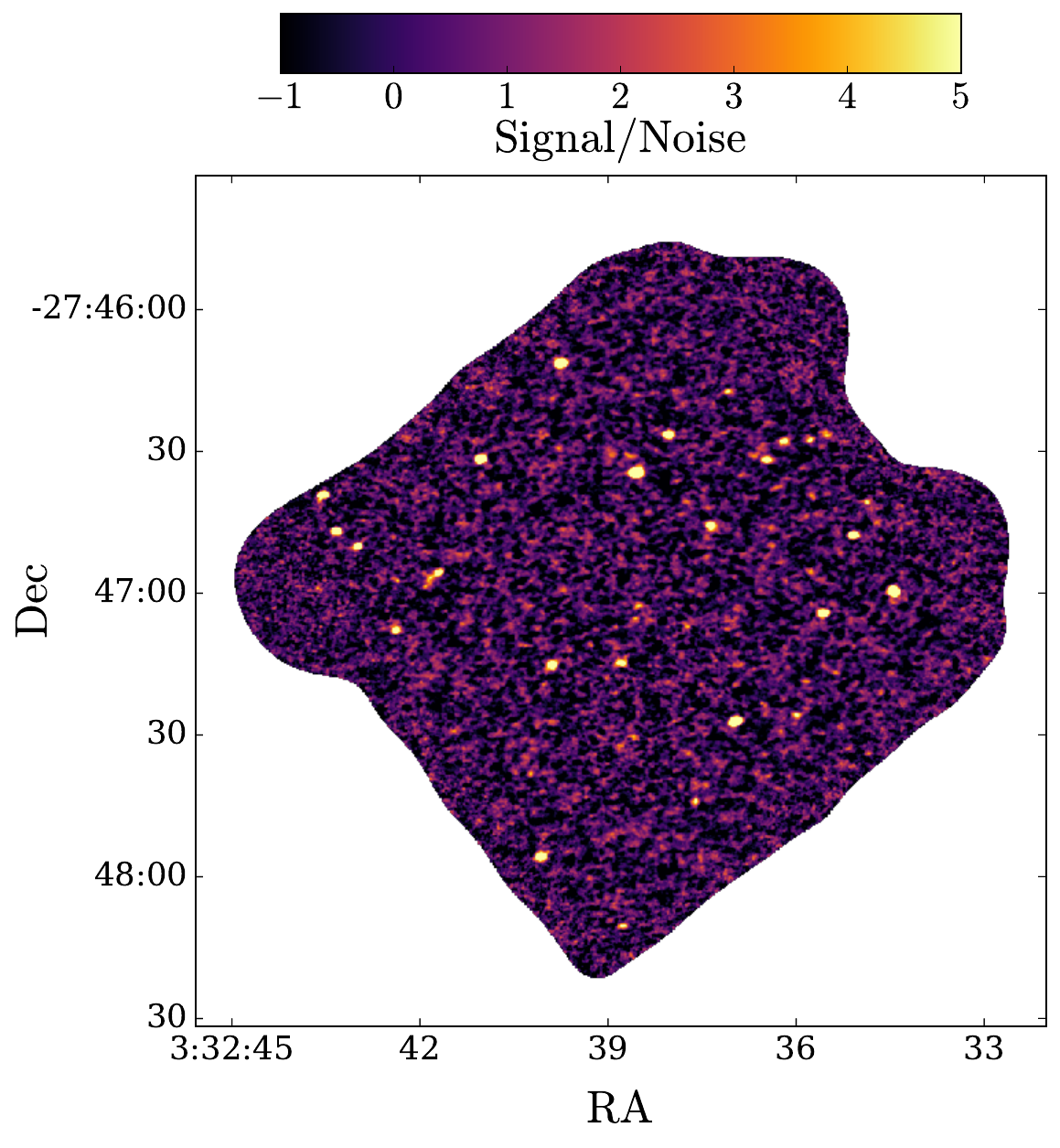}
\end{center}
\caption{{\it Left:} Signal map of the deepest region of the HUDF after combining all of the available archival ALMA Band-6 data given in Table~\ref{table:obs}. This region is defined by the contour where the primary beam reaches 0.2, covering a total of 4.2\,arcmin$^2$.  {\it Middle:} Corresponding noise map. {\it Right:} S/N map, from dividing the signal map by the noise map.}
\label{fig:snr}
\end{figure*}

\section{Results for sources}
\label{sec:results}

\subsection{Cataloguing 1-mm sources}
\label{sec:catalogue}

The central deep map shown in Fig.~\ref{fig:snr} contains all existing ALMA Band-6 data in this region and thus should be the best map currently achievable; for reference, the deepest part goes down to 4.6\,$\umu$Jy\,beam$^{-1}$, which is 50\,per cent deeper than any previous map.  Over an area of 1.5\,arcmin$^2$ our new map has an rms below 9.0\,$\umu$Jy\,beam$^{-1}$, which is 5\,per cent better than the deepest previously available map of this size (from ASPECS), and it has a noise level below 35\,$\umu$Jy\,beam$^{-1}$ over the full $4.2\,{\rm arcmin}^2$. It is therefore of interest to see if any new galaxies are detected in this new map.

We ran the simple peak-finding algorithm \textsc{find\_peaks}, available in the \textsc{photutils} \textsc{python} module, on both the S/N map contained within the region of interest, and on the negative of the same S/N map. One approach to setting the detection threshold is to use the most significant negative peak to set the level above which we might expect all positive peaks to be real galaxies; for reference, the most negative peak was found to have a S/N of $-4.1$, and there are 35 positive peaks with a S/N higher than $+4.1$. However, we can also lower the threshold to include more real sources at the expense of being less confident about the reality of each one.  

\begin{figure}
\begin{center}
\includegraphics[width=0.5\textwidth]{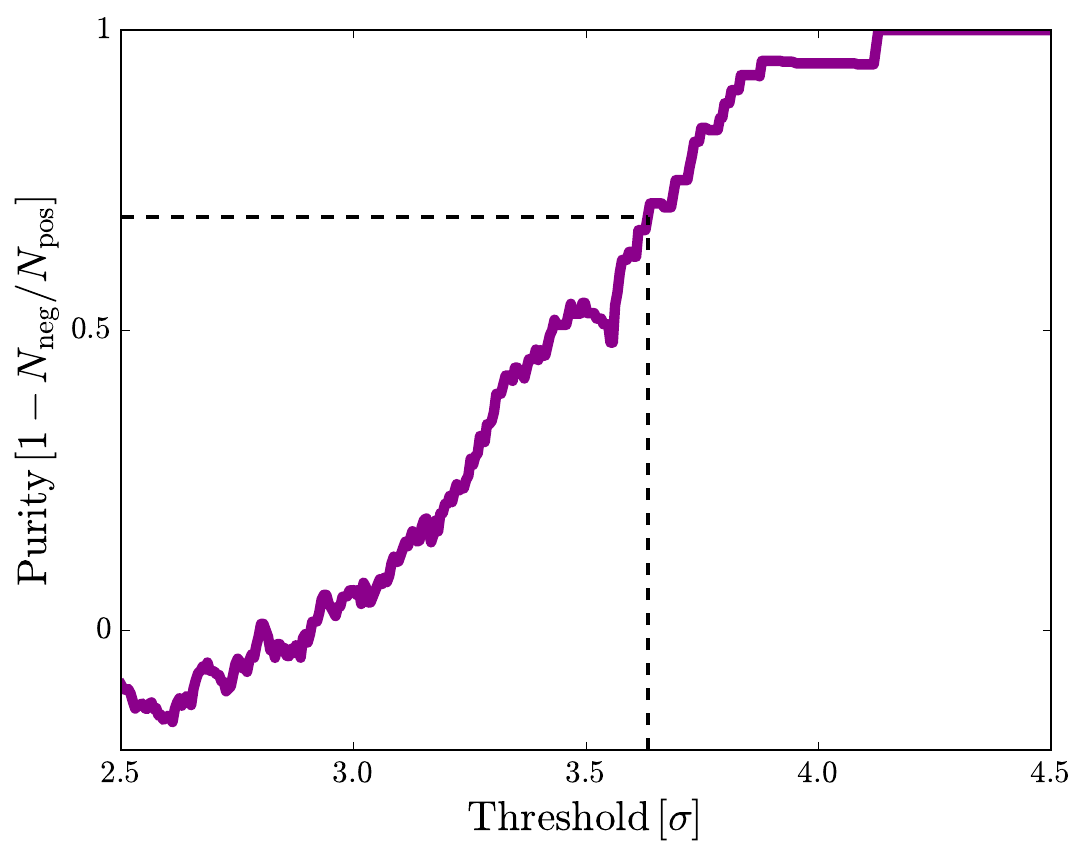}
\end{center}
\caption{Purity, i.e.\ 1 minus the the total number of positive to negative peaks greater than a given S/N map, plotted as a function of S/N. A purity of 0.7 corresponds to a S/N threshold of 3.6, which we use to extract sources from our deep map (Fig.~\ref{fig:snr}).}
\label{fig:purity}
\end{figure}

As an alternative means of setting the threshold, we looked at how the ratio of the total number of positive to negative peaks greater than a given S/N thresholds varies. We can define the `purity' to be $1\,{-}\,{\cal N}_{\rm neg}/{\cal N}_{\rm pos}$, such that a value of 0 means there are as many negative as positive peaks and 1 is reached when there are no more significant negative peaks. In Fig.~\ref{fig:purity} we plot this purity as a function of S/N threshold. The choice of a threshold is of course a trade off (a smaller threshold will result in more false positives), and to include more candidate sources (with the understanding that not all may be real) we choose a purity of 0.7, which corresponds to a peak ${\rm S/N}\,{=}\,3.6$. There are 45 sources above this threshold. The catalogue from ASPECS used a `fidelity' threshold (the differential ratio of galaxies in S/N bins) that resulted in a least significant source with ${\rm S/N}\,{=}\,3.3$, while \citet{Dunlop2017} used a S/N threshold of 3.5 but removed sources with no {\it HST\/} counterpart. Our choice of threshold is therefore similar to (although slightly more conservative than) what was used in previous studies, and we can also use our JADES catalogue to investigate possible false positives.

\begin{figure}
\begin{center}
\includegraphics[width=0.5\textwidth]{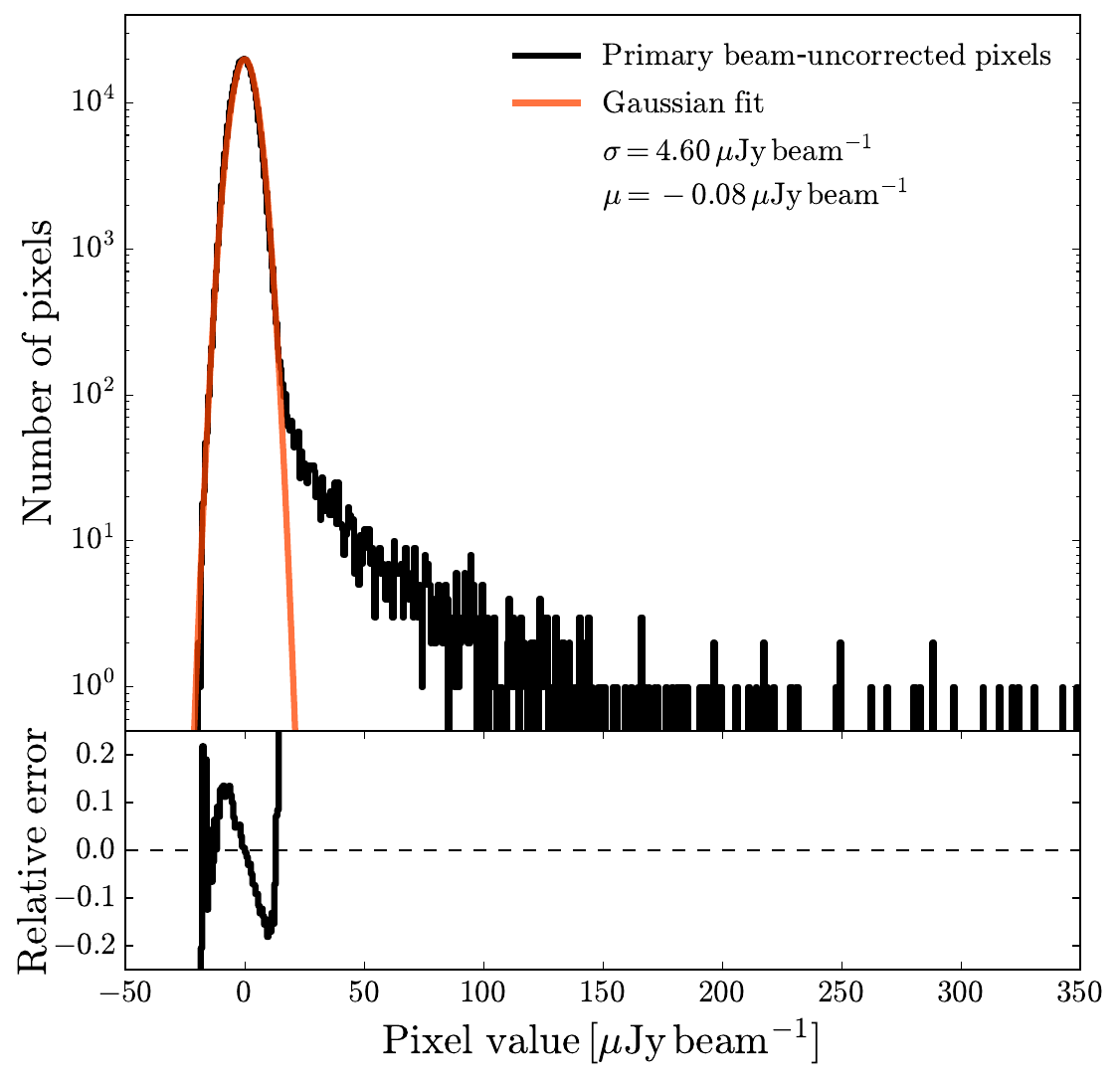}
\end{center}
\caption{Pixel histogram of our map.  The top panel shows the distribution of pixel values in our primary-beam-uncorrected map (black) and a Gaussian fit to the distribution (red), after masking all positive pixels brighter than the most negative pixel (here $-19\,\mu$Jy\,beam$^{-1}$). The best-fit mean and standard deviation are given in the top-right corner. The bottom panel shows the relative residual between the best-fit Gaussian and the real data (defined as data\,${/}\,$model\,$-$\,1), from which we can see that the Gaussian model deviates from the data by no more than 20\,per cent out to pixel values of $20\,\mu$Jy\,beam$^{-1}$, after which real sources dominate and the distribution is no longer Gaussian.}
\label{fig:pixel_hist}
\end{figure}

In Fig.~\ref{fig:pixel_hist} we show the distribution of pixel values (in units of $\mu$Jy\,beam$^{-1}$) in our primary beam-uncorrected map. Here, instead of using the primary beam provided by {\tt CASA}, we scale our own noise map by the minimum noise value to calculate the primary beam map, since this is the map used to search for sources. We can see that the distribution appears quite Gaussian at the negative end, while real sources heavily skew the positive distribution. Our source extraction relies on the Gaussianity of the noise, and so we fit a Gaussian to the pixel distribution, after masking all positive pixels brighter than the most negative pixel (here $-19\,\mu$Jy\,beam$^{-1}$). We find a mean value of $-0.08\,\mu$Jy\,beam$^{-1}$ (very close to zero, as expected) and a width of $4.60\,\mu$Jy\,beam$^{-1}$. Our expectation is that the width should be equal to the minimum noise value used to create our primary beam estimate, which here is $4.58\,\mu$Jy\,beam$^{-1}$, so we find good agreement. Lastly, we calculate the relative residual between the Gaussian fit and the actual pixel distribution (defined as data\,${/}\,$model\,$-$\,1), and show this in the bottom panel of Fig.~\ref{fig:pixel_hist}. We find that the Gaussian model deviates from the actual data by no more than 20\,per cent for pixel values less than $20\,\mu$Jy\,beam$^{-1}$, after which real sources dominate and the distribution is no longer Gaussian. Of course, we still expect real sources to contribute to the map below $20\,\mu$Jy\,beam$^{-1}$ in this `P(D)' analysis \citep[see e.g.][and references therein]{Vernstrom2014}, which could be the cause of the deviation.

\begin{figure*}
\begin{center}
\includegraphics[width=\textwidth]{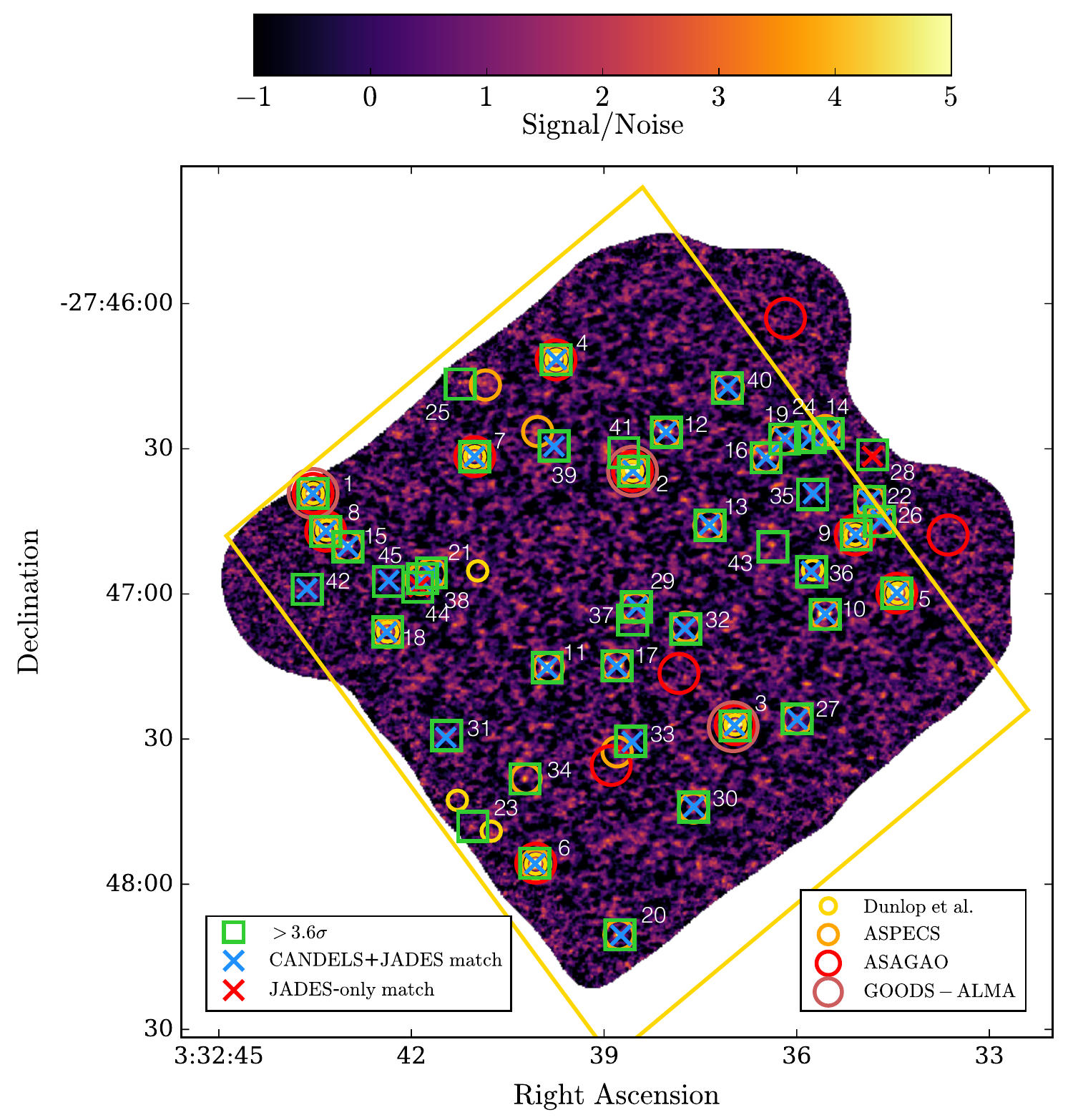}
\end{center}
\caption{Signal-to-noise ratio map of the deep central region, covering $4.2\,{\rm arcmin}^2$, with peaks ${>}\,3.6\,\sigma$ indicated as green boxes. Galaxies found by \citet{Dunlop2017} are shown as yellow circles, galaxies from the ASPECS survey \citep{Gonzalez-lopez2020} are orange circles and galaxies from the ASAGAO survey \citep{Hatsukade2018} are red circles. A few of the brightest galaxies are also detected in the GOODS-ALMA survey \citep{Gomez-guijarro2022} and shown as brown circles. The gold contour shows the footprint of the deepest {\it HST\/} data from the CANDELS/HUDF09 survey, while the entire region is covered by the JADES survey. Galaxies with counterparts in both the CANDELS catalogue and the JADES catalogue are indicated with a blue cross and galaxies with only a JADES counterpart are indicated with a red cross. There are no galaxies with a CANDELS/HUDF09 counterpart but no JADES counterpart. Numbers refer to the labels in Tables~\ref{table:sources} and \ref{table:sources_ids}.}
\label{fig:sources}
\end{figure*}

The positions of sources extracted from this search are shown in Fig.~\ref{fig:sources}, with positions and peak S/N values given in Table~\ref{table:sources}. Appendix~\ref{sec:appendixB} contains 5\,arcsec$\,{\times}\,$5\,arcsec cutouts overlaid on {\it JWST\/} F356W images.  For single-dish surveys, DSFGs are almost always unresolved, but this is not the case for ALMA data. Hence we need to decide how to quote brightness values when some DSFGs are resolved. For the flux densities, we follow a procedure similar to the ASAGAO survey; for each source we fit a 2-dimensional Gaussian profile, fixing the position to the peak pixel and the amplitude to the value of the peak pixel, but allowing the size, ellipticity and position angle to vary. We then calculate the number of beams contained within each source as
\begin{equation}
{\cal N}_{\mathrm{b}} = \frac{a  b}{\theta_{\mathrm{maj}}  \theta_{\mathrm{min}}},
\end{equation}
\noindent
where $a$ and $b$ are the best-fit major and minor FWHM values, and $\theta_{\mathrm{maj}}$ and $\theta_{\mathrm{min}}$ are the synthesized beam major and minor FWHM, respectively. If the number of beams is greater than 1, we calculate the integrated flux density at 243\,GHz, or $1230\,\umu$m, as 
\begin{equation}
S_{1230} = S_{\mathrm{peak}} {\cal N}_{\mathrm{b}},
\end{equation}
\noindent
otherwise the integrated flux density is simply the peak pixel value. We also flag fits where the minor/major axis ratio is less than 0.5 as bad fits, and use peak pixel values for these sources. Uncertainties are taken from the noise map, and uncertainties from the fits are propagated to the integrated flux densities. All of our 1-mm flux densities are listed in Table~\ref{table:sources}. In this table we sort our sources by $S_{1230}$, using the prefix ALMA-HUDF.

As a check, we compare the 1-mm flux densities extracted here to the flux densities given in the four previously published surveys. For simplicity, in this comparison we simply match published sources to our new catalogue using a search radius of 1\,arcsec. The mean frequency of the map from \citet{Dunlop2017} is 221\,GHz (see Table~\ref{table:obs}), so we correct their flux densities to the mean frequency of our map (243\,GHz) assuming a modified blackbody SED with spectral index $\beta\,{=}\,1.8$, a dust temperature of 25\,K, and using each galaxy's spectroscopic/photometric redshift when available, otherwise a redshift of 1.5. Similarly, the mean frequency of the map from GOODS-ALMA \citep{Gomez-guijarro2022} is 265\,GHz, so we follow the same procedure and apply a correction factor of 0.8. The mean frequency of the remaining maps are effectively identical to ours, so we do not apply any further corrections.

\begin{figure}
\begin{center}
\includegraphics[width=0.5\textwidth]{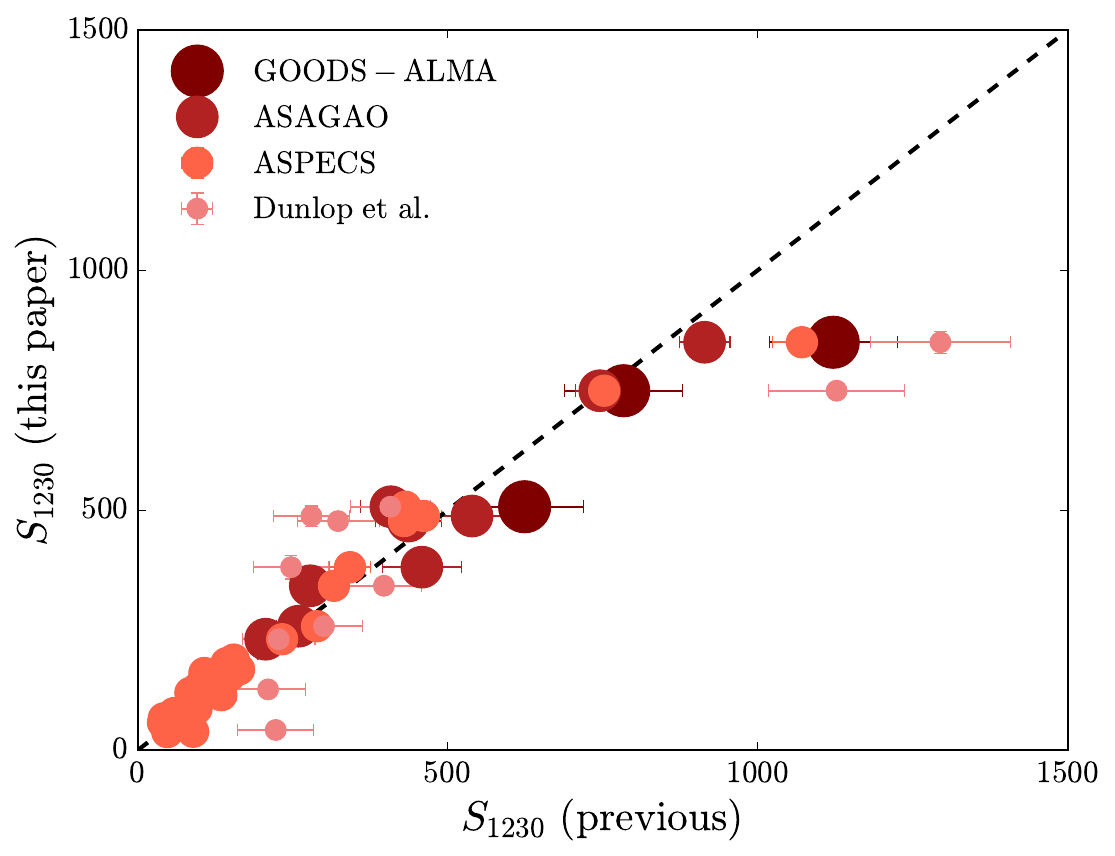}
\end{center}
\caption{Comparison of the flux densities measured in this work with flux densities from \citet{Dunlop2017}, ASPECS \citep{Gonzalez-lopez2020}, ASAGAO \citep{Hatsukade2018} and GOODS-ALMA \citep{Gomez-guijarro2022}.}
\label{fig:fluxes}
\end{figure}

The flux densities of matched peaks are shown in Fig.~\ref{fig:fluxes}, and the cross-matched IDs are given in Table~\ref{table:sources_ids}. We find generally good agreement with all of the previously-published flux densities after applying the above corrections. The flux densities from \citet{Dunlop2017} appear to have more scatter compared to later studies; however, those data were taken in ALMA's early observing cycles when calibration was more uncertain. We also see that the brightest source in our map (ALMA ID 1) is somewhat fainter than what was reported in the ASPECS and GOODS-ALMA surveys, but this source is right at the edge of the ASPECS primary beam (and thus our primary beam as well), so we expect that there may be additional systematic uncertainties that are not being taken into account for this source.

In Fig.~\ref{fig:sources} we show our detected sources compared to those found by \citet{Dunlop2017}, the ASAGAO survey \citep{Hatsukade2018} and the ASPECS survey \citep{Gonzalez-lopez2020}, as well as a few sources from the wider GOODS-ALMA survey \citep{Gomez-guijarro2022}. It appears that most of the detections in our combined map coincide with published sources, but there are 13 new DSFGs. Of these 13 new DSFGs, six have a peak S/N$\,{>}\,4.1$, meaning that they have been detected at a purity level of 100\,per cent. There are also a few published detections that do not appear here; typically, these are low-significance sources that could have been positive noise excursions. 

In Appendix~\ref{sec:appendix} we perform the same source extraction procedure on the larger but shallower map covering the ASAGAO footprint (excluding the central deep region, where we have already detected all of the sources in this shallow map), now fixing the detection threshold to a peal S/N$\,{>}\,4.5$ (as was done to make the ASAGAO catalogue). We find that the purity at this threshold is 0.9 (which is therefore fairly conservative) and we find a total of 27 additional galaxies, nine of which are new. We use a similar algorithm to measure flux densities, and find similar flux densities compared to those published by ASAGAO \citep{Hatsukade2018} and GOODS-ALMA \citep{Gomez-guijarro2022}. As a test, we also extract peaks from the central region of the shallow map and measured their flux densities, and find good agreement with the flux densities measured in our deep map.

\begin{table*}
\centering
\caption{Positions, peak S/N values, and flux densities ($S_{1230}$) for the 45 sources found in the ALMA 1-mm combined image at ${\rm S/N_{peak}}\,{>}\,3.6$. Here flux densities are measured by fitting Gaussian profiles to the sources in order to estimate the number of beams per source, then scaling the peak pixel value by the number of beams; for sources consistent with 1 beam or less, we set the flux density to be the peak pixel value. For the redshift column, galaxies with spectroscopic redshifts from \citet{Boogaard2023} are indicated by an asterisk and galaxies with spectroscopic redshifts from \citet{Rieke2023} are indicated by a dagger, which we use when fitting SEDs. For galaxies with matches in the JADES catalogue and sufficient photometry to fit SEDs (S/N$\,{>}\,$5 in the F356W band) we provide stellar masses and photometric redshifts (when no spectroscopic redshift is available) from McLeod et al. (in prep.). The median photometric redshift uncertainty of our ALMA sample is $\Delta z / (1\,{+}\,z)\,{\simeq}\,0.03$, while the relative uncertainties in stellar masses are small. Corresponding identifications in other published catalogues are given in Table~\ref{table:sources_ids}. ALMA ID 45 is at the position of a blended low-$z$ and high-$z$ galaxy. We assume that this ALMA source corresponds to the high-$z$ galaxy, the photometric redshift and stellar mass are derived using only long-wavelength {\it JWST\/} photometry where the high-$z$ galaxy is detected.}
\label{table:sources}
\begin{tabular}{lccccc}
\hline
Name & RA Dec [J2000] & S/N$_{\mathrm{peak}}$ & $S_{1230}$ [$\mu$Jy] & $z$ & $\log \left( M_{\ast} / \mathrm{M}_{\odot} \right)$\\
\hline
ALMA-HUDF-1 & 3:32:43.53 $-$27:46:39.2 & 36.4 & 850$\pm$23 & 2.696$^{\ast}$ & 10.1\\
ALMA-HUDF-2 & 3:32:38.54 $-$27:46:34.6 & 81.2 & 749$\pm$10 & 2.543$^{\ast}$ & \phantom{0}9.7\\
ALMA-HUDF-3 & 3:32:36.96 $-$27:47:27.2 & 47.8 & 507$\pm$12 & 2.47\phantom{0$^{\ast}$} & 10.7\\
ALMA-HUDF-4 & 3:32:39.75 $-$27:46:11.6 & 25.3 & 488$\pm$21 & 1.551$^{\ast}$ & 11.1\\
ALMA-HUDF-5 & 3:32:34.44 $-$27:46:59.8 & 31.7 & 477$\pm$17 & 1.414$^{\ast}$ & 10.7\\
ALMA-HUDF-6 & 3:32:40.08 $-$27:47:55.6 & 16.8 & 381$\pm$25 & 1.997$^{\ast}$ & 10.7\\
ALMA-HUDF-7 & 3:32:41.01 $-$27:46:31.6 & 30.0 & 342$\pm$13 & 2.454$^{\ast}$ & 10.5\\
ALMA-HUDF-8 & 3:32:43.33 $-$27:46:47.0 & 21.0 & 258$\pm$12 & 2.695$^{\ast}$ & 10.7\\
ALMA-HUDF-9 & 3:32:35.08 $-$27:46:47.8 & 21.2 & 231$\pm$12 & 2.580$^{\ast}$ & 10.7\\
ALMA-HUDF-10 & 3:32:35.56 $-$27:47:04.2 & 17.6 & 189$\pm$12 & 3.601$^{\ast}$ & 10.0\\
ALMA-HUDF-11 & 3:32:39.88 $-$27:47:15.2 & 10.0 & 181$\pm$19 & 1.095$^{\ast}$ & 10.5\\
ALMA-HUDF-12 & 3:32:38.03 $-$27:46:26.6 & 21.1 & 168$\pm$9\phantom{0} & 3.711$^{\ast}$ & 11.0\\
ALMA-HUDF-13 & 3:32:37.35 $-$27:46:45.8 & \phantom{0}8.7 & 161$\pm$20 & 1.845$^{\ast}$ & \phantom{0}9.9\\
ALMA-HUDF-14 & 3:32:35.51 $-$27:46:26.8 & \phantom{0}4.5 & 154$\pm$36 & 1.382$^{\ast}$ & 10.0\\
ALMA-HUDF-15 & 3:32:42.99 $-$27:46:50.2 & 15.7 & 134$\pm$9\phantom{0} & 1.037$^{\ast}$ & 10.7\\
ALMA-HUDF-16 & 3:32:36.48 $-$27:46:31.8 & 10.2 & 133$\pm$14 & 1.096$^{\ast}$ & \phantom{0}9.4\\
ALMA-HUDF-17 & 3:32:38.80 $-$27:47:14.8 & \phantom{0}8.6 & 130$\pm$16 & 1.848$^{\ast}$ & 10.5\\
ALMA-HUDF-18 & 3:32:42.37 $-$27:47:07.8 & 10.8 & 126$\pm$13 & 1.317$^{\ast}$ & 11.0\\
ALMA-HUDF-19 & 3:32:36.19 $-$27:46:28.0 & \phantom{0}9.0 & 119$\pm$14 & 2.574$^{\ast}$ & 10.6\\
ALMA-HUDF-20 & 3:32:38.75 $-$27:48:10.4 & \phantom{0}6.0 & 114$\pm$19 & 2.823$^{\ast}$ & \phantom{0}9.8\\
ALMA-HUDF-21 & 3:32:41.69 $-$27:46:55.6 & 11.7 & \phantom{0}98$\pm$8\phantom{0} & 1.996$^{\ast}$ & 10.3\\
ALMA-HUDF-22 & 3:32:34.86 $-$27:46:40.8 & \phantom{0}5.5 & \phantom{0}96$\pm$19 & 1.098$^{\ast}$ & 10.2\\
ALMA-HUDF-23 & 3:32:41.04 $-$27:47:48.0 & \phantom{0}4.0 & \phantom{0}89$\pm$22 & \dots & \phantom{0}\dots\\
ALMA-HUDF-24 & 3:32:35.78 $-$27:46:27.6 & \phantom{0}6.0 & \phantom{0}86$\pm$16 & 1.093$^{\ast}$ & 10.4\\
ALMA-HUDF-25 & 3:32:41.24 $-$27:46:16.6 & \phantom{0}3.7 & \phantom{0}81$\pm$22 & \dots & \phantom{0}\dots\\
ALMA-HUDF-26 & 3:32:34.71 $-$27:46:45.0 & \phantom{0}3.8 & \phantom{0}78$\pm$22 & 1.552$^{\ast}$ & 10.1\\
ALMA-HUDF-27 & 3:32:35.99 $-$27:47:25.8 & \phantom{0}5.8 & \phantom{0}77$\pm$15 & 2.643$^{\ast}$ & 10.3\\
ALMA-HUDF-28 & 3:32:34.82 $-$27:46:31.2 & \phantom{0}3.7 & \phantom{0}76$\pm$21 & \dots & \phantom{0}\dots\\
ALMA-HUDF-29 & 3:32:38.50 $-$27:47:02.6 & \phantom{0}4.7 & \phantom{0}66$\pm$15 & 0.948$^{\ast}$ & 11.0\\
ALMA-HUDF-30 & 3:32:37.61 $-$27:47:44.0 & \phantom{0}6.3 & \phantom{0}66$\pm$10 & 1.542$^{\ast}$ & \phantom{0}9.7\\
ALMA-HUDF-31 & 3:32:41.45 $-$27:47:29.2 & \phantom{0}4.1 & \phantom{0}58$\pm$14 & 0.621$^{\dagger}$ & \phantom{0}9.0\\
ALMA-HUDF-32 & 3:32:37.73 $-$27:47:07.2 & \phantom{0}4.4 & \phantom{0}57$\pm$14 & 0.667$^{\ast}$ & \phantom{0}9.8\\
ALMA-HUDF-33 & 3:32:38.59 $-$27:47:30.4 & \phantom{0}4.1 & \phantom{0}54$\pm$14 & 2.642$^{\dagger}$ & \phantom{0}9.6\\
ALMA-HUDF-34 & 3:32:40.23 $-$27:47:38.2 & \phantom{0}4.3 & \phantom{0}50$\pm$11 & \dots & \phantom{0}\dots\\
ALMA-HUDF-35 & 3:32:35.75 $-$27:46:39.4 & \phantom{0}3.8 & \phantom{0}46$\pm$14 & 2.07\phantom{0$^{\ast}$} & \phantom{0}9.8\\
ALMA-HUDF-36 & 3:32:35.77 $-$27:46:55.4 & \phantom{0}3.9 & \phantom{0}42$\pm$12 & 1.721$^{\dagger}$ & \phantom{0}9.8\\
ALMA-HUDF-37 & 3:32:38.56 $-$27:47:05.4 & \phantom{0}4.4 & \phantom{0}41$\pm$9\phantom{0} & \dots & \phantom{0}\dots\\
ALMA-HUDF-38 & 3:32:41.83 $-$27:46:56.8 & \phantom{0}4.9 & \phantom{0}38$\pm$8\phantom{0} & 1.999$^{\ast}$ & 10.1\\
ALMA-HUDF-39 & 3:32:39.78 $-$27:46:29.4 & \phantom{0}3.8 & \phantom{0}38$\pm$11 & \dots & \phantom{0}\dots\\
ALMA-HUDF-40 & 3:32:37.08 $-$27:46:17.4 & \phantom{0}6.0 & \phantom{0}38$\pm$7\phantom{0} & 2.227$^{\ast}$ & \phantom{0}9.0\\
ALMA-HUDF-41 & 3:32:38.69 $-$27:46:30.8 & \phantom{0}3.9 & \phantom{0}34$\pm$9\phantom{0} & \dots & \phantom{0}\dots\\
ALMA-HUDF-42 & 3:32:43.62 $-$27:46:59.0 & \phantom{0}4.1 & \phantom{0}33$\pm$8\phantom{0} & 1.569$^{\dagger}$ & \phantom{0}9.7\\
ALMA-HUDF-43 & 3:32:36.37 $-$27:46:50.2 & \phantom{0}3.9 & \phantom{0}32$\pm$8\phantom{0} & \dots & \phantom{0}\dots\\
ALMA-HUDF-44 & 3:32:41.90 $-$27:46:58.6 & \phantom{0}4.2 & \phantom{0}32$\pm$8\phantom{0} & \dots & \phantom{0}\dots\\
ALMA-HUDF-45 & 3:32:42.35 $-$27:46:57.4 & \phantom{0}4.3 & \phantom{0}30$\pm$7\phantom{0} & 2.42\phantom{0$^{\ast}$} & \phantom{0}9.9\\
\hline
\end{tabular}
\end{table*}

\begin{table*}
\centering
\caption{Cross-matched identifications between our catalogue of sources and previous studies. Near-infrared matches are from JADES \citep{Rieke2023} and CANDELS \citep{Guo2013}. Matches with other ALMA Band-6 surveys are from \citet{Dunlop2017}, ASAGAO \citep{Hatsukade2018}, GOODS-ALMA \citep{Gomez-guijarro2022}, and ASPECS \citep{Gonzalez-lopez2020}. ALMA ID 44 is not detected in our F356W-selected catalogue, but has a counterpart in the JADES catalogue, which we provide below. The JADES ID corresponding to ALMA ID 45 is a blend of a low-$z$ and a high-$z$ galaxy, and we assume that our ALMA source corresponds to the galaxy at high redshift.}
\label{table:sources_ids}
\begin{tabular}{lcccccc}
\hline
Name & JADES ID & CANDELS ID & Dunlop et al. ID & ASAGAO ID & GOODS-ALMA ID & ASPECS ID\\
\hline
ALMA-HUDF-1 & 208820 & J033243.52$-$274639.0 & UDF2 & 4 & A2GS9 & C06\\
ALMA-HUDF-2 & 209117 & J033238.54$-$274634.0 & UDF3 & 5 & A2GS25 & C01\\
ALMA-HUDF-3 & 204232 & J033236.96$-$274727.2 & UDF5 & 12 & A2GS41 & C02\\
ALMA-HUDF-4 & 211273 & J033239.73$-$274611.2 & UDF8 & 16 & \dots & C05\\
ALMA-HUDF-5 & 207012 & J033234.43$-$274659.5 & UDF6 & 13 & \dots & C03\\
ALMA-HUDF-6 & 202563 & J033240.05$-$274755.4 & UDF11 & 15 & \dots & C10\\
ALMA-HUDF-7 & 209357 & J033241.02$-$274631.4 & UDF4 & 10 & \dots & C04\\
ALMA-HUDF-8 & 208030 & J033243.32$-$274646.7 & UDF7 & 14 & \dots & C11\\
ALMA-HUDF-9 & 208000 & J033235.07$-$274647.5 & UDF13 & 23 & \dots & C07\\
ALMA-HUDF-10 & 206834 & J033235.55$-$274703.8 & \dots & \dots & \dots & C09\\
ALMA-HUDF-11 & 205449 & J033239.88$-$274715.0 & \dots & \dots & \dots & C16\\
ALMA-HUDF-12 & 209777 & J033238.02$-$274626.2 & \dots & \dots & \dots & C08\\
ALMA-HUDF-13 & 208134 & J033237.35$-$274645.4 & \dots & \dots & \dots & C18\\
ALMA-HUDF-14 & 209492 & J033235.48$-$274626.6 & \dots & \dots & \dots & C23\\
ALMA-HUDF-15 & 207739 & J033242.98$-$274649.9 & \dots & \dots & \dots & C13\\
ALMA-HUDF-16 & 209285 & J033236.44$-$274631.5 & \dots & \dots & \dots & C12\\
ALMA-HUDF-17 & 205379 & J033238.79$-$274714.7 & \dots & \dots & \dots & C17\\
ALMA-HUDF-18 & 206183 & J033242.37$-$274707.6 & UDF16 & \dots & \dots & C15\\
ALMA-HUDF-19 & 209617 & J033236.17$-$274627.6 & \dots & \dots & \dots & C19\\
ALMA-HUDF-20 & 201501 & J033238.72$-$274810.3 & \dots & \dots & \dots & C24\\
ALMA-HUDF-21 & 207227 & J033241.68$-$274655.4 & \dots & \dots & \dots & C14a\\
ALMA-HUDF-22 & 208277 & J033234.85$-$274640.4 & \dots & \dots & \dots & C25\\
ALMA-HUDF-23 & \dots & \dots & \dots & \dots & \dots & \dots\\
ALMA-HUDF-24 & 209480 & J033235.77$-$274627.4 & \dots & \dots & \dots & C20\\
ALMA-HUDF-25 & \dots & \dots & \dots & \dots & \dots & \dots\\
ALMA-HUDF-26 & 208267 & J033234.67$-$274644.5 & \dots & \dots & \dots & C26\\
ALMA-HUDF-27 & 204579 & J033235.98$-$274725.6 & \dots & \dots & \dots & C21\\
ALMA-HUDF-28 & 129574 & \dots & \dots & \dots & \dots & \dots\\
ALMA-HUDF-29 & 206703 & J033238.48$-$274702.4 & \dots & \dots & \dots & C33\\
ALMA-HUDF-30 & 203384 & J033237.61$-$274744.0 & \dots & \dots & \dots & C22\\
ALMA-HUDF-31 & 204483 & J033241.45$-$274729.3 & \dots & \dots & \dots & \dots\\
ALMA-HUDF-32 & 206205 & J033237.73$-$274706.9 & \dots & \dots & \dots & C32\\
ALMA-HUDF-33 & 204449 & J033238.55$-$274730.2 & \dots & \dots & \dots & \dots\\
ALMA-HUDF-34 & \dots & \dots & \dots & \dots & \dots & C27\\
ALMA-HUDF-35 & 208812 & J033235.73$-$274639.0 & \dots & \dots & \dots & \dots\\
ALMA-HUDF-36 & 124908 & J033235.76$-$274655.0 & UDF15 & \dots & \dots & \dots\\
ALMA-HUDF-37 & \dots & \dots & \dots & \dots & \dots & \dots\\
ALMA-HUDF-38 & 207221 & J033241.83$-$274657.0 & \dots & \dots & \dots & C14b\\
ALMA-HUDF-39 & 265959 & J033239.76$-$274629.3 & \dots & \dots & \dots & \dots\\
ALMA-HUDF-40 & 210730 & J033237.07$-$274617.1 & \dots & \dots & \dots & C31\\
ALMA-HUDF-41 & \dots & \dots & \dots & \dots & \dots & \dots\\
ALMA-HUDF-42 & 207079 & J033243.61$-$274658.7 & \dots & \dots & \dots & \dots\\
ALMA-HUDF-43 & \dots & \dots & \dots & \dots & \dots & \dots\\
ALMA-HUDF-44 & 267661 & \dots & \dots & \dots & \dots & \dots\\
ALMA-HUDF-45 & 207277 & J033242.35$-$274657.0 & \dots & \dots & \dots & \dots\\
\hline
\end{tabular}
\end{table*}

\subsection{Matching mm-selected sources to near-IR-selected sources}

The expected uncertainty in our ALMA positions is $\delta \mathrm{RA} = \delta \mathrm{Dec} = 0.6\,{\times}\,{\rm FWHM}\,{\div}\,{\rm (S/N)}$ \citep{Ivison2007}, so we expect the $1\,\sigma$ radial uncertainty to be $\delta r \,{\approx}\,$0.76\,arcsec/(S/N) (using the geometric mean of the elliptical ALMA beam). Since the probability density of finding a source a distance $r$ from its true position is proportional to $r \mathrm{e}^{-r^{2}/2\delta r^{2}}$, one must go out to a distance of 2.5$\delta r$ in order to find a correct match with 95\,per cent certainty. For a given ALMA detection, we thus search the JADES catalogue out to a distance of 1.9\,arcsec/(S/N), where S/N here is the peak S/N of each source found above our detection threshold. For the most significant ALMA sources this search radius is unphysically small (much less than 1 pixel), so we apply a minimal search radius of 0.3\,arcsec. 

We preform a similar counterpart search with the CANDELS catalogue, including the deep infrared data from the HUDF09 survey \citep{Bouwens2011}. Here, we simply apply a uniform search radius of 0.6\,arcsec, as we found that the {\it HST\/} F160W morphologies of our DSFGs were often clumpy, leading to unrealistic offsets with respect to our ALMA centroids. We checked by eye that all identified CANDELS counterparts were indeed the same galaxy as the JADES counterparts. It should be noted that \citet{Dunlop2017} found a systematic offset between the CANDELS and ALMA astrometry, which was resolved by applying an offset of about 0.25\,arcsec south to the CANDELS positions; the same offset was applied here. In Table~\ref{table:sources_ids} we provide the JADES and CANDELS IDs for these matches, and in Appendix~\ref{sec:appendixB} we show the positions of matched sources in our ALMA cutouts overlaid over {\it JWST\/} F356W imaging.

At the position of our ALMA source ID 45 we found that there were two blended galaxies in the longer-wavelength {\it JWST\/} images, yet at shorter {\it JWST\/} wavelengths one of these galaxies was undetected. In the JADES catalogue there was only one ID for these two sources (ID 207277) that had a spectroscopic redshift of 0.332. This spectroscopic redshift likely corresponds to the galaxy brighter at short wavelengths, while our ALMA source is probably the second galaxy that is only detected at longer wavelengths. We therefore use the longer wavelength photometry (where this galaxy is detected) to fit an SED, which results in a photometric redshift of 2.42. Throughout the rest of the paper we report results derived from this fit. 

Lastly, our nominal search radius did not return a counterpart to ALMA ID 38, yet this source is located slightly less than 0.5\,arcsec from a $z\,{=}\,1.998$, $M_{\ast}\,{\simeq}\,10^{10}\,$M$_{\odot}$ JADES galaxy (the JADES ID is 207221). Moreover, the spectroscopic redshift of this source was independently determined to be $z\,{=}\,1.999$ by \citet{Boogaard2023}, nearly equal to that reported by JADES, so we assign JADES ID 207221 as the counterpart. There is another ALMA source (ID 44) located less than 2\,arcsec south of source 38, whose closest {\it JWST\/} counterpart is in the JADES catalogue and not in our F356W-selected catalogue (the JADES ID is 267661) since it is very close to JADES ID 207221 and suffered blending issues. The reported JADES photometric redshift for ID 44 is 2.02, very close to the spectroscopic redshift of the JADES galaxy found near ID 38, thus it is likely that they are at the same redshift and are undergoing a merger, resulting in complicated morphologies that could easily lead to large offsets between our ALMA centroids and the corresponding JADES positions. We therefore classify ALMA ID 44 as having a JADES counterparts, although we do not fit an SED to it due to the blending issues.

Sources found to have counterparts within their given search radii in both the CANDELS catalogue and the JADES catalogue are marked in Fig.~\ref{fig:sources} with blue crosses. There are 37 ALMA-detected galaxies with counterparts in both catalogues, seven of which are ALMA sources that have not been previously reported. There are an additional two ALMA-detected galaxies with a counterpart only in the JADES catalogue, both of which are ALMA sources that have not been previously reported. We do not find any CANDELS counterparts with no corresponding JADES counterpart, which highlights one of the benefits of the deeper catalogue of the HUDF made possible with {\it JWST} (although the improvement is not dramatic, since CANDELS gets close to identifying all of our mm sources). 

This leaves six ALMA-detected sources with no optical or near-infrared counterpart; one of these has previously-published ALMA identifications, the remaining five do not. It is worth pointing out that two of the sources with no counterparts are close to the edge of the map where the noise is the largest (IDs 23 and 25), but the other four are found near the centre of the map where the data are much deeper (IDs 34, 37, 41 and 43).

The number of beams in our ALMA map can be estimated as the area of the map (4.2\,arcmin$^2$) divided by the beam area ($2\pi\theta_{\rm maj}\theta_{\rm min} / 8\ln2$), which results in about 8500. From Gaussian statistics, we would expect around 1.5 beams to randomly have a significance greater than $3.6\,\sigma$; however, there are really twice as many statistically independent noise samples due to correlations with the beam \citep[e.g.,][]{Condon1997,Condon1998,Dunlop2017}, so really we would expect three false positive detections. The measured number of negative peaks more significant than 3.6 is 14, which is larger than the expectation from a Gaussian distribution -- the negative peaks might not be perfectly Gaussian distributed, but this is not surprising given the non-uniform $uv$ coverage. On the other hand, the number of ALMA-detected sources with no optical or near-infrared counterpart (six) is more comparable to the number of negative peaks, and so we cannot rule out the possibility that some of them are false positives.

\begin{figure*}
\begin{center}
\includegraphics[width=0.75\textwidth]{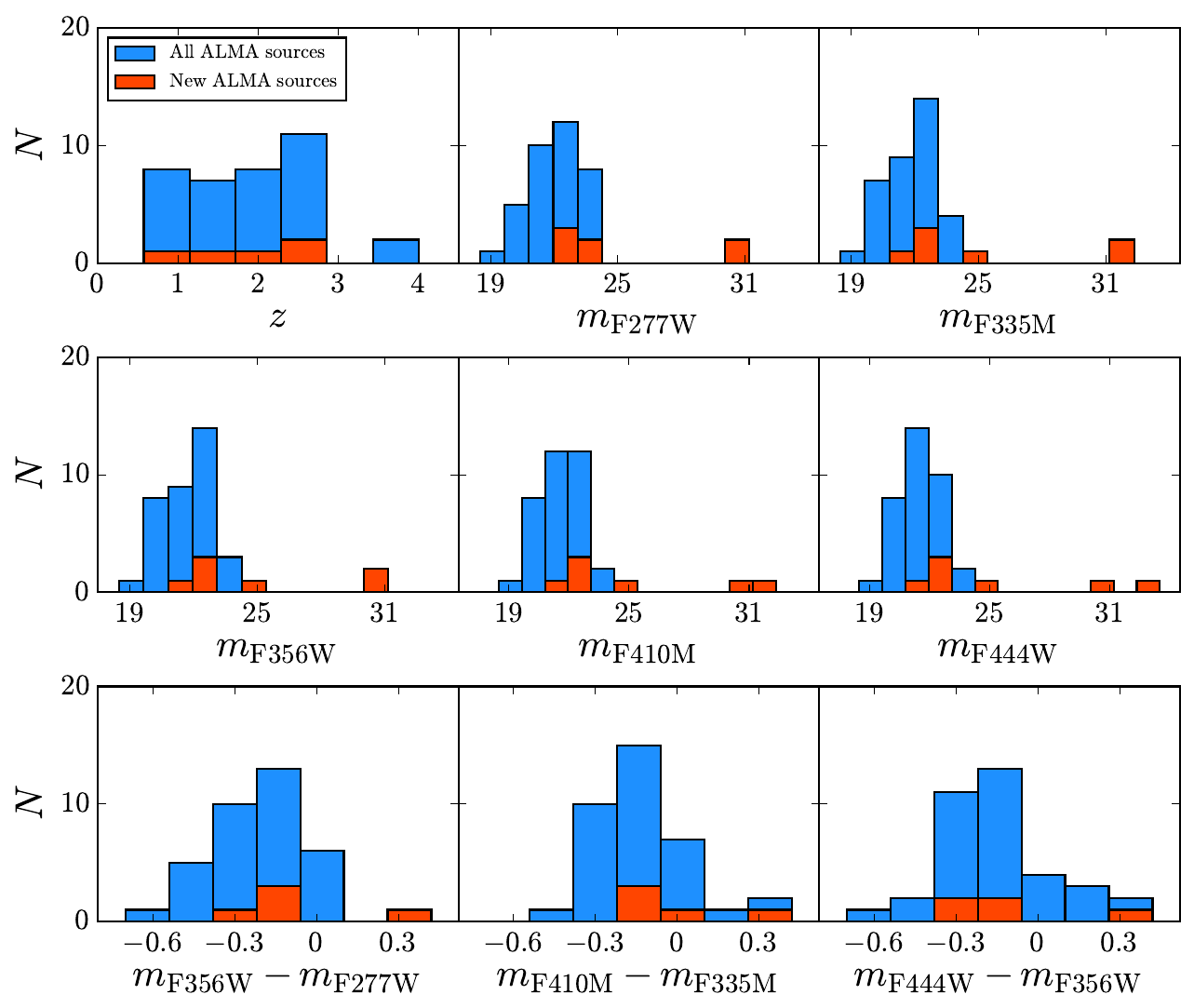}
\end{center}
\caption{Distribution of ALMA source properties matched to our JADES catalogue (\citealt{Rieke2023}, McLeod et al.\ in prep.). The first (top left) panel shows the distribution of redshifts (spectroscopic when available, otherwise photometric). The next five panels show distributions of NIRCam magnitudes for the ALMA sources in the JADES catalogue. The final three panels show distributions of three selected NIRCam colours. The distributions from all 39/45 ALMA galaxies with {\it JWST\/} detections are shown in blue, while the distributions from the subset of eight new ALMA galaxies with {\it JWST\/} detections are shown in red. In the redshift panel (top left) there are only 36 ALMA galaxies with available redshifts, and five new ALMA galaxies with redshifts. In the bottom row we also only plot the five galaxies with available redshifts because the other three have quite uncertain colours.}
\label{fig:histograms}
\end{figure*}

In Fig.~\ref{fig:histograms} we show the distributions of redshifts, magnitudes, and colours from our JADES catalogue. For the redshifts and colours, we filter out two galaxies which have a S/N$\,{<}\,$5 in the F356W band and ALMA ID 44 due to blending issues as the photometry is not reliable. For the remaining galaxies, the median redshift uncertainty is $\Delta z / (1\,{+}\,z)\,{\simeq}\,0.03$, so we expect them to be accurate. We display the distributions of all 39/45 ALMA galaxies with {\it JWST\/} detections and also highlight the distributions of the eight ALMA galaxies that were not found in previous surveys. The first (top left) panel in Fig.~\ref{fig:histograms} shows the distribution of redshifts (spectroscopic where available, otherwise photometric). We see that the redshift distribution is flat around $z\,{\simeq}\,2$, and our new galaxies appear to follow this trend. Next we show the distributions of the magnitudes of the sources in the NIRCam images (F277W, F335M, F356W, F410M and F444W). We see that most ALMA sources have magnitudes ranging from 19 to 25, with the two faintest sources extending out to ${>}\,$30 (these are the two sources that we do not fit SEDs to). Lastly we show distributions for three NIRCam colours; most ALMA galaxies are fairly red (i.e.\ brighter at longer wavelengths), and our new, fainter ALMA galaxies tend to follow the same distribution.

\subsection{Stellar mass-redshift distribution}

As described in Section~\ref{M_star-z_data}, the photometric redshifts and stellar masses for most JADES galaxies have been estimated using the extensive multiwavelength imaging available for the HUDF. It is therefore of interest to see how the photometric redshifts and stellar masses of our mm-selected galaxies compare to the typical galaxies found in this field. For our list of sources with a JADES counterpart, we filter out all sources with S/N$\,{<}\,$5 in the F356W band, since these sources will not have reliable SED fits, and ALMA ID 44, which is heavily blended with ALMA ID 38. For the remaining sources, we check to see if there is a spectroscopic redshift, and otherwise use the photometric redshift.

\begin{figure*}
\begin{center}
\includegraphics[width=\textwidth]{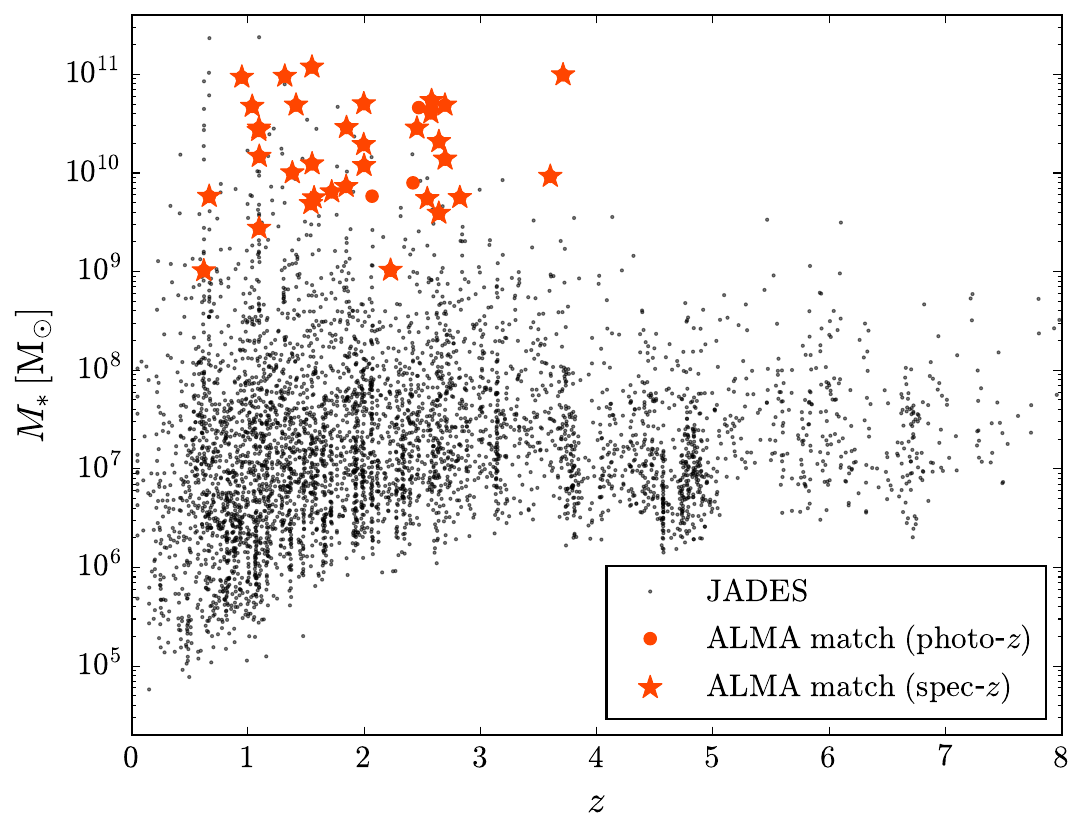}
\end{center}
\caption{Stellar mass versus redshift for JADES galaxies (McLeod et al.\ in prep.), with JADES galaxies detected in our $S_{1230}$ image highlighted in red (circles indicate photometric redshifts, stars indicate spectroscopic redshifts). The apparent vertical features are a result of the grid used for the photometric redshift fitting.}
\label{fig:z_vs_Mstar}
\end{figure*}

In Fig.~\ref{fig:z_vs_Mstar} we plot stellar mass versus redshift for all JADES galaxies (McLeod et al.\ in prep.), and highlight our ALMA Band-6-selected sources with good SED fits in red. We find that most of our ALMA-selected sources are high-stellar-mass galaxies between $z\,{=}\,1$ and 3, similar to what was found in earlier ALMA surveys of the HUDF \citep[e.g.,][]{Dunlop2017,McLure2018,Aravena2020}. In particular, there are 53 galaxies with $M_{\ast}\,{>}\,10^{10}\,$M$_{\odot}$, 22 of which have been selected in our ALMA image. Also of note is that at 1.23\,mm we are sensitive primarily to $z\,{>}\,1$ objects, since in our sample of 36 galaxies with spectroscopic or photometric redshifts, only three are at $z\,{<}\,1$.  In a similar vein, it may also be worth noting that all of the $\log(M_{\ast}/{\rm M}_{\odot})\,{>}\,10.3$ galaxies at $2\,{<}\,z\,{<}\,3$ are detected in our ALMA image.

\begin{figure*}
\begin{center}
\includegraphics[width=\textwidth]{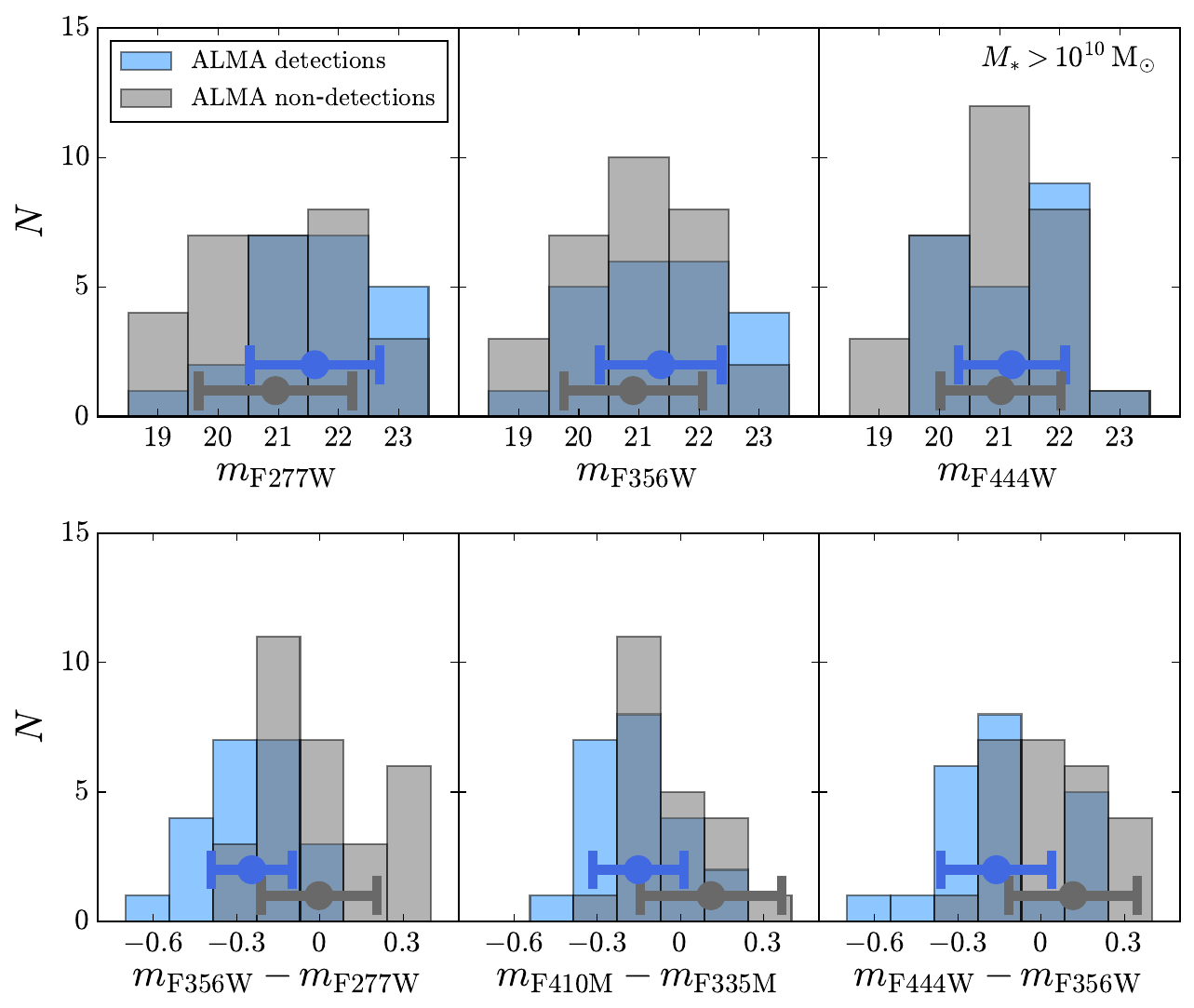}
\end{center}
\caption{{\it Top row:} Distributions of three select NIRCam magnitudes for all ALMA-detected galaxies with {\it JWST\/} counterparts and with stellar masses above 10$^{10}\,$M$_{\odot}$, compared to all {\it JWST\/} galaxies with stellar masses above 10$^{10}\,$M$_{\odot}$ that are not detected by ALMA. {\it Bottom row:} Same as top row but for three select NIRCam colours. In all panels, the means and standard deviations of each subsample are shown as the points with error bars.
While we do not see a significant difference between the two samples in terms of magnitude, the mm-bright galaxies detected by ALMA with high stellar masses tend to have redder NIRCam colours compared to ALMA-undetected galaxies with equally large stellar masses.}
\label{fig:highMstar_histograms}
\end{figure*}

In order to investigate the difference between galaxies with $M_{\ast}\,{>}\,10^{10}\,$M$_{\odot}$ that we have detected with ALMA versus those that we have not detected with ALMA, in Fig.~\ref{fig:highMstar_histograms} we show the distributions of three select NIRCam magnitudes and three NIRCam colours for the subsamples of 22 ALMA-detected galaxies and 31 ALMA-undetected galaxies. We find that there is no discernible difference in magnitudes between the two samples; however, mm-bright ALMA sources tend to have redder NIRCam colours than the galaxies with similarly high stellar masses that we have {\it not\/} detected with ALMA.

\subsection{Stacking on near-IR-selected positions}\label{sec:stacking}

One major benefit of the deep ancillary catalogues available in the HUDF is the ability to stack on the positions of undetected galaxies in different stellar mass and redshift bins, thus estimating the statistical properties of fainter mm-emitting sources that are not individually detectable in our image. Such an analysis was done using the original ASPECS map and {\it HST\/} catalogue \citep{magnelli2020}, but here the major improvements are a deeper 1-mm map and a deeper catalogue of infrared-selected galaxies from {\it JWST}.

To do this, we follow a procedure similar to \textsc{Simstack} \citep{Viero2013}.  However, since we adapt this for a non-circular beam and non-uniform noise distribution, it is worth describing what we do in a little detail. First, we mask out all galaxies that we have detected in our ALMA map (see Table~\ref{table:sources}), with a source mask set to be $3\times{\rm FWHM}$ in diameter. We also restrict this stacking analysis to the deep central map shown in Fig.~\ref{fig:sources}. Then we subtract the weighted mean of the image -- this is crucial, since the `stack' is really the covariance between a map and catalogue \citep[see section~3 of][]{Marsden2009} and will give a biased result unless the map has zero mean.

Next, we define a grid of four stellar mass bins, logarithmically spaced between $\log \left( M_{\ast}/\mathrm{M}_{\odot} \right) \,{=}\,7.4$ and $\log \left( M_{\ast}/\mathrm{M}_{\odot} \right) \,{=}\,11.4$, with a width of $\Delta \log \left( M_{\ast}/\mathrm{M}_{\odot} \right) \,{=}\,1$. We also define a grid of four redshift bins between 0 and 8, with a width of 2. For the stacking catalogue, we again turn to JADES, using stellar masses and redshifts given by McLeod et al. (in prep.); here redshifts are spectroscopic if available, otherwise photometric. These bins are chosen because they are expected to contain the galaxies comprising the vast majority of the CIB (see Section~\ref{sec:consistency} for more details).

For each redshift and stellar mass bin we produce a `hits' map, which is simply a copy of our ALMA map with pixel values set to the number of JADES galaxies within the bin contained within each pixel. We convolve each hits map with the ALMA beam, thereby producing a model image of the sky where the only free parameter of each map is its amplitude, which in this case can be interpreted as the best-fit flux density of the galaxies within the given stellar mass and redshift bin. As was done with the data, we also subtract the weighted mean of the convolved maps.

As described in \citet{Viero2013}, in general there are correlations between redshift bins simply due to the presence of large-scale structure, and to take these into account one would have to fit for all of the amplitudes in the defined bins simultaneously. In our case, redshift bins have a width of 2, so these correlations are expected to be negligible. We therefore fit for the amplitude in each bin independently, which reduces to solving a simple weighted linear regression between the ALMA map and maps of `hits' in the JADES catalogue for each bin. The solution is
\begin{equation}\label{eq:best-fit-S}
    \hat{S}_{\alpha}=\frac{\sum_j w_j D_j N^j_{\alpha}}{\sum_j w_j (N^j_{\alpha})^2},
\end{equation}
\noindent
where $j$ labels each map pixel, $\alpha$ denotes the redshift/stellar mass bin, $N_{\alpha}$ is the hits map of bin $\alpha$ convolved with the ALMA beam with the weighted mean subtracted, $D$ is the data map with the weighted mean subtracted and $w_j\,{=}\,1/\sigma_j^2$ are the inverse-variance weights. The sum is performed over all the pixels in the map. Equation~\ref{eq:best-fit-S} is effectively the covariance between the map and the catalogue, but with the ALMA beam taken into account.

To estimate the uncertainties in $\hat{S}_{\alpha}$, for each redshift and stellar mass bin we generate 1000 catalogues with the same number of sources in each bin but with random positions, and evaluate Eq.~\ref{eq:best-fit-S} for each one. We find the distribution of values to be well-described by a Gaussian, so we take the standard deviation of the random catalogue flux densities to be the $1\,\sigma$ error in $\hat{S}_{\alpha}$.
In order to visually show our results, we also evaluate Eq.~\ref{eq:best-fit-S} after shifting the hits map $N_{\alpha}$ relative to the data map $D$ within boxes of 10\,arcsec; this is effectively the cross-correlation between the two maps, where the central pixel is the zero-lag cross-correlation (or the covariance), which is the value we are most interested in.

\begin{figure*}
\begin{center}
\includegraphics[width=0.70\textwidth]{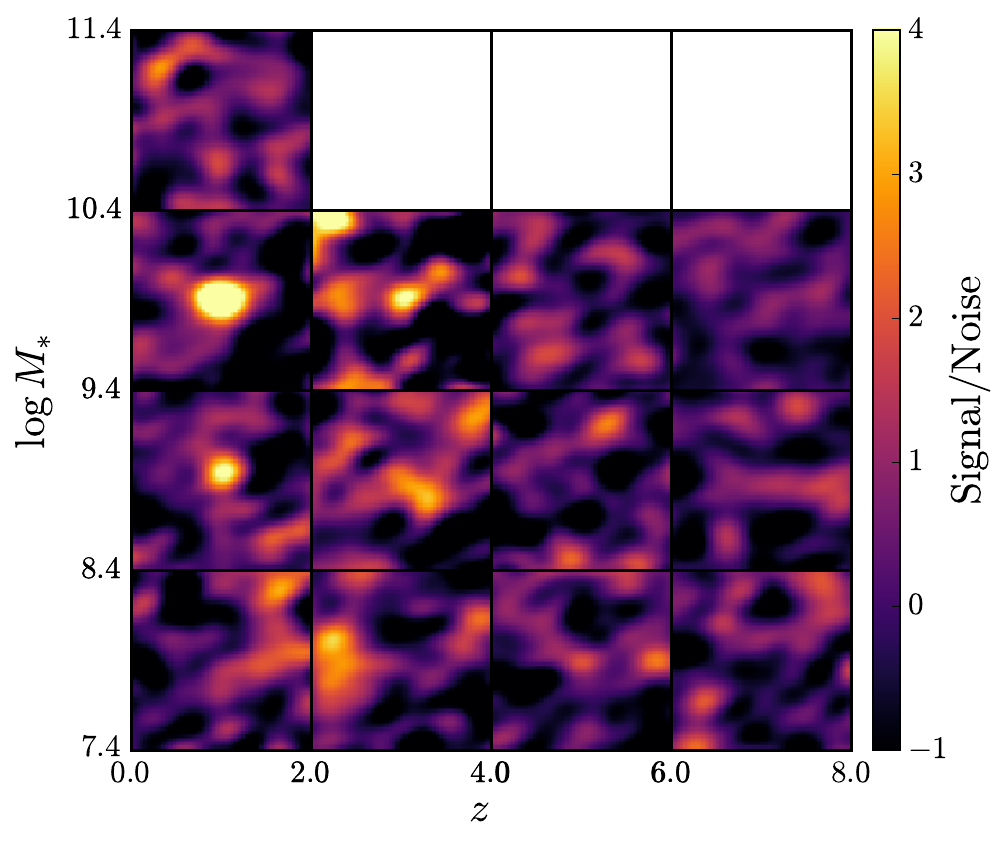}
\includegraphics[trim=1.1cm 0 -1.1cm 0,width=0.60\textwidth]{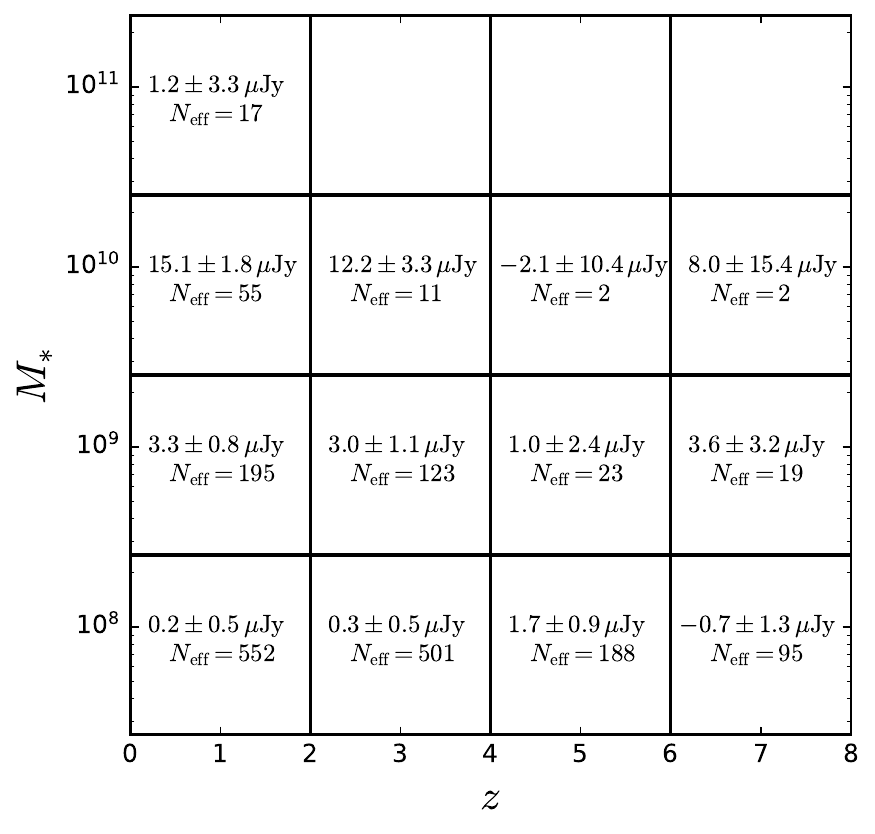}
\end{center}
\caption{{\it Top:} Stacks at the positions of JADES galaxies in redshift and stellar mass (from McLeod et al.\ in prep.) bins, after masking all of our detected ALMA sources. The cutouts are 10\,arcsec$\,{\times}\,10\,$arcsec. The colour represents S/N, where the signal is the variance-weighted mean and the noise is the uncertainty in the variance-weighted mean. {\it Bottom:} Central pixel values and uncertainties of the top panel (i.e.\ our best estimate of the average 1230-$\umu$m flux density of galaxies within each bin), with the number of JADES galaxies contributing to each bin also given.  For a definition of the quantity $N_{\rm eff}$, see Eq.~\ref{eq:Neff}. The $z\,{=}\,$0--2, $M_{\ast}\,{=}\,10^{9.4}$--$10^{10.4}$ bin can be split into four smaller sub-bins: for $z\,{=}\,$0--1, $M_{\ast}\,{=}\,10^{9.9}$--$10^{10.4}$M$_{\odot}$, we find $S_{\nu}\,{=}\,(22.1\pm7.9)\,\mu$Jy and $N_{\rm eff}\,{=}\,4$; for $z\,{=}\,$0--1, $M_{\ast}\,{=}\,10^{8.9}$--$10^{9.4}$M$_{\odot}$, we find $S_{\nu}\,{=}\,(15.3\pm3.6)\,\mu$Jy and $N_{\rm eff}\,{=}\,15$; for $z\,{=}\,$1--2, $M_{\ast}\,{=}\,10^{9.9}$--$10^{10.4}$M$_{\odot}$, we find $S_{\nu}\,{=}\,(18.2\pm3.8)\,\mu$Jy and $N_{\rm eff}\,{=}\,11$; and for $z\,{=}\,$1--2, $M_{\ast}\,{=}\,10^{8.9}$--$10^{9.4}$M$_{\odot}$. we find $S_{\nu}\,{=}\,(12.1\pm2.9)\,\mu$Jy and $N_{\rm eff}\,{=}\,24$.}
\label{fig:stack}
\end{figure*}

The top panel of Figure~\ref{fig:stack} shows our results from the cross-correlation, while in the bottom panel we list the central pixel values, which are the best-fit flux densities of the galaxies within each bin. For discussing the CIB contributions we want to know the pixel sum of the best-fit average surface brightness for each bin (weighted by the inverse-variance), which we calculate as
\begin{equation}
\begin{split}
    \hat{I}_{\alpha} & =  \frac{\hat{S}_{\alpha}}{2 \pi \sigma_{\rm maj} \sigma_{\rm min}} \frac{\sum_j w_j N_{\alpha}^j}{\sum_j w_j} \\
    & = \frac{\hat{S}_{\alpha} N_{\alpha,\mathrm{eff}}}{A}.
\end{split}
\end{equation}
\noindent
Here $\sigma_{\rm maj}$ and $\sigma_{\rm min}$ are the beam major and minor axes (in standard deviation units), respectively, the quantity $A$ is the solid angle of the map and $N_{\alpha,\mathrm{eff}}$ is the effective total number of sources from catalogue $\alpha$ in the map, weighted by the noise,
\begin{equation}
    N_{\alpha,\mathrm{eff}} = \frac{\sum_j w_j N_{\alpha}^j}{\sum_j w_j} \frac{A}{2 \pi \sigma_{\rm maj} \sigma_{\rm min}}.
    \label{eq:Neff}
\end{equation}
\noindent
This is just the average number of sources per beam (weighted by the noise), times the number of beams in the map. In Fig.~\ref{fig:stack} we provide the values of $N_{\alpha,\mathrm{eff}}$ along with $\hat{S}_{\alpha}$. Note that when we calculate the weighted mean surface brightness of the sky from our model, the weighted mean should not be subtracted from $N_{\alpha}^j$.

It is worth noting that a simpler stacking technique (just summing pixel values at the positions of {\it JWST} galaxies; see \citealt{Marsden2009}) produces similar results, merely with slightly less significance. Furthermore, we also tried stacking directly on the dirty map, but we found that the resulting flux densities changed by ${<}1$\,per cent.

\begin{figure}
\begin{center}
\includegraphics[width=0.5\textwidth]{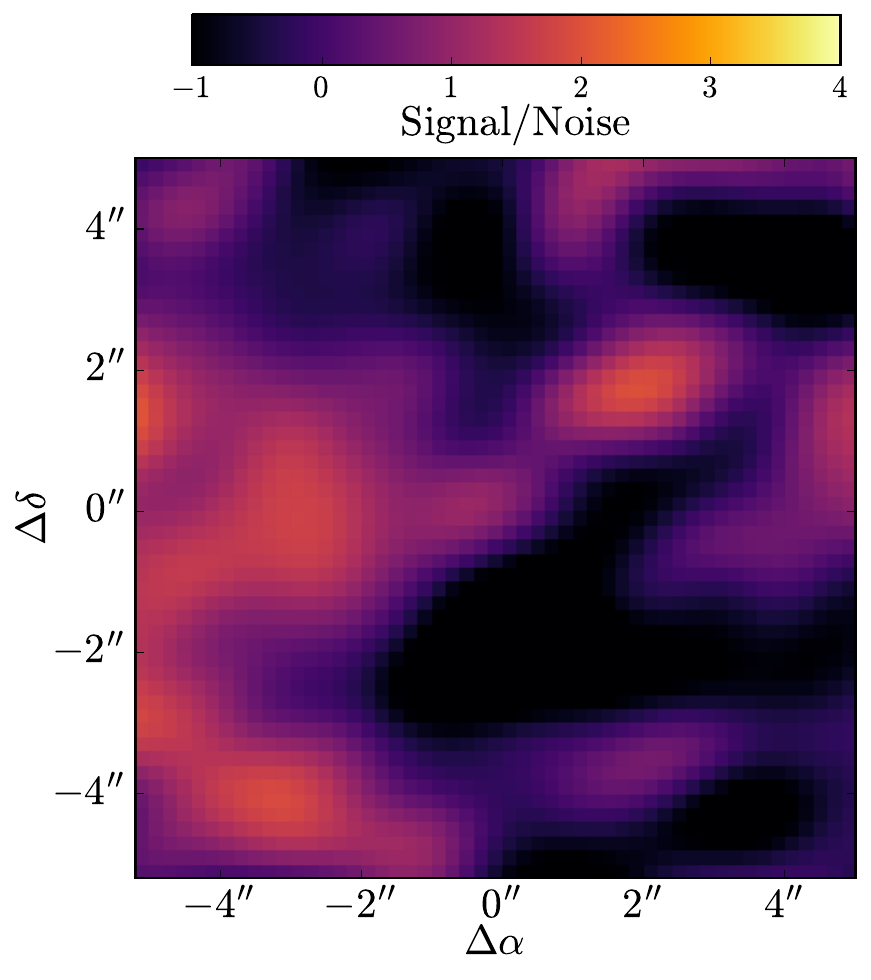}
\end{center}
\caption{Stack at the positions of 21 JADES sources flagged as stars that happen to lie within our ALMA 1-mm map; as expected, we do not find a significant signal.}
\label{fig:stack_stars}
\end{figure}

As a completely separate null test, we also stacked our ALMA map at the positions of 21 sources from the JADES catalogue flagged as stars that happen to fall within our ALMA map. This stack is shown in Fig.~\ref{fig:stack_stars} and we can see that it is consistent with noise.

In Fig.~\ref{fig:stack} we see that many high stellar mass bins are blank; this is because we have either detected all of the galaxies within these bins, or there were no galaxies within these bins to begin with. We also see that there are stacked peaks ${\gtrsim}\,3\,\sigma$ in all bins with stellar masses between $10^{8.4}\,{\rm M}_{\odot}$ and $10^{10.4}\,{\rm M}_{\odot}$ and with redshifts between 0 and 4. Across the lowest stellar mass the stacks tend to be positive but with error bars overlapping 0, meaning that galaxies with stellar masses ${<}\,10^{8.4}\,$M$_{\odot}$ barely contribute to our ALMA map (a topic that is explored further in Section~\ref{sec:consistency}). Finally, we note that the $z\,{=}\,$0--2, $M_{\ast}\,{=}\,10^{9.4}$--$10^{10.4}$M$_{\odot}$ bin has a very high S/N, thus we could extract more information from this bin by splitting it into smaller sub-bins: for $z\,{=}\,$0--1, $M_{\ast}\,{=}\,10^{9.9}$--$10^{10.4}$M$_{\odot}$, we find $S_{\nu}\,{=}\,(22.1\pm7.9)\,\mu$Jy and $N_{\rm eff}\,{=}\,4$; for $z\,{=}\,$0--1, $M_{\ast}\,{=}\,10^{8.9}$--$10^{9.4}$M$_{\odot}$, we find $S_{\nu}\,{=}\,(15.3\pm3.6)\,\mu$Jy and $N_{\rm eff}\,{=}\,15$; for $z\,{=}\,$1--2, $M_{\ast}\,{=}\,10^{9.9}$--$10^{10.4}$M$_{\odot}$ we find $S_{\nu}\,{=}\,(18.2\pm3.8)\,\mu$Jy and $N_{\rm eff}\,{=}\,11$; and for $z\,{=}\,$1--2, $M_{\ast}\,{=}\,10^{8.9}$--$10^{9.4}$M$_{\odot}$, we find $S_{\nu}\,{=}\,(12.1\pm2.9)\,\mu$Jy and $N_{\rm eff}\,{=}\,24$. These values are consistent with higher stellar mass and lower redshift galaxies being brighter at 1\,mm.

In terms of stellar mass, the main conclusions of these stacking results are that: (1) there are about 60 galaxies with stellar masses around $10^{10}\,{\rm M}_{\odot}$ lying just below our $3.6\,\sigma$ ALMA threshold, which have flux densities around 15\,$\umu$Jy; (2) there are more than 300 galaxies with stellar masses around $10^{9}\,{\rm M}_{\odot}$ that can also be statistically detected in the ALMA map, with individual flux densities of a few $\umu$Jy; and (3) galaxies with stellar masses around $10^{8}\,{\rm M}_{\odot}$ or below
have flux densities that are too low to be detected in the ALMA map, even statistically.

\section{The cosmic infrared background in the HUDF}
\label{sec:cib}
The absolute intensity of extragalactic light has been studied at all wavelengths from $\gamma$-rays to the radio \citep{Hill2018}.  Peaking at around $200\,\umu$m, the CIB is the brightness of the extragalactic sky at infrared wavelengths, averaged over the whole sky, after subtracting all contributions originating from the Solar System and the Milky Way. This absolute value tells us about the history of star formation and has been measured using the FIRAS instrument onboard the \textit{COBE} satellite \citep{Fixsen1998}.  The 1-mm background lies on the long-wavelength side of the CIB and we can try to use our new map to estimate what fraction can be accounted for in DSFGs. 

\subsection{Resolved source contribution to the cosmic infrared background}

An important question is whether the intensity of the CIB can be recovered by summing the contribution from known galaxies, or if there exists an additional population of sources or a genuinely diffuse component of the CIB.
This question has been addressed by many previous studies at wavelengths around 1\,mm \citep[e.g.][]{Penner2011,Viero2013,fujimoto2016,Dunlop2017,Hatsukade2018,Gonzalez-lopez2020,Gomez-guijarro2022,chen2023}, with results either in the tens of per cent range, or requiring a model fit to the number counts and extrapolated below the flux densities of the faintest galaxies detected. Here we explore this question with our new ALMA map at 1.23\,mm and our new JADES catalogue, to see if we can recover the total CIB intensity solely with directly-detected galaxies. To start with, the sum of our detected source flux densities (${>}\,3.6\,\sigma$) is (7.33$\,{\pm}\,$0.10)\,mJy, and these sources are detected within an area of $1.15\,{\times}\,10^{-3}\,{\rm deg}^2$.  Therefore we have resolved a total intensity of $(6.35\,{\pm}\,0.09)\,{\rm Jy}\,{\rm deg}^{-2}$ in individually-detected DSFGs.

In addition to this, our stacking analysis (Fig.~\ref{fig:stack}) demonstrates that our map is also statistically sensitive to fainter galaxies. Multiplying $\hat{S}_{\alpha}$ by $N_{\rm eff}$ in each bin shown in Fig.~\ref{fig:stack}, and summing, yields a flux density of $(2.6\,{\pm}\,0.5)$\,mJy, or an intensity of $(2.4\,{\pm}\,0.5)\,{\rm Jy}\,{\rm deg}^{-2}$ (where the area used in our stack is $1.11\,{\times}\,10^{-3}\,{\rm deg}^2$, slightly less than the full map due to our source mask); this stacking result is larger than has previously been possible, due to the combination of a deeper ALMA map and larger catalogue from JWST.  Lastly, noting that there are no galaxies with $S_{1230}\,{>}\,0.85$\,mJy present in our map, we could also add a contribution from brighter sources.  The GOODS-ALMA survey \citep{Gomez-guijarro2022} covered a much larger area with ALMA at 1\,mm (about 72\,arcmin$^2$) and found 22 galaxies with $S_{1230}\,{>}\,0.85$\,mJy, corresponding to an intensity of $(1.31\,{\pm}\,0.02)\,{\rm Jy}\,{\rm deg}^{-2}$ (including the factor of 0.8 to convert their flux densities from 268\,GHz to 243\,GHz) and so we can add this to our resolved CIB contribution as well.

In total, we directly or statistically detect a total CIB intensity of $(8.7\,{\pm}\,0.5)\,{\rm Jy}\,{\rm deg}^{-2}$ in our map, and by adding in a contribution from brighter sources, we estimate that the true value of the CIB is $(10.0\,{\pm}\,0.5)\,{\rm Jy}\,{\rm deg}^{-2}$. We note that if we instead perform our stacking analysis on the full map (without masking bright detected sources) we obtain a CIB estimate of $(8.7\,{\pm}\,0.8)\,{\rm Jy}\,{\rm deg}^{-2}$, consistent with our approach of detecting bright objects then masking them to add the stacking result.  Now we have to determine what fraction of the background we have accounted for. While we have not performed any deboosting corrections for the individual flux densities measured in our catalogue, the fact that we find a nearly identical result when stacking on the full map with no masking shows that any bias from the flux density boosting of weaker S/N sources has very little effect on our conclusions about the total CIB intensity.

\begin{figure}
\begin{center}
\includegraphics[width=0.5\textwidth]{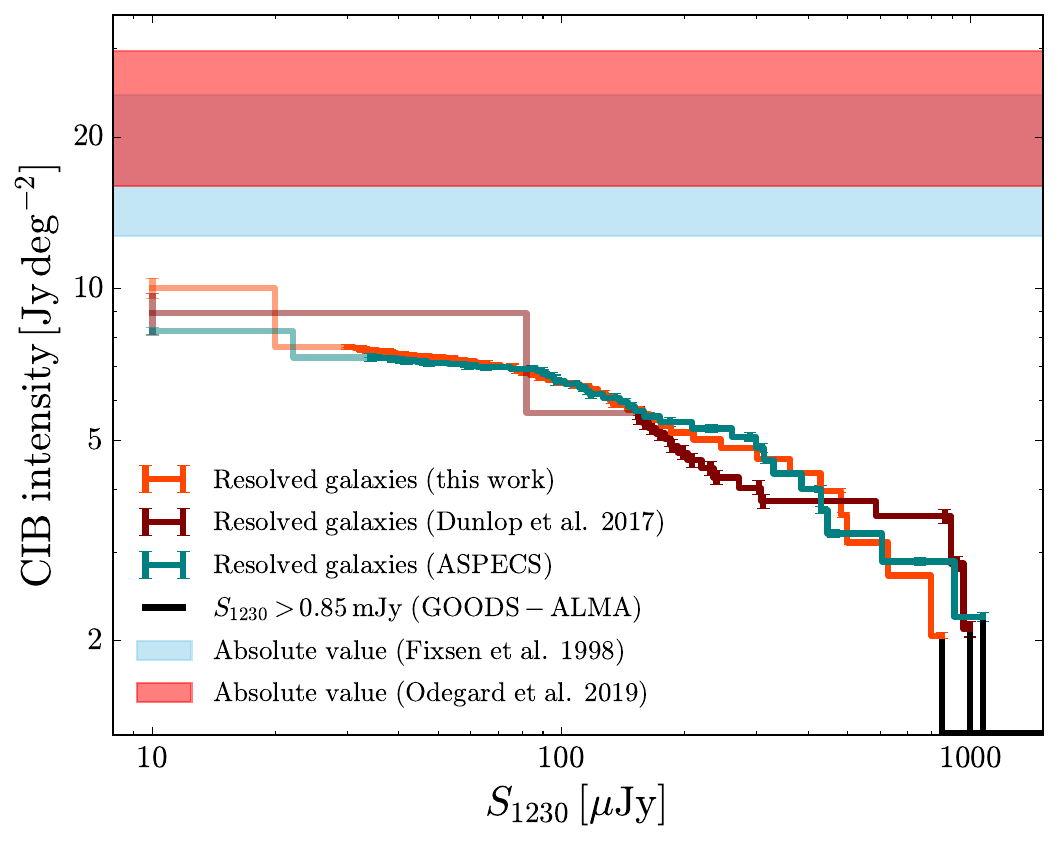}
\end{center}
\caption{CIB intensity estimates at 1.23\,mm (or 243\,GHz). The stepped lines show running sums of individual galaxy flux densities from this work (red), \citet[][brown]{Dunlop2017} and ASPECS \citep[][teal]{Gonzalez-lopez2020}. The contribution to the CIB from $S_{1230}\,{>}\,0.85\,$mJy sources ($1.3\,{\rm Jy}\,{\rm deg}^{-2}$, estimated from the GOODS-ALMA survey) has been added to each line, since $0.85\,$mJy is roughly the brightest source in the HUDF. For each study, the contribution from stacking on undetected near-IR-selected sources is added as a line of faded colour, arbitrarily placed at 10\,$\mu$Jy.  The coloured bands show two different 68\,per cent confidence estimates of the absolute value of the background.}
\label{fig:cib_sum}
\end{figure}

\subsection{The absolute value of the cosmic infrared background}
Estimating the absolute level of the CIB at 1\,mm is subject to larger uncertainties than our estimate of the amount
contributed by sources.
The spectrum of the CIB was measured by \textit{COBE}-FIRAS \citep{Fixsen1998}, but with fairly large systematics-dominated uncertainties.  More recently \citet{Odegard2019} used {\it Planck\/} in combination with FIRAS to estimate the total intensity of the CIB in the {\it Planck}-High-Frequency Instrument channels, including at 217 and 353\,GHz (the closest {\it Planck\/} frequencies to the mean frequency of our ALMA image, 243\,GHz). The best-fit CIB spectral shape from \citet{Fixsen1998} is a modified blackbody function of the form
\begin{equation}\label{eq:nu_eff}
    I(\nu) = A \left( \frac{\nu}{\nu_0} \right)^\beta B_{\nu}(T_{\rm d}),
\end{equation}
where $A$ is a constant, $\nu_0$ is a fiducial frequency, $\beta\,{=}\,0.65$, $B_{\nu}$ is the Planck function and $T_{\rm d}\,{=}\,18.5\,$K.
The uncertainties in the three fit parameters $A$, $\beta$ and $T_{\rm d}$ are significantly correlated (the correlation coefficients reported in the paper are larger than 95\,per cent), meaning that there is effectively only one free parameter in the fit. Lacking the data needed to do the fit ourselves, we simply fix $\beta$ and $T_{\rm d}$ (thus fixing the shape of the CIB spectrum), but take into account the uncertainties in the amplitude throughout our calculations.

To interpolate the CIB intensity to the frequency of our ALMA map, we first estimate what the transmission function of our ALMA map is, and use that to estimate the effective frequency of the image (which may be different from the mean frequency of 243\,GHz). Each ALMA observation consists of two side bands 4\,GHz wide, whose central frequencies are separated by 12\,GHz, and for each observation we have already computed the rms of the observatory-produced MFS image ($\sigma_i$; see Table~\ref{table:obs}). Assuming each observation's individual transmission function is flat (in $\nu$) across the two sidebands, we can calculate the mean transmission function of all of the relevant observations used to produce our final map, weighted by each observation's inverse variance. To each weight we must also include a factor of the ratio of the given observation's beam area (set by $\theta_{{\rm maj},i}$ and $\theta_{{\rm min},i}$) to the final image's beam area (set by $\theta_{\rm maj}$ and $\theta_{\rm min}$), thus the transmission function weight for observation $i$ is
\begin{equation}
    w_i = \frac{1}{\sigma_i^2} \frac{\theta_{{\rm maj},i} \theta_{{\rm min},i}}{\theta_{\rm maj}\theta_{\rm min}}.
\end{equation}
\noindent
To obtain a transmission function, $T(\nu)$, appropriate for the average of the entire map, we only consider the observations from surveys that uniformly covered the HUDF, namely programmes 2012.1.00173.S (from \citealt{Dunlop2017}), 2015.1.00098.S (ASAGAO), 2015.1.00543.S and 2017.1.00755.S (GOODS-ALMA) and 2016.1.00324.L (ASPECS). We find that the resulting transmission function is dominated by ASPECS and is largely flat across Band~6, with slightly more weight towards lower frequencies.

With $T(\nu)$ in hand, we determine the effective frequency of the CIB in our image using the modified blackbody parameters fit to the CIB spectrum from FIRAS \citep{Fixsen1998}.  Specifically, we calculate
\begin{equation}
    \nu_{\rm eff} = \frac{\int T(\nu) I(\nu) \nu d\nu} {\int T(\nu) I(\nu) d\nu},
\end{equation}
with $I(\nu)$ from Eq.~\ref{eq:nu_eff} and the constants $A$ and $\nu_0$ cancel out in the ratio of integrals. We find a value of $\nu_{\rm eff}\,{=}\,241.4\,$GHz, which is close to the ALMA data combined central frequency.

We finally estimate the amplitude of the CIB at $\nu_{\rm eff}$ from Eq.~\ref{eq:nu_eff}, using two different approaches to obtain the amplitude $A$. The first approach is to use $A\,{=}\,(1.3\pm0.4)\,{\times}\,10^{-5}$, which is a direct result from \citet{Fixsen1998}, assuming that the shape is fixed. The second approach is to fit a power law between the CIB intensities from \citet{Odegard2019} at 217 and 353\,GHz (there is an exact solution because there are two measurements and two free parameters), then use the fit to interpolate the intensity at 241.4\,GHz. To propagate uncertainties, we draw 10{,}000 Gaussian random numbers from the measured amplitudes and calculate the corresponding CIB value at 241.4\,GHz.

We find absolute CIB values of $19_{-5}^{+6}\,{\rm Jy}\,{\rm deg}^{-2}$ using the fit from \citet{Fixsen1998} and $22_{-6}^{+7}\,{\rm Jy}\,{\rm deg}^{-2}$ using the interpolation from \citet{Odegard2019}, where the error bars are the 68\,per cent confidence intervals of the posterior distributions and the central values are the means of the distributions.

We show these estimates graphically in Fig.~\ref{fig:cib_sum}.  Specifically we plot with a blue band the 68\,per cent range of the absolute value of the CIB estimated using FIRAS data alone \citep{Fixsen1998} and with a pink band we plot the estimate using FIRAS data in combination with {\it Planck} for foreground removal \citep{Odegard2019}.  Although the \citet{Odegard2019} results substantially shrink some uncertainties compared to the \citet{Fixsen1998} results, that is really only the case at shorter wavelengths; at 1.2\,mm the uncertainties are similar, but the background is actually a little higher in the more recent paper.  The difference between the two estimates indicates that the background at these wavelengths is still quite uncertain.

In Fig.~\ref{fig:cib_sum} we also present the contribution of sources to the CIB, by plotting the cumulative intensity of resolved galaxies from this work as a function of their flux density, including the estimated contribution from $S_{1230}\,{>}\,0.85$\,mJy galaxies and also the stacking-estimate contribution from galaxies in the JADES catalogue. For reference we show the same analysis using the results from \citet{Dunlop2017} and ASPECS \citep{Gonzalez-lopez2020}, where in both works an estimate of the contribution to the CIB from stacking was performed. Our new results are similar to those from previous studies, with our new analysis detecting the faintest sources and finding the highest total background value in the HUDF region.

\subsection{Consistency between resolved sources and the absolute value of the cosmic infrared background}
\label{sec:consistency}

Clearly the absolute value measurements of the CIB are larger than what we find by summing the flux densities of known galaxies (detected by both ALMA and {\it JWST\/}). It is therefore important to consider whether or not the two kinds of measurement are consistent. 

There are fluctuations in the CIB \citep[e.g.,][]{PlanckXXX} whose amplitude depends on the area observed. Studies of the CIB at submm wavelengths found that $\delta I/I\,{\simeq}\,15\,$per cent on scales around 10\,arcmin \citep{Viero2009,Viero2013b}, which is larger than the area of our deep ALMA map.  To find a better estimate of the expected amplitude of these fluctuations for maps the size of our HUDF image, we use the Simulated Infrared Dusty Extragalactic Sky (SIDES) mock catalogue \citep{Bethermin2017,Gkogkou2023}. Briefly, SIDES uses dark matter halos in a simulated light cone to obtain clustered positions on the projected sky, then attaches stellar masses and far-infrared SEDs to the halos using a two star-formation-mode model of galaxy evolution. The total simulated area of the SIDES simulation is 117\,deg$^2$, but here we only use a 1\,deg$^2$ tile from the full simulation.

Since each simulated galaxy has an SED, we first estimate their 243-GHz flux densities by integrating their SEDs through the transmission function derived above. We next take 100 random patches from the 1\,deg$^2$ tile, each with the area of our image of the HUDF (here 4.2\,arcmin$^2$). The CIB intensity of each random patch is then just the sum of the flux densities of the galaxies in the patch divided by the area. However, since we find no galaxies with $S_{1230}\,{>}\,0.85$\,mJy in our real ALMA map, then we subtract these from our random patches as well; this assumes that there are no additional fluctuations from the population of $S_{1230}\,{>}\,0.85$\,mJy galaxies (i.e.\ we simply add a constant for the bright part of the background).  Additionally, we neglect the contribution from galaxies at stellar masses where we were unable to detect any statistical signal in the real data.

Following this procedure, we find that the mean CIB value within 4.2\,arcmin$^2$ patches of the SIDES simulation (after subtracting the contribution from $S_{1230}\,{>}\,0.85$\,mJy galaxies) is 11.1\,Jy\,deg$^{-2}$, comparable to the value of $(8.6\,{\pm}\,0.5)\,{\rm Jy}\,{\rm deg}^{-2}$ that we have measured. The standard deviation of the SIDES simulation patches is 1.9\,Jy\,deg$^{-2}$, thus fluctuations are $\delta I/I\,{=}\,17\,$per cent.  We must therefore take this into account when comparing the mean CIB value from FIRAS averaged over nearly the whole sky to the single 4.2\,arcmin$^2$ patch we have observed.

We now estimate the probability of measuring a CIB value of $(10.0\,{\pm}\,0.5)\,{\rm Jy}\,{\rm deg}^{-2}$ assuming that the true value is either $19_{-5}^{+6}\,{\rm Jy}\,{\rm deg}^{-2}$ \citep{Fixsen1998} or $23_{-8}^{+6}\,{\rm Jy}\,{\rm deg}^{-2}$ \citep{Odegard2019} and that fluctuations are 17\,per cent. We take our 10{,}000 CIB absolute value realizations discussed above and additionally draw 10{,}000 Gaussian-distributed values for a fluctuation amplitude, then multiply these together. We draw 10{,}000 Gaussian-distributed numbers for our measured CIB value, and take the difference between these and the possible absolute values. Finally, our statistic is the fraction of the area of the resulting posterior distribution that is less than zero, which can be interpreted as the probability of obtaining our actual measured CIB value or less while taking into account both the measurement uncertainties and intrinsic CIB fluctuations.
We find that the probability of measuring a CIB value of $(10.0\,{\pm}\,0.5)\,{\rm Jy}\,{\rm deg}^{-2}$, given the absolute CIB measurement from \citet{Fixsen1998} is 8.9\,per cent, or 5.3\,per cent given the absolute CIB measurement from \citet{Odegard2019}.

Assuming that the absolute value measurements from FIRAS are correct and that our measurement is also correct, we can calculate the required level of statistical excursion (in units of $\sigma\,{=}\,0.17$) corresponding to the HUDF. Using the same 10{,}000 random values for the absolute FIRAS values and the measured values, we find that the CIB fluctuation at the position of the HUDF must be $-2.4_{-1.4}^{+0.4}\sigma$ using the CIB value from \citet{Fixsen1998}, or $-2.9_{-1.2}^{+0.3}\sigma$ from \citet{Odegard2019} (here the central values are the means of the posterior distributions and the error bars are 68\,per cent confidence intervals). What this means is that the variance in HUDF fields is large enough that our results can explain the whole of the CIB provided that the HUDF happens to be a relatively mild (${\simeq}\,2\,\sigma$) underdense direction on the sky.\footnote{The HUDF was not selected entirely randomly; however, there is no particular reason to believe that the criteria used for its selection would make it likely to have a lower than average background \citep{Beckwith2006}.}

As a final check, we use the simulated catalogue of galaxies from SIDES to estimate the fraction of the CIB emitted by galaxies with stellar masses between 10$^{7.4}$\,M$_{\odot}$ and 10$^{11.4}$\,M$_{\odot}$ and with redshifts between 0 and 8 (namely the parameter space over which we stacked on undetected JADES galaxies).  For each of the 100 random patches described above, we also calculate the sum of the flux densities from all galaxies within our stacking range divided by the sum of all of the galaxies in the HUDF-sized region. We find that the average ratio is 97\,per cent, with a standard deviation of 1\,per cent. Thus if we accept that SIDES provides a reasonable model for counts at these wavelengths, we do not expect that we are missing a significant contribution to our ALMA measurement of the CIB from even fainter and lower stellar mass galaxies.

Turning back to the data, if we include sources detected by ALMA, the contribution to the CIB from 10$^{10.4}$--10$^{11.4}$\,M$_{\odot}$ galaxies is $(3.5\,{\pm}\,0.1)\,{\rm Jy}\,{\rm deg}^{-2}$, from 10$^{9.4}$--10$^{10.4}$\,M$_{\odot}$ galaxies is $(3.2\,{\pm}\,0.1)\,{\rm Jy}\,{\rm deg}^{-2}$, from 10$^{8.4}$--10$^{9.4}$\,M$_{\odot}$ galaxies is $(1.1\,{\pm}\,0.2)\,{\rm Jy}\,{\rm deg}^{-2}$ and from 10$^{7.4}$--10$^{8.4}$\,M$_{\odot}$ galaxies is $(0.5\,{\pm}\,0.4)\,{\rm Jy}\,{\rm deg}^{-2}$.  If we stack on the next lowest mass bin (10$^{6.4}$--10$^{7.4}$\,M$_{\odot}$) over the full redshift range we obtain a CIB contribution of $(-0.5\,{\pm}\,0.5)\,{\rm Jy}\,{\rm deg}^{-2}$.  This is consistent with the
CIB having essentially converged over the stellar mass range that we have probed.

A potential issue here is the completeness of the JADES catalogue within our stacking range. For our 1000 random SIDES patches we also keep track of the total number of galaxies with stellar masses and redshifts within our stacking region. We find that the average number of galaxies is 1890 (with a standard deviation of 150). In the JADES catalogue there are 1856 galaxies in our ALMA image with stellar masses and redshifts within our stacking range, which is consistent with the total number of galaxies expected from the SIDES simulation. For reference, there are 1561 galaxies from the CANDELS catalogue that are in our ALMA image footprint within the same stacking range. Adding JADES has helped us to find more of the CIB within our ALMA map, and it seems that going even deeper in the optical/near-infrared will not add significantly to the source-derived CIB estimate.

What these numbers indicate is that our estimate of the 1.23-mm CIB (from individually-detected galaxies, together with statistical stacking results) appears to contain essentially all of the possible galaxies that would contribute to the CIB, and that it is genuinely lower than what the mean value is estimated to be. However, the chances that our small patch of the extragalactic sky falls on a negative fluctuation of the CIB are not small enough to rule out the hypothesis that we have indeed recovered essentially the entire CIB from these known galaxies.

\section{Improvement over previous studies}
\label{sec:improvements}
The deepest previously-published ALMA survey of the HUDF at 1\,mm is ASPECS and so a question of interest is how much of an improvement have we achieved by including all of the additional data. Qualitatively, looking at Table~\ref{table:obs} we can see that ASPECS is by far the deepest of the individual surveys. The maps from \citet{Dunlop2017}, ASAGAO and GOODS-ALMA overlap with the entire ASPECS map and so by combining them with ASPECS the result must be deeper. Additionally, by including more individual pointings, we have been able to make some regions even deeper still.

To quantify the difference between ASPECS and our combined image, we downloaded the ASPECS map produced using \textsc{CASA} in the MFS mode and made public by the ASPECS team,\footnote{\url{https://almascience.nrao.edu/alma-data/lp/ASPECS}} then ran the primary-beam-corrected image through our algorithm for generating the noise map (see Section~\ref{sec:data_uv}). The pixel size of the ASPECS map is 0.2\,arcsec, the same as our $uv$-combined map, and the beamsizes are very similar ($1.49\,{\rm arcsec}\,{\times}\,1.07\,{\rm arcsec}$ versus $1.53\,{\rm arcsec}\,{\times}\,1.08\,{\rm arcsec}$), so we simply computed the ratio of the two noise maps to assess the amount by which the noise improves with the additional data. Unsurprisingly we find that across the entire ASPECS region the noise (meaning here the rms after masking sources) in our map is smaller, ranging from about 5\,per cent smaller in the central region to about 50\,per cent smaller near the edges and around the deepest individual pointings. The ASPECS map was made going out to a primary beam value of 0.1 and covers a total area of 4.2\,arcmin$^2$, 3.3\,arcmin$^2$ of which has a noise level less than 35\,$\umu$Jy\,beam$^{-1}$, whereas our map was made going out to a primary beam value of 0.2 and all 4.2\,arcmin$^2$ has a noise level less than 35\,$\umu$Jy\,beam$^{-1}$. Thus by combining the different data sets we not only reduce the noise across the majority of the map by 5\,per cent, but we also expand the area where the map is at its most sensitive.

\citet{Gonzalez-lopez2020} searched the ASPECS map for sources down to a fidelity of 0.5, and their lowest significance source had a S/N of 3.3. Their catalogue contains a total of 35 sources (four of which are not detected in our map), whereas we find 45 sources; it is worth noting that all 45 of our sources are contained within the area mapped by ASPECS. Thus the extra depth we have been able to achieve by combining archival ALMA data with ASPECS has led to a significant increase in detected sources.  

In addition to detecting more sources, we are able to recover more of the CIB through stacking thanks to our deeper JADES catalogue. If we stack on the 1661 galaxies from the CANDELS catalogue with stellar masses between 10$^{7.4}$\,M$_{\odot}$ and 10$^{11.4}$\,M$_{\odot}$ and with redshifts between 0 and 8 (after masking detected ALMA sources) we obtain $(1.5\,{\pm}\,0.4)\,{\rm Jy}\,{\rm deg}^{-2}$, compared to $(2.4\,{\pm}\,0.5)\,{\rm Jy}\,{\rm deg}^{-2}$ using the JADES catalogue. 

Now turning to the wider HUDF region, \citet{Hatsukade2018} also combined the ASAGAO survey with the survey from \citet{Dunlop2017} and the first GOODS-ALMA survey \citep{Franco2018}; their final map has a mean rms of about $75\,\umu{\rm Jy}\,{\rm beam}^{-1}$, with a beamsize of $0.59\,{\rm arcsec}\,{\times}\,0.53\,{\rm arcsec}$ and a pixel size of 0.1\,arcsec after applying a 250\,k$\lambda$ taper. Our shallow and wide combined map (presented in Appendix~\ref{sec:appendix}) contains additional GOODS-ALMA data that were not available when the ASAGAO map was constructed, as well as more individual pointings, while maintaining a similar beamsize ($0.87\,{\rm arcsec}\,{\times}\,0.64\,{\rm arcsec}$) and pixel size (0.12\,arcsec). We thus also downloaded the ASAGAO 250\,k$\lambda$-tapered map\footnote{\url{https://sites.google.com/view/asagao26/alma-data?authuser=0}} to quantitatively check the improvement. After running the ASAGAO map through our noise algorithm, we find that the new GOODS-ALMA data reduces the noise by 10--20\,per cent (ignoring the deep  central region), while some individual pointings go about 50\,per cent deeper.

\begin{figure}
\begin{center}
\includegraphics[width=0.5\textwidth]{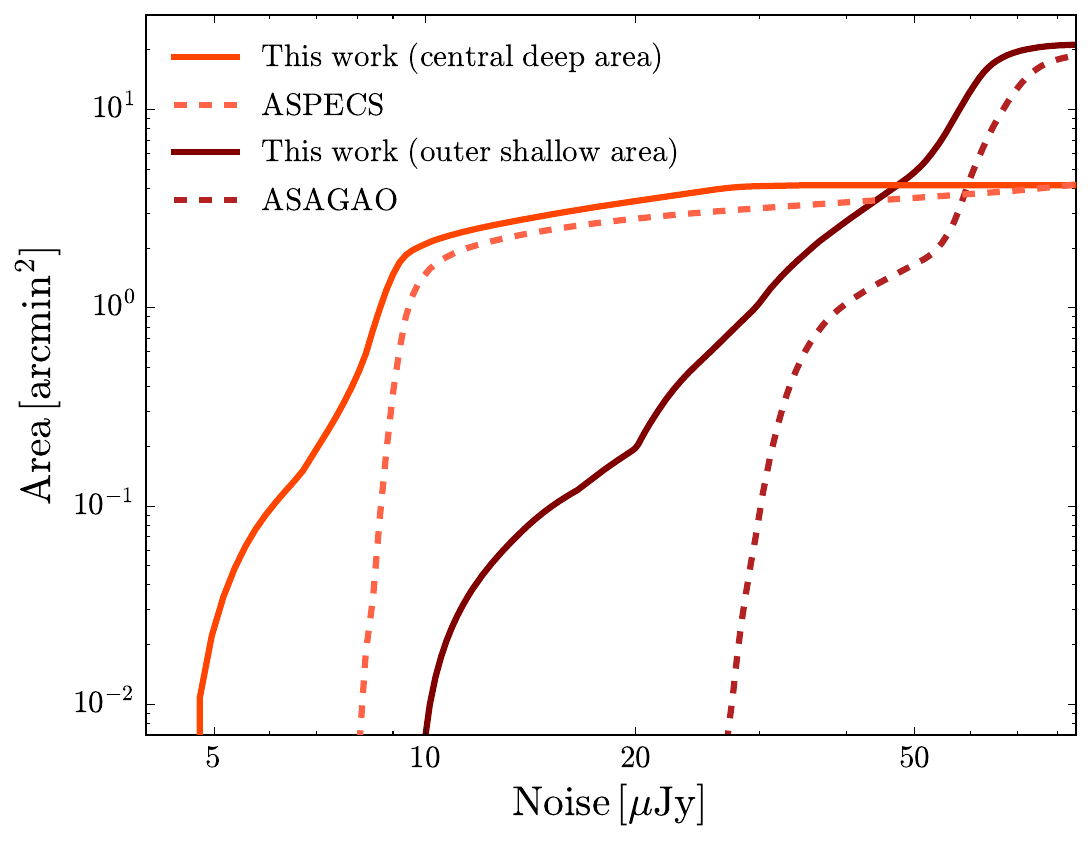}
\end{center}
\caption{Cumulative map area as a function of (1$\,\sigma$) noise level. Our combined deep central map has a sensitivity about 50\,per cent better than the previously-published ASPECS map over roughly 0.1\,arcmin$^2$. Our combined shallow large map is also about 50\,per cent deeper than the previously-published ASAGAO map over roughly 0.1\,arcmin$^2$, and 10--20\,per cent deeper over the remaining area (excluding the deep central region). For all the maps shown here, we have calculated the noise using the same approach described in Section~\ref{sec:data_uv}.}
\label{fig:area_sensitivity}
\end{figure}

To present these improvements quantitatively, in Fig.~\ref{fig:area_sensitivity} we show the cumulative map area as a function of noise level for our combined deep central map, the ASPECS map, our combined shallow large map, and the ASAGAO map. For the combined shallow large map and the ASAGAO map, we mask the footprint of the deep central map, since we did not search that area for sources. We emphasize that the noise levels for each map have been calculated using the same approach. We can see that both of our combined maps are about 50\,per cent deeper than the previously-published maps out to about 0.1\,arcmin$^2$, and remain deeper out to the primary beam cutoffs that we have defined.

\section{Conclusions}
\label{sec:conclusion}

We have produced a series of 1-mm maps of the HUDF by combining all of the previously-published survey data in the $uv$ plane in various ways, reducing the noise compared to previous studies. We specifically constructed a deep map covering 4.2\,arcmin$^2$ and a shallower map covering 25.4\,arcmin$^2$. Our deep map has a pixel rms that ranges from 5 to 50\,per cent lower than in the best previous study, with an area of about 1.5\,arcmin reaching below $9\,\umu{\rm Jy}\,{\rm beam}^{-1}$ and a minimum of about $4.6\,\umu{\rm Jy}\,{\rm beam}^{-1}$ reached in some regions. Our shallow map has a pixel rms that ranges from 10 to 50\,per cent lower than in the best previous wider and shallower study. We make all of our maps publicly available\footnote{\url{https://doi.org/10.5683/SP3/YWBVWH}}

We searched our deep map for sources down to a signal-to-noise threshold of 3.6, finding a total of 45 peaks in the S/N map, 13 of which are new.  Nearly all (39/45) of these ALMA sources have near-IR counterparts detected by {\it JWST\/} and {\it HST\/}.  We additionally find 27 sources in our wider map, nine of which are new. The JADES data enable stellar masses and photometric redshifts to be estimated for the ALMA source counterparts, and we find that they are all relatively high $M_\ast$ galaxies. Compared with ALMA-undetected galaxies at similar $M_{\ast}$, the ALMA-detected galaxies typically have redder {\it JWST\/} colours.  With our larger sample of mm-selected sources in the HUDF, other studies investigating the statistical properties of the faintest star-forming galaxies could be carried out. For example, \citet{Aravena2020} studied the SFR-stellar mass relation of ASPECS galaxies and \citet{Boogaard2023} looked at the morphologies of ASPECS galaxies in {\it JWST\/}'s MIRI filters; expanding to a larger sample size and using improved SED fits from {\it JWST\/} photometry could lead to more robust conclusions.

Since the vast majority of near-IR-selected galaxies are not directly detected in our 1-mm map, we performed a stacking analysis on their positions.  We found significant average signals for all galaxies in the range $z\,{=}\,0$ to 3 and with stellar masses between $10^{9.4}\,{\rm M}_{\odot}$ and $10^{10.4}\,{\rm M}_{\odot}$, as well as a roughly $3\,\sigma$ signal from $10^{8.4}\,{\rm M}_{\odot}$ to $10^{9.4}\,{\rm M}_{\odot}$ stellar mass galaxies between $z\,{=}\,0$ and 2. The evidence is that $M_\ast\,{\sim}\,10^{10}\,{\rm M}_\odot$ ALMA-\textit{undetected} galaxies have a 1-mm flux density around $15\,\umu$Jy and would be individually detected in even deeper integrations.  The stacking results also show the value of our new data products for performing similar statistical analyses on sets of galaxies detected in other wavebands.

We used our galaxy detections, as well as our stacking analysis, to estimate the level of the CIB at 1.23\,mm that we have resolved.  We thus account for a background level of $(10.0\,{\pm}\,0.5)\,{\rm Jy}\,{\rm deg}^{-2}$, with an expectation that even fainter galaxies will hardly change this number.  There is still a large uncertainty in the total background estimate, and we also stress that the variance expected in a region as small as the HUDF is around 20\,per cent.  We do not recover all
of the background, and there are a number of possible resolutions: (1) HUDF is a 2--$3\,\sigma$ negative fluctuation in the CIB; (2) the absolute level of the CIB has been overestimated; (3) there is a new population of galaxies that contribute at 1\,mm, but are not detected by {\it JWST\/}; or (4) a genuinely diffuse component of the background exists.  Given that the simple explanation (1) cannot be excluded, then the suggestion is that deep ALMA+{\it JWST\/} observations may account for all of the mm-wave background.

The HUDF is probably the best-studied extragalactic field, and it is crucial to probe this field at longer (${>}\,500\,\umu$m) wavelengths in order to understand how the earliest galaxies began forming their stars, complementing the data obtained by {\it HST\/}, {\it JWST\/} and other telescopes at shorter wavelengths.  The 4.2\,arcmin$^2$ map presented here is the deepest such image of the HUDF available to-date, yet it is clear that there are still many more galaxies left to detect. Deeper ALMA observations of the HUDF will inevitably uncover these galaxies, providing a more complete understanding of galaxy evolution at early times.

\section*{Acknowledgements}

This research used the Canadian Advanced Network For Astronomy Research (CANFAR) operated in partnership by the Canadian Astronomy Data Centre and The Digital Research Alliance of Canada, with support from the National Research Council of Canada, the Canadian Space Agency, CANARIE and the Canadian Foundation for Innovation. RH and DS acknowledge the support of the Natural Sciences and Engineering Research Council of Canada (NSERC). This paper makes use of the ALMA data ADS/JAO.ALMA\#2012.1.00173.S,
2013.1.00718.S, 2013.1.01271.S, 2015.1.00098.S, 2015.1.00543.S, 2015.1.00664.S, 2015.1.00821.S, 2015.1.00870.S, 2015.1.01096.S, 2015.1.01379.S, 2015.1.01447.S, 2015.A.00009.S, 2016.1.00324.L, 2016.1.00721.S, 2016.1.00967.S, 2017.1.00190.S, 2017.1.00755.S, 2018.1.00567.S and 2018.1.01044.S. ALMA is a partnership of ESO (representing its member states), NSF (USA) and NINS (Japan), together with NRC (Canada), MOST and ASIAA (Taiwan) and KASI (Republic of Korea), in cooperation with the Republic of Chile. The Joint ALMA Observatory is operated by ESO, AUI/NRAO and NAOJ.  This research made use of \textsc{Photutils}, an Astropy package for detection and photometry of astronomical sources.
This work is based in part on observations taken by the CANDELS Multi-Cycle Treasury Program with the NASA/ESA \textit{HST}, which is operated by the Association of Universities for Research in Astronomy, Inc., under NASA contract NAS5-26555.
This work is based in part on observations made with the NASA/ESA/CSA \textit{James Webb Space Telescope}.
We thank Leindert Boogaard, Dan Eisenstein and Ian Smail for helpful suggestions. For the purpose of open access, the author has applied a Creative Commons Attribution (CC BY) licence to any Author Accepted Manuscript version arising from this submission.

\section*{Data Availability}

All of the data products described in this paper are publicly available at \url{https://doi.org/10.5683/SP3/YWBVWH}. The raw ALMA data used to produce these data products are publicly available at the ALMA archive. The {\it JWST\/} images from the JADES Collaboration used in this paper are also publicly available, and the catalogue derived using JADES images are described and made available in a separate paper.

\bibliographystyle{mnras}
\bibliography{ALMA_HUDF_stack}

\appendix

\section{Sources in the wider region}
\label{sec:appendix}

\begin{table*}
\centering
\caption{Archival ALMA projects downloaded from the ALMA archive and used to create the larger and shallower combined 1-mm image.}
\label{table:obs_outer}
\begin{threeparttable}
\begin{tabular}{lccccc}
\hline
Project ID & Target name(s) & Frequency range & \phantom{0}Map rms$^{\rm a}$ & Synthesized beam$^{\rm b}$ & Sky coverage$^{\rm c}$\phantom{0}\\
 & & [GHz] & \phantom{0}[$\umu$Jy\,beam$^{-1}$] & [arcsec$\,{\times}\,$arcsec] & [arcmin$^2$]\phantom{0}\\
\hline
2015.1.00098.S$^{\rm d}$ & HUDF-JVLA-ALMA & 244.3--271.8 & \phantom{0}57 & 0.20$\,{\times}\,$0.16 & 34.7\phantom{0}\\
2015.1.00543.S$^{\rm d}$ & GOODS-S & 255.1--274.7 & 130 & 0.26$\,{\times}\,$0.22 & 54.2\phantom{0}\\
2015.1.00664.S & KMOS3DGS4-27882 & 255.2--274.8 & \phantom{0}63 & 0.15$\,{\times}\,$0.14 & \phantom{0}0.23\\
2015.1.00821.S & z7\_GSD\_3811 & 218.2--236.6 & \phantom{0}23 & 0.70$\,{\times}\,$0.51 & \phantom{0}0.32\\
2015.1.00870.S & 19842 & 223.2--242.8 & \phantom{0}42 & 0.58$\,{\times}\,$0.49 & \phantom{0}0.30\\
2015.1.01379.S & GMASS\_0953 & 212.3--216.8 & \phantom{0}26 & 0.66$\,{\times}\,$0.54 & \phantom{0}0.36\\
2016.1.00721.S & z6\_GSD\_1388 & 250.9--269.3 & \phantom{0}23 & 0.78$\,{\times}\,$0.53 & \phantom{0}0.24\\
2016.1.00967.S & ZFOURGE\_CDSF\_4409 & 243.8--262.1 & \phantom{0}28 & 1.24$\,{\times}\,$0.84 & \phantom{0}0.26\\
2017.1.00755.S$^{\rm d}$ & GOODS-S & 255.1--274.9 & 120 & 1.31$\,{\times}\,$0.82 & 54.1\phantom{0}\\
2017.1.00190.S & z7\_GSD\_3811 & 218.4--236.9 & \phantom{0}12 & 0.61$\,{\times}\,$0.49 & \phantom{0}0.32\\
2018.1.00567.S & All ASAGAO pointings & 244.2--262.9 & \phantom{0}29 & 0.65$\,{\times}\,$0.47 & \phantom{0}5.15\\
2018.1.01044.S & cdfs\_535, cdfs\_769  & 223.1--242.9 & \phantom{0}27 & 0.63$\,{\times}\,$0.51 & \phantom{0}0.61\\
\hline
\end{tabular}
\begin{tablenotes}
\item $^{\rm a}$ Approximate map rms estimated using the observatory MFS data products. For programmes with a single tuning, this is the rms of the primary-beam-uncorrected map after masking all known sources. For programmes with multiple tunings, we follow the same procedure and then estimate the rms of the weighted mean of the images as $1/\big( \Sigma_i 1/\sigma_i^2 \big)$. The actual rms from combining images at different tunings in $uv$ space using the MFS mode will generally be less than this estimate.
\item $^{\rm b}$ Maximum synthesized beam across all frequencies of a given data set. The actual synthesized beam from combining data taken in different tunings using the MFS mode may differ slightly from this estimate.
\item $^{\rm c}$ The sky coverage overlapping with our data selection criteria, which may be less than the total sky coverage of the given programme.
\item $^{\rm d}$ Programmes used to make a mosaic with uniform noise.
\end{tablenotes}
\end{threeparttable}
\end{table*}

\begin{figure*}
\begin{center}
\includegraphics[width=0.33\textwidth]{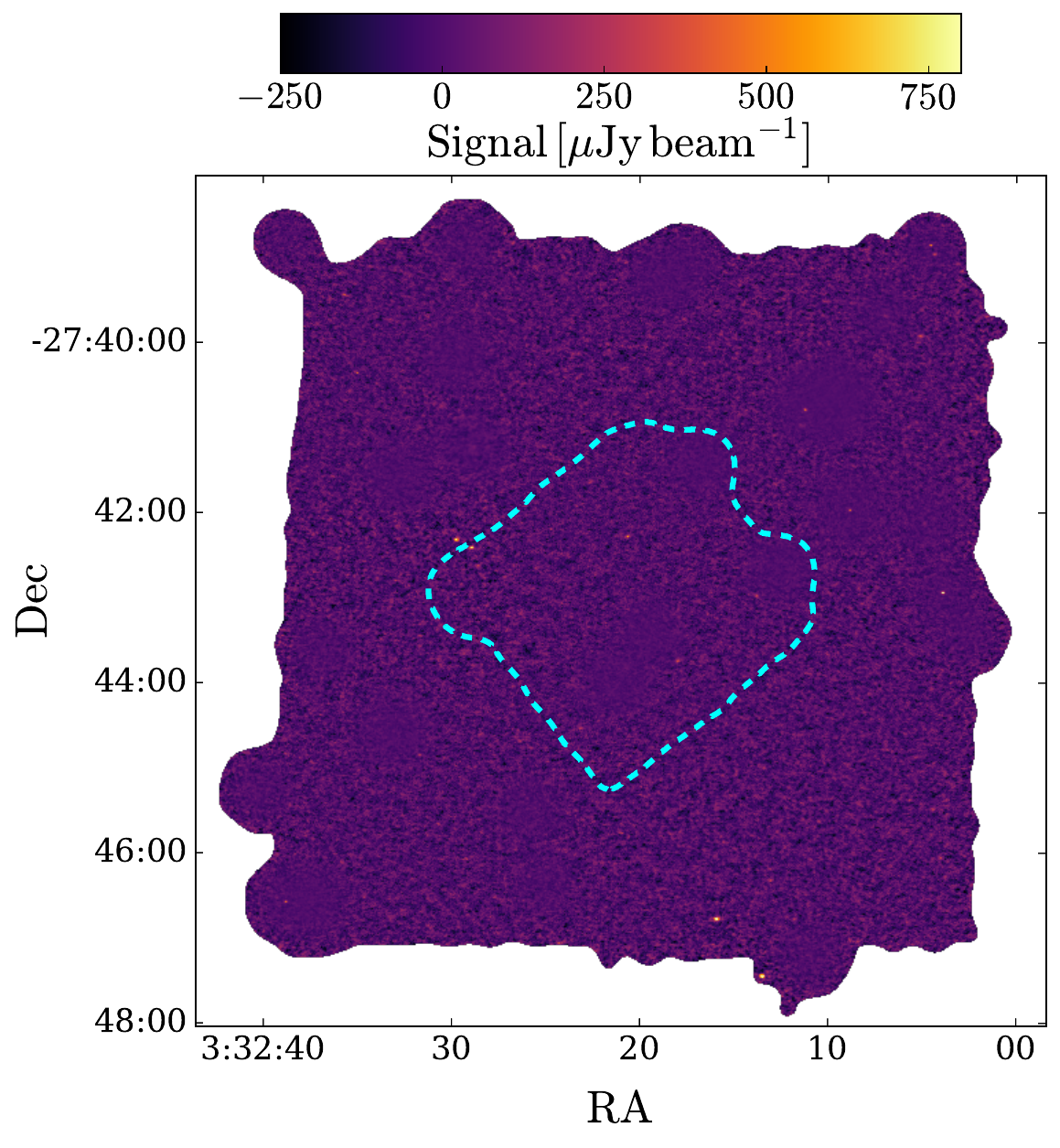}
\includegraphics[width=0.33\textwidth]{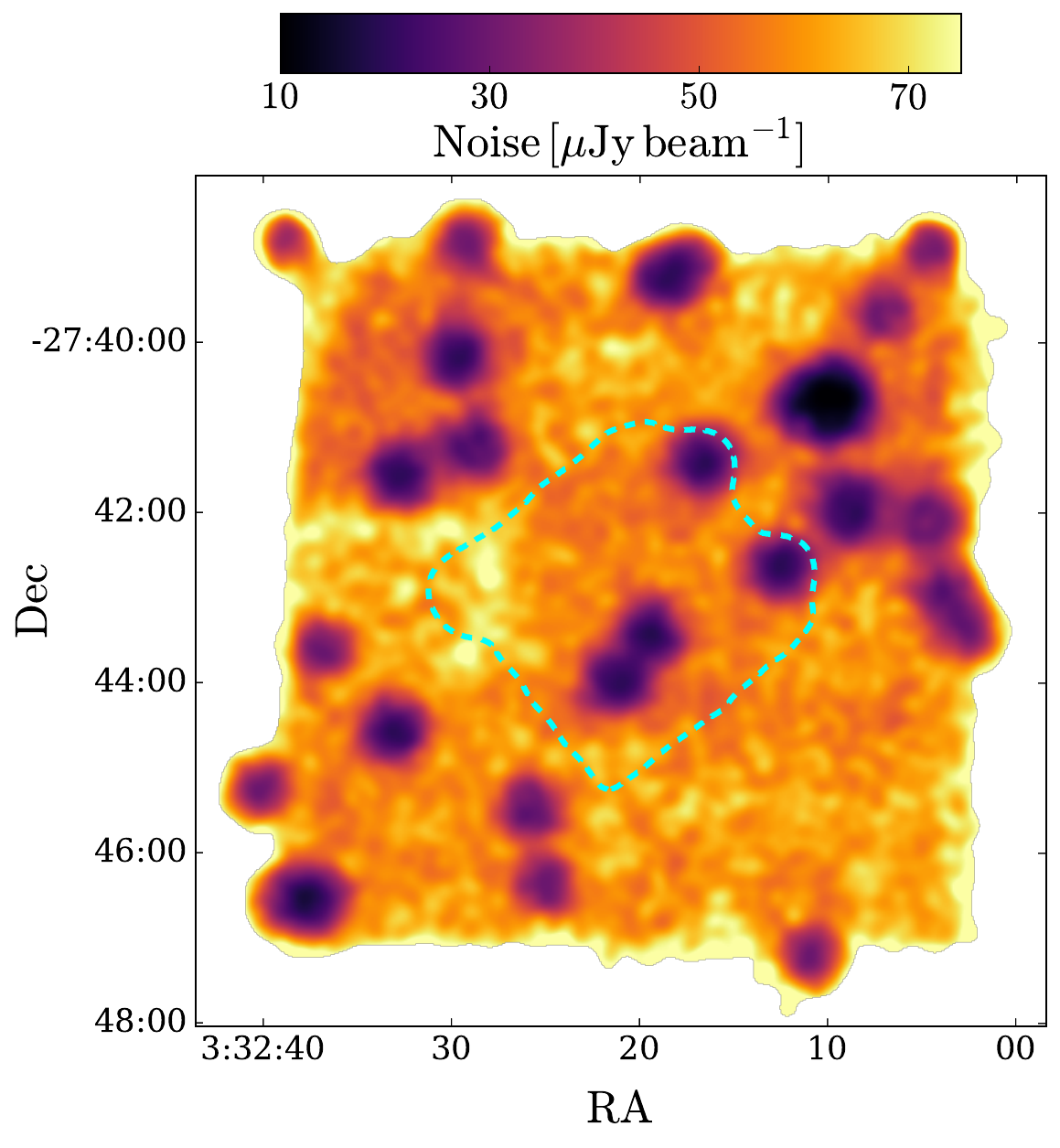}
\includegraphics[width=0.33\textwidth]{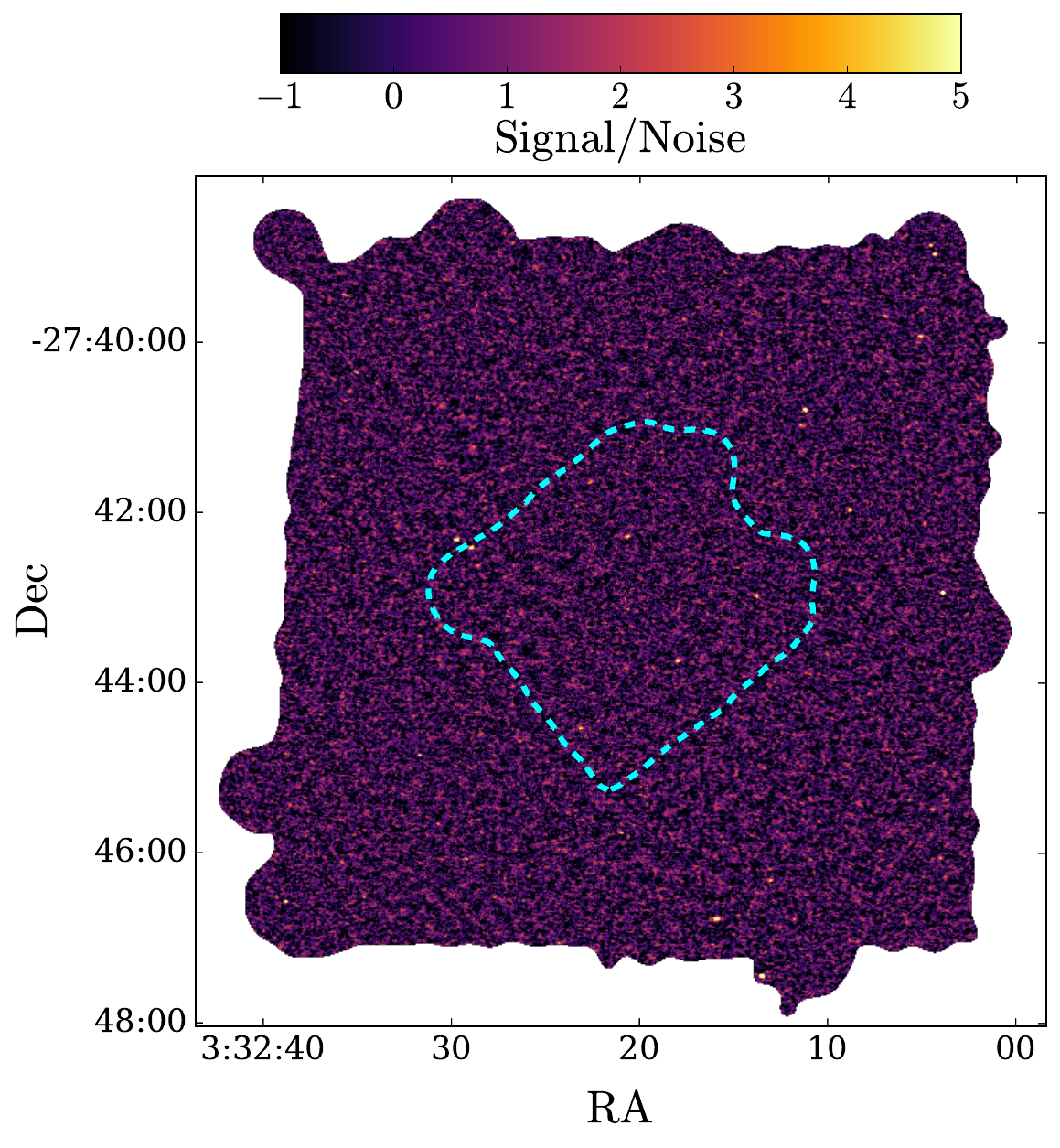}
\end{center}
\caption{{\it Left:} Signal map of the shallower, larger region of the HUDF after combining all of the available archival ALMA Band-6 data given in Table~\ref{table:obs_outer}. This region is defined by the contour where the primary beam reaches 0.4, covering a total of 25.4\,arcmin$^2$. {\it Middle:} Corresponding noise map. {\it Right:} S/N map, made by dividing the signal map by the noise map. In all panels, the cyan contour shows the footprint of our combined deeper central map, which we exclude from our source extraction procedure in the shallower region.}
\label{fig:snr_outer}
\end{figure*}

\begin{figure*}
\begin{center}
\includegraphics[width=\textwidth]{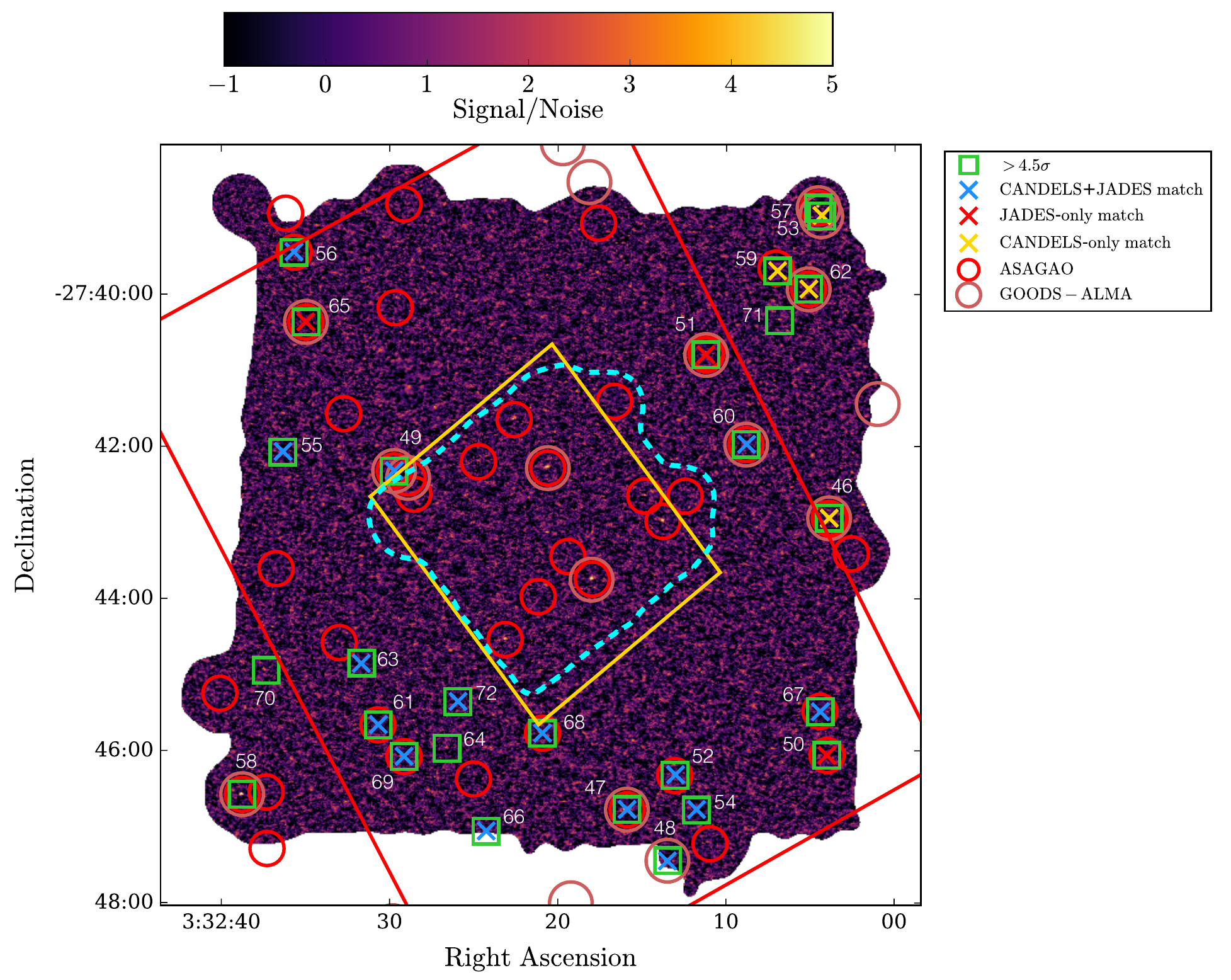}
\end{center}
\caption{S/N map of the larger and shallower combined ALMA data, covering 25.4\,arcmin$^2$, with peaks ${>}\,4.5\,\sigma$ (where the fidelity is 0.9) indicated as green boxes. Galaxies from the ASAGAO survey \citep{Hatsukade2018} are shown as red circles and galaxies found in the GOODS-ALMA survey \citep{Gomez-guijarro2022} are shown as brown circles. Galaxies with counterparts in both the CANDELS catalogue and the JADES catalogue are indicated with a blue cross, galaxies with only a CANDELS counterpart are indicated with a gold cross and galaxies with only a JADES counterpart are indicated with a red cross. The dashed cyan contour shows the footprint of our combined deeper central map, which we exclude from our source extraction procedure in this appendix, while the gold contour shows the footprint of the deepest {\it HST\/} data from the CANDELS survey and the red contour shows the footprint of the JADES survey.}
\label{fig:sources_outer}
\end{figure*}

\begin{figure}
\begin{center}
\includegraphics[width=0.5\textwidth]{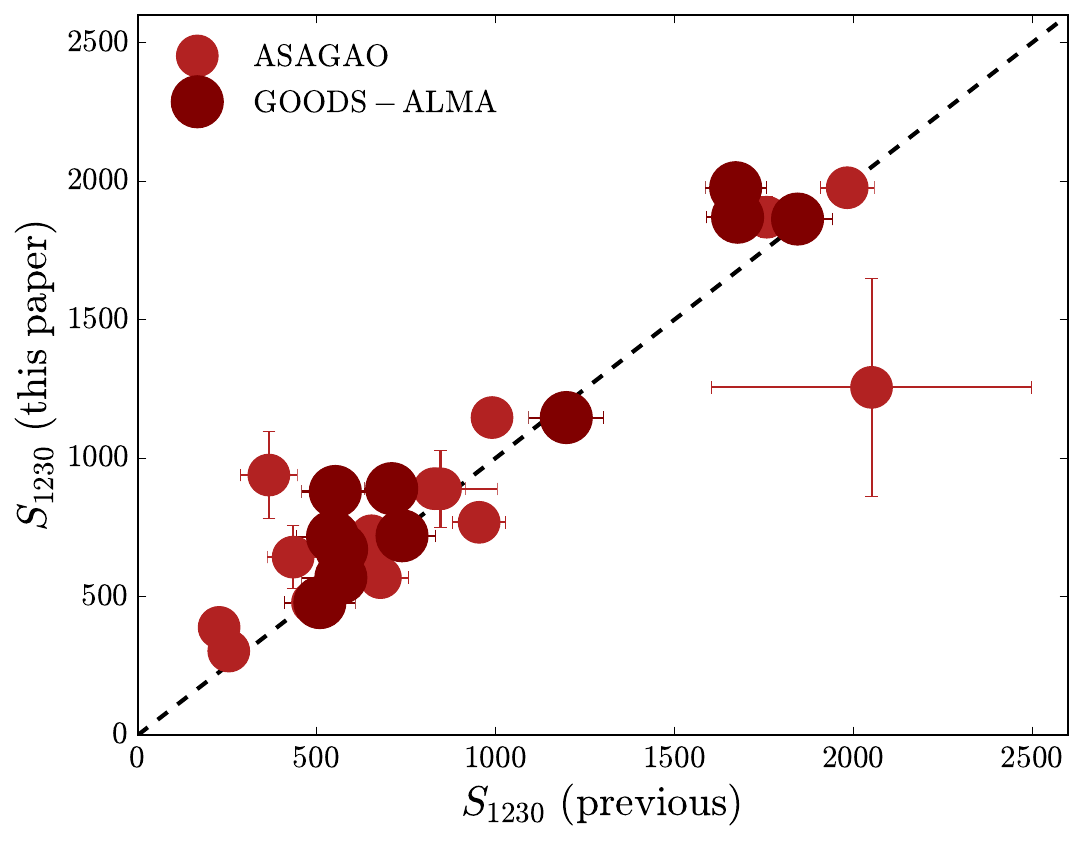}
\end{center}
\caption{Comparison of the flux densities measured in the outer region of the map versus flux densities from ASAGAO \citep{Hatsukade2018} and GOODS-ALMA \citep{Gomez-guijarro2022}.}
\label{fig:fluxes_outer}
\label{lastpage}
\end{figure}

\begin{table*}
\centering
\caption{Positions, peak S/N values, and flux densities ($S_{1230}$) for the 27 sources found in the ALMA 1-mm combined image at ${\rm S/N_{peak}}\,{>}\,4.5$. Here flux densities are measured by fitting Gaussian profiles to the sources in order to estimate the number of beams per source, then scaling the peak pixel value by the number of beams. For sources consistent with 1 beam or less and for sources with peak pixels ${>}\,1000\,\mu$Jy we set the flux density to be the peak pixel value. For the extended galaxy ALMA-HUDF-69, we use an aperture with a radius of 2\,arcsec to calculate a flux density. For the redshift column, galaxies with spectroscopic redshifts from \citet{Boogaard2023} are indicated by an asterisk, galaxies with spectroscopic redshifts from \citet{Rieke2023} are indicated by a dagger, galaxies with spectroscopic redshifts from \citet{Santini2015} are indicated by a double dagger ($\ddag$), and galaxies with photometric redshifts from \citet{Santini2015} are indicated by a section symbol ($\S$). When fitting SEDs, we use spectroscopic redshifts when available. For galaxies with matches in the JADES catalogue and sufficient photometry to fit SEDs (S/N$\,{>}\,$5 in the F356W band) we provide stellar masses and photometric redshifts (when no spectroscopic redshift is available) from McLeod et al. (in prep.). For galaxies outside of the JADES footprint that have a counterpart in the CANDELS catalogue, we provide stellar masses from \citet{Santini2015}, and mark these with a paragraph symbol ($\P$). The median photometric redshift uncertainty of our ALMA sample is $\Delta z / (1\,{+}\,z)\,{\simeq}\,0.03$, while the relative uncertainties in stellar masses are small.}
\label{table:sources_outer}
\begin{tabular}{lccccc}
\hline
Name & RA Dec [J2000] & S/N$_{\mathrm{peak}}$ & $S_{1230}$ [$\mu$Jy] & $z$ & $\log \left( M_{\ast} / \mathrm{M}_{\odot} \right)$\\
\hline
ALMA-HUDF-46 & 3:32:28.51 $-$27:46:58.3 & 76.3 & 1976$\pm$26\phantom{0} & 2.309$^{\ddag}$ & 10.9$^{\P}$\\
ALMA-HUDF-47 & 3:32:35.73 $-$27:49:16.2 & 25.1 & 1870$\pm$75\phantom{0} & 2.576$^{\ddag}$ & 11.2\\
ALMA-HUDF-48 & 3:32:34.27 $-$27:49:40.4 & 22.7 & 1863$\pm$82\phantom{0} & 3.58 & 10.2\\
ALMA-HUDF-49 & 3:32:44.04 $-$27:46:35.9 & 16.1 & 1146$\pm$71\phantom{0} & 2.698$^{\ast}$ & 10.5\\
ALMA-HUDF-50 & 3:32:28.59 $-$27:48:50.5 & \phantom{0}6.3 & \phantom{0}939$\pm$157 & 3.594$^{\dagger}$ & 10.4\\
ALMA-HUDF-51 & 3:32:32.91 $-$27:45:41.0 & 36.7 & \phantom{0}890$\pm$26\phantom{0} & 1.16 & 9.6\\
ALMA-HUDF-52 & 3:32:34.01 $-$27:49:00.0 & \phantom{0}6.7 & \phantom{0}889$\pm$140 & 2.63 & 10.9\\
ALMA-HUDF-53 & 3:32:28.78 $-$27:44:35.3 & 12.3 & \phantom{0}879$\pm$75\phantom{0} & 4.84$^{\S}$ & 10.5$^{\P}$\\
ALMA-HUDF-54 & 3:32:33.25 $-$27:49:16.5 & \phantom{0}5.0 & \phantom{0}828$\pm$173 & 3.56 & 9.5\\
ALMA-HUDF-55 & 3:32:48.00 $-$27:46:27.0 & \phantom{0}4.7 & \phantom{0}811$\pm$181 & 2.03 & 10.2\\
ALMA-HUDF-56 & 3:32:47.60 $-$27:44:52.5 & 11.8 & \phantom{0}768$\pm$65\phantom{0} & 1.930$^{\ddag}$ & 10.7\\
ALMA-HUDF-57 & 3:32:28.91 $-$27:44:31.6 & 18.0 & \phantom{0}719$\pm$44\phantom{0} & \dots & \dots\\
ALMA-HUDF-58 & 3:32:49.46 $-$27:49:08.9 & 20.2 & \phantom{0}717$\pm$39\phantom{0} & \dots & \dots\\
ALMA-HUDF-59 & 3:32:30.36 $-$27:45:01.4 & \phantom{0}4.9 & \phantom{0}680$\pm$144 & 2.94$^{\S}$ & 10.7$^{\P}$\\
ALMA-HUDF-60 & 3:32:31.48 $-$27:46:23.4 & 24.6 & \phantom{0}671$\pm$30\phantom{0} & 2.225$^{\dagger}$ & 10.9\\
ALMA-HUDF-61 & 3:32:44.60 $-$27:48:36.2 & \phantom{0}6.0 & \phantom{0}642$\pm$115 & 2.593$^{\dagger}$ & 10.7\\
ALMA-HUDF-62 & 3:32:29.25 $-$27:45:10.0 & 10.0 & \phantom{0}568$\pm$57\phantom{0} & 1.83$^{\S}$ & 10.3$^{\P}$\\
ALMA-HUDF-63 & 3:32:45.19 $-$27:48:06.9 & \phantom{0}7.2 & \phantom{0}553$\pm$83\phantom{0} & 3.94 & 9.5\\
ALMA-HUDF-64 & 3:32:42.15 $-$27:48:47.2 & \phantom{0}4.7 & \phantom{0}493$\pm$112 & \dots & \dots\\
ALMA-HUDF-65 & 3:32:47.17 $-$27:45:25.5 & \phantom{0}9.3 & \phantom{0}478$\pm$52\phantom{0} & 3.62 & 10.2\\
ALMA-HUDF-66 & 3:32:40.76 $-$27:49:26.4 & \phantom{0}4.8 & \phantom{0}441$\pm$91\phantom{0} & 2.130$^{\ddag}$ & 10.6\\
ALMA-HUDF-67 & 3:32:28.83 $-$27:48:29.9 & \phantom{0}6.1 & 1255$\pm$388 & 1.687$^{\dagger}$ & 11.0\\
ALMA-HUDF-68 & 3:32:38.74 $-$27:48:40.1 & \phantom{0}6.2 & \phantom{0}389$\pm$63\phantom{0} & 2.64 & 10.9\\
ALMA-HUDF-69 & 3:32:43.68 $-$27:48:51.0 & \phantom{0}5.1 & \phantom{0}303$\pm$60\phantom{0} & 2.40 & 10.3\\
ALMA-HUDF-70 & 3:32:48.60 $-$27:48:10.3 & \phantom{0}4.7 & \phantom{0}300$\pm$64\phantom{0} & \dots & \dots\\
ALMA-HUDF-71 & 3:32:30.29 $-$27:45:24.9 & \phantom{0}4.6 & \phantom{0}252$\pm$55\phantom{0} & \dots & \dots\\
ALMA-HUDF-72 & 3:32:41.76 $-$27:48:25.1 & \phantom{0}4.5 & \phantom{0}179$\pm$44\phantom{0} & 1.57 & 10.1\\
\hline
\end{tabular}
\end{table*}

\begin{table*}
\centering
\caption{Cross-matched IDs between our catalogue of sources and previous studies. Near-infrared matches are from JADES \citep{Rieke2023} and CANDELS \citep{Guo2013}. Matches with other ALMA Band-6 surveys are given from ASAGAO \citep{Hatsukade2018} and GOODS-ALMA \citep{Gomez-guijarro2022}. One ALMA source (ID 51, indicated by an asterisk) does not have a JADES/CANDELS ID (likely due to blending with another nearby galaxy), so we provide the ID from from McLeod et al. (in prep.) instead. Lastly, we note that ALMA ID 49 is UDF1 in the \citet{Dunlop2017} catalogue.}
\label{table:sources_outer_ids}
\begin{tabular}{lcccc}
\hline
Name & JADES ID & CANDELS ID & ASAGAO ID & GOODS-ALMA ID\\
\hline
ALMA-HUDF-46 & \dots & J033228.50$-$274658.1 & 2 & A2GS1\\
ALMA-HUDF-47 & 196290 & J033235.71$-$274916.0 & 3 & A2GS3\\
ALMA-HUDF-48 & 193893 & J033234.28$-$274940.4 & \dots & A2GS2\\
ALMA-HUDF-49 & 209026 & J033244.02$-$274635.7 & 1 & A2GS21\\
ALMA-HUDF-50 & 198459 & \dots & 33 & \dots\\
ALMA-HUDF-51 & 38804$^{\ast}$ & \dots & 7 & A2GS28\\
ALMA-HUDF-52 & 197581 & J033233.99$-$274859.6 & 31 & \dots\\
ALMA-HUDF-53 & \dots & J033228.77$-$274434.9 & \dots & A2GS37\\
ALMA-HUDF-54 & 96357 & J033233.24$-$274916.0 & \dots & \dots\\
ALMA-HUDF-55 & 209737 & J033247.99$-$274626.6 & \dots & \dots\\
ALMA-HUDF-56 & 217193 & J033247.60$-$274452.0 & 6 & \dots\\
ALMA-HUDF-57 & \dots & \dots & 20 & A2GS33\\
ALMA-HUDF-58 & \dots & \dots & 17 & A2GS38\\
ALMA-HUDF-59 & \dots & J033230.35$-$274501.1 & \dots & \dots\\
ALMA-HUDF-60 & 209962 & J033231.45$-$274623.1 & 8 & A2GS19\\
ALMA-HUDF-61 & 199505 & J033244.59$-$274835.8 & 19 & \dots\\
ALMA-HUDF-62 & \dots & J033229.23$-$274509.7 & 11 & A2GS23\\
ALMA-HUDF-63 & 201793 & J033245.20$-$274806.8 & \dots & \dots\\
ALMA-HUDF-64 & \dots & \dots & \dots & \dots\\
ALMA-HUDF-65 & 214839 & \dots & 9 & A2GS40\\
ALMA-HUDF-66 & 195280 & J033240.74$-$274926.1 & \dots & \dots\\
ALMA-HUDF-67 & 199996 & J033228.82$-$274829.7 & 44 & \dots\\
ALMA-HUDF-68 & 198545 & J033238.73$-$274839.8 & 29 & \dots\\
ALMA-HUDF-69 & 198451 & J033243.67$-$274850.8 & 26 & \dots\\
ALMA-HUDF-70 & \dots & \dots & \dots & \dots\\
ALMA-HUDF-71 & \dots & \dots & \dots & \dots\\
ALMA-HUDF-72 & 200418 & J033241.75$-$274824.9 & \dots & \dots\\
\hline
\end{tabular}
\end{table*}

Here we present a shallower but larger ALMA 1-mm map, in an area defined by the ASAGAO footprint, minus the footprint of the deeper inner region. We will also give a catalogue of sources found in this wider map.  The total area of the shallower map is 25.4\,arcmin$^2$, so after subtracting the area of the inner region we end up with an effective area of 21.2\,arcmin$^2$. The typical pixel rms is $60\,\umu{\rm Jy}\,{\rm beam}^{-1}$ outside of the individual pointings and $20\,\umu{\rm Jy}\,{\rm beam}^{-1}$ within the individual pointings.

Table~\ref{table:obs_outer} outlines the programmes used to produce the combined image, Fig.~\ref{fig:snr_outer} shows the resulting signal, noise and S/N maps, Table~\ref{table:sources_outer} provides positions, peak S/N values, flux densities, redshifts and stellar masses for all of the sources found with a peak ${\rm S/N}\,{>}\,4.5$ (where the fidelity reaches a value of 0.9), ignoring the deeper central region where we can find fainter sources; for reference, the most significant negative peak found in this search has a (nagative) S/N of 5.1. In this table we also sort our sources by $S_{1230}$, using the prefix ALMA-HUDF, and here beginning with the number 46 in order to be consecutive after the numbering of the main catalogue of sources in the deeper region. Finally, Table~\ref{table:sources_outer_ids} provides IDs matched to other catalogues. One ALMA source (ID 51) is not included in the JADES and CANDELS catalogues (likely due to blending with another nearby galaxy), so we list the ID from McLeod et al. (in prep.) instead.

Flux densities were measured using the same method described in Section~\ref{sec:catalogue}, with one adjustment. We found that a 2D Gaussian was a poor fit to high S/N sources (presumably mostly because of the combination of data sets with different resolution), which resulted in flux densities a factor of about 2 larger than reported in previous studies. We therefore set a threshold of 1000\,$\mu$Jy above which we do not fit a 2D Gaussian, and instead just take the peak pixel value. We found that this resulted in flux densities consistent with previous studies. Lastly, our source ALMA-HUDF-69 has a counterpart ID 44 in the ASAGAO catalogue, where it is reported that the peak pixel flux density is nearly 6 times fainter than the integrated flux density, thus it is likely that this galaxy is also highly resolved in our ALMA image. We find that a 2D Gaussian is a poor fit to the ALMA data, so we simply use an aperture with a radius of 2\,arcsec to measure its flux density. As an additional test, we extracted peaks from the central region of the shallow map and measured their flux densities in order to compare them with our deep map; we found good agreement between the two measurements for all ${>}\,4.5\sigma$ sources in the shallow map.

In Fig.~\ref{fig:sources_outer} we show the positions of our sources, along with sources found in the ASAGAO \citep{Hatsukade2018} and GOODS-ALMA \citep{Gomez-guijarro2022} surveys. Blue crosses indicate sources with a CANDELS and JADES match, gold crosses indicate sources with a CANDELS-only match (these galaxies all lie outside of the JADES footprint) and red crosses indicate galaxies with a JADES-only match. Figure~\ref{fig:fluxes_outer} compares our photometry to the photometry of the ASAGAO and GOODS-ALMA surveys for all overlapping sources. In Appendix~\ref{sec:appendixB} we show cutouts of these ALMA sources overlaid on {\it JWST\/} F356W imaging. 

We find 27 galaxies in our defined region, nine of which are not in a previously-published catalogue, with one of these new sources having a peak S/N$\,{>}\,5.1$, i.e. more significant than the most significant negative peak. Of these nine, six have a JADES/CANDELS counterpart, while the remaining two do not; however, these two galaxies lie outside of both the deep {\it HST\/} footprint and the JADES footprint.  It may also be worth noting that the fourth brightest source (ALMA-HUDF-51) in this supplementary catalogue lies just beyond the boundary of our deep map, but is detected by \citep{Dunlop2017} and given the name UDF1 in their paper.

We repeated the analyses of the sources in the wide map that we performed for the deep map.  In particular we looked at: the redshift, magnitude and colour distributions of the previously known versus new sources, as in Fig.~\ref{fig:histograms}; the stellar mass versus redshift distribution, as in Fig.~\ref{fig:z_vs_Mstar}; and the magnitude and colour distributions of the ALMA galaxies versus ALMA-undetected galaxies of similar stellar mass, as in Fig.~\ref{fig:highMstar_histograms}.  In each case the results were similar to those from the deep map, but with fewer sources.

It would also be possible to carry out a stacking analysis in the wider region.  However, for a fixed number density of optical/near-IR galaxies, the stacking uncertainty in a bin, $\delta \hat{S}_{\alpha}$, depends on the survey area and depth of the 1-mm map in a way that essentially scales like the square root of the total observing time. We can see that in the following way.  Let us assume that we have a catalogue with a fixed number density of sources, $n_{\rm s}$, and we are stacking on two maps, one with area $A_1$ and noise per pixel (or beamsized-pixel, say) $\sigma_1$, and the other with area $A_2$ and noise per pixel $\sigma_2$.  The total numbers of sources to stack on in the two maps are then $N_1=n_{\rm s}A_1$ and $N_2=n_{\rm s}A_2$.  The error on the stack will be the noise in a pixel reduced by the square root of the number of samples, hence the ratio if errors between the two maps will be
\begin{equation}
    \frac{\delta \hat{S}_1}{\delta \hat{S}_2} = \frac{\sigma_1}{\sigma_2}\left(\frac{N_2}{N_1}\right)^{1/2} = \frac{\sigma_1}{\sigma_2}\left(\frac{n_{\rm s}A_2}{n_{\rm s}A_1}\right)^{1/2} = \frac{\sigma_1}{\sigma_2}\left(\frac{A_2}{A_1}\right)^{1/2} .
    \label{eq:area_depth_time}
\end{equation}
But the time taken to map an area to fixed depth is proportional to the area, while the time to map a fixed area to a given depth is proportional to the inverse square of the depth.  In other words the integration times are $t_1\propto A_1/\sigma_1^2$ and $t_2\propto A_2/\sigma_1^2$, which means that the ratio in Eq.~\ref{eq:area_depth_time} is $\delta \hat{S}_1/\delta \hat{S}_2=(t_2/t_1)^{1/2}$.  For the case at hand, since so much more integration time went into the deep map, then stacking results from the wider map add will have much larger uncertainties and provide very little additional information.

\section{Source cutouts}
\label{sec:appendixB}

\begin{figure*}
\begin{center}
\includegraphics[width=0.75\textwidth]{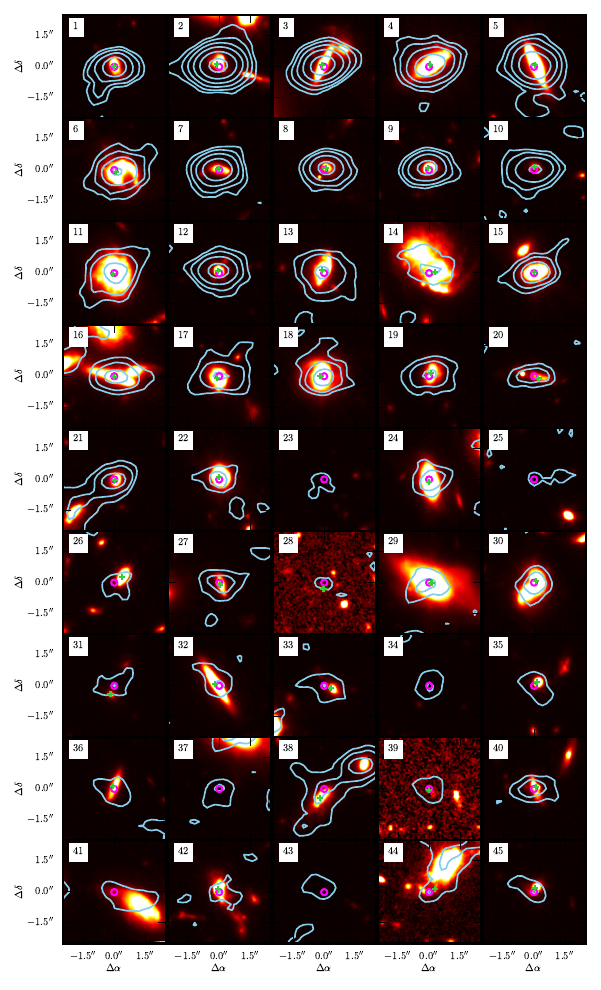}
\end{center}
\caption{5\,arcsec$\,{\times}\,$5\,arcsec cutouts. The background shows {\it JWST\/} F356W imaging, while the contours are from our deep ALMA map in steps of 2$^n\sigma$, with $n\,{=}\,1$, 2, 3, 4, 5 and 6. ALMA centroids are indicated by magenta circles and the positions of {\it JWST\/} counterparts are indicated by green crosses. ALMA IDs 28 and 39 all have a F356W flux density signal-to-noise ratios less than 5, and we do not use their photometry to fit SEDs.}
\label{fig:cutouts}
\end{figure*}

\begin{figure*}
\begin{center}
\includegraphics[width=0.75\textwidth]{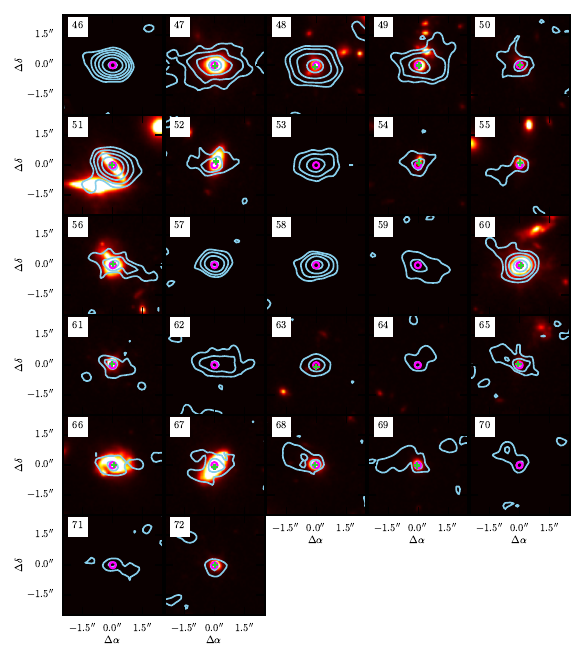}
\end{center}
\caption{Same as Fig.~\ref{fig:cutouts}, but for our sources in the wider map. For ALMA sources found outside the JADES footprint the background is blank.}
\label{fig:cutouts_outer}
\end{figure*}

Here we provide 5\,arcsec$\,{\times}\,$5\,arcsec cutouts of our ALMA sources overlaid on {\it JWST\/} F356W images. In each panel of Fig.~\ref{fig:cutouts} we show ALMA contours from our deep map in steps of 2$^n\sigma$, with $n\,{=}\,1$, 2, 3, 4, 5 and 6. ALMA centroids are indicated by magenta circles and the positions of {\it JWST\/} counterparts are indicated by green crosses. The {\it JWST\/} images were lightly smoothed by a Gaussian filter with a FWHM of 1.8 pixels for presentation purposes. Fig.~\ref{fig:cutouts_outer} shows the same thing but for our wider shallow map; note that some of the ALMA sources found in this wider map are outside of the JADES footprint, so we leave their backgrounds blank.

\section{Alternative data treatment and tests}
\label{sec:alternative}
\subsection{Data combination in the image plane}

\begin{figure*}
\begin{center}
\includegraphics[width=0.49\textwidth]{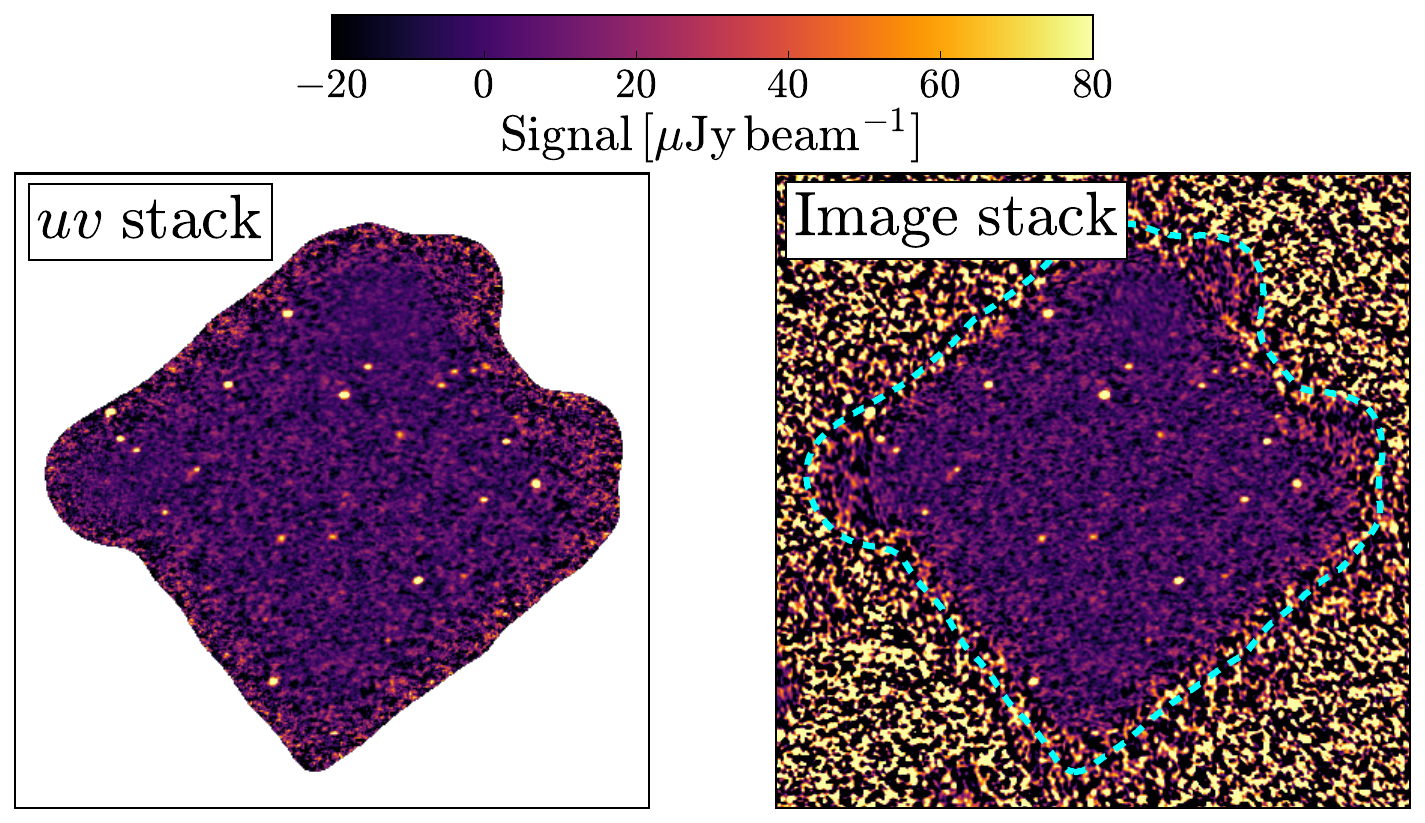}
\includegraphics[width=0.49\textwidth]{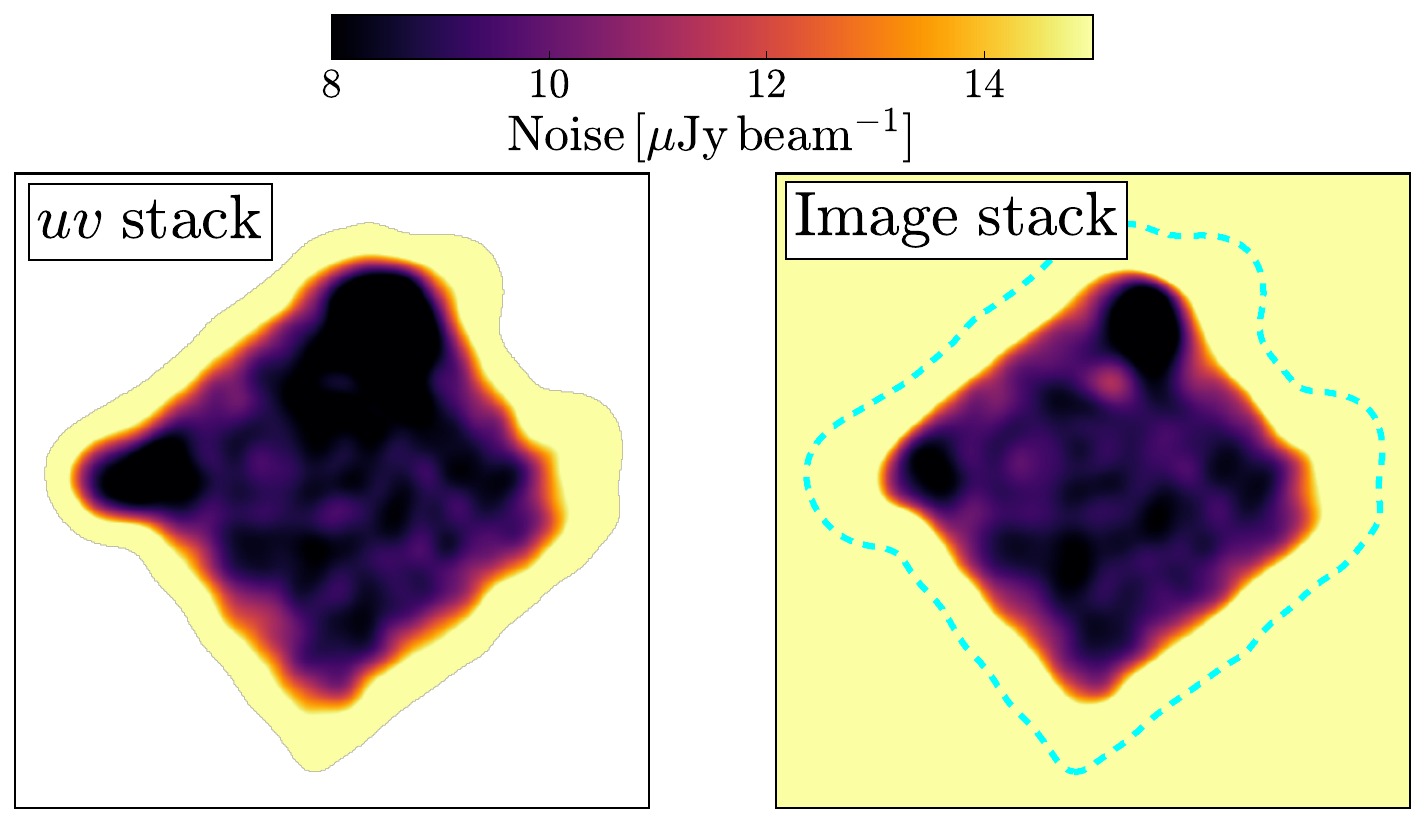}
\includegraphics[width=0.49\textwidth]{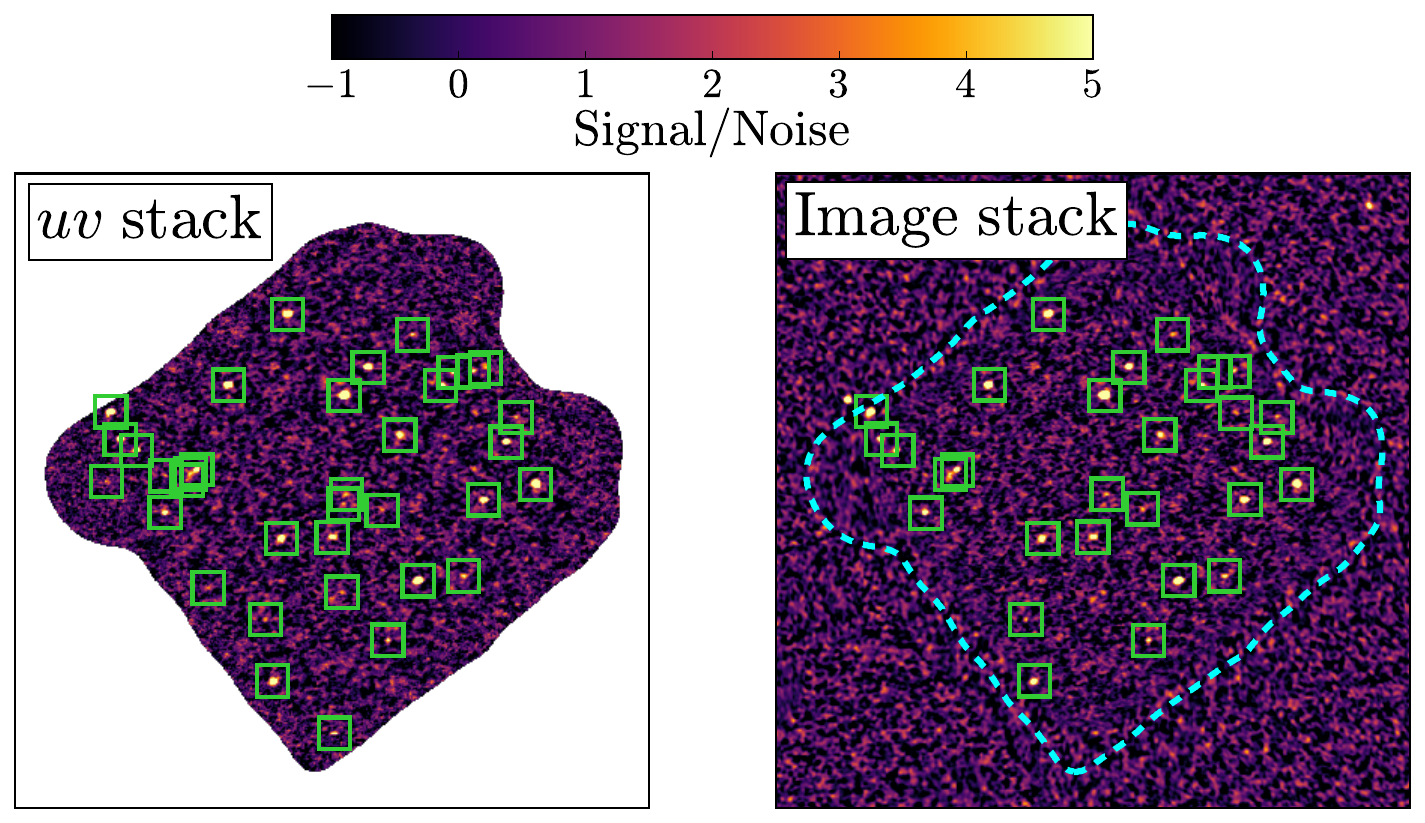}
\end{center}
\caption{Results from combining all of the data in Table~\ref{table:obs} in the $uv$ plane before imaging, and imaging the same data individually before combining. Signal maps are shown in the two top left panels, noise maps are shown in the two top right panels and S/N maps are shown in the bottom two panels. In each case, the $uv$ combination is on the left and the real-space image combination is on the right. Sources with a peak S/N$\,{>}\,$4 are marked with green squares in the S/N maps.}
\label{fig:test_images}
\end{figure*}

In addition to our $uv$-combined images, we also tried combining archival ALMA data-product images (each one being produced individually using \textsc{tclean} in MFS mode) in real space. This provides a consistency check for how well we are able to combine in $uv$ space the data spanning a wide frequency range and large array configuration range. For this test we focus on the deep central mosaic (following the ASPECS footprint), where the data sets included are listed in Table~\ref{table:obs}.

The first step in this process is to convolve each image to the same resolution. A common beamsize was selected as the largest beam across all of the data sets, and for each image the convolution kernel required to produce this common beam was found numerically using the \textsc{create\_matching\_kernel} function from the \textsc{python} \textsc{photutils} module \citep{Bradley2022}. Each image was then convolved with this kernel. We expect that convolution to a lower resolution will effectively increase the pixel noise (when in units of mJy\,beam$^{-1}$). The largest beam across all of our data comes from the ASPECS pilot programme (2013.1.00718.S, see \citealt{aravena2016}), which has a beam of about $1.5\,{\rm arcsec}\,{\times}\,0.8\,{\rm arcsec}$. The most sensitive map (across a significantly larger area) comes from the full ASPECS survey (2016.1.00324.L, see \citealt{Gonzalez-lopez2020}), where the beam is about $1.2\,{\rm arcsec}\,{\times}\,0.8\,{\rm arcsec}$, or the second-largest beam across all of our images. We found that degrading the resolution of the full ASPECS map to match the ASPECS pilot map ultimately produces a map with a higher rms (after combining all the data) compared to removing the ASPECS pilot map from the combination and convolving each image to match that of the full ASPECS map; our final combined image therefore does not contain data from the ASPECS pilot, although this amounts to only a small loss of data (see Table~\ref{table:obs}).

We next created a pixel grid for the image. The size of the pixels in the grid was chosen as the largest pixel size within the set of 61 images (in this case 0.2\,arcsec), and the grid was set to span 5\,arcmin, centred at 03$^{\rm h}$32$^{\rm m}$39.0$^{\rm s}$ $-$27$^{\circ}$47$^{\prime}$29.1$^{\prime\prime}$. Next, each of the 61 convolved images (minus those from the ASPECS pilot) were reprojected onto this grid using the \textsc{python} function \textsc{reproject\_interp}, part of the \textsc{reproject} module \citep{Robitaille2020}.

The next step was to create a noise map for each reprojected and convolved image. To do this, we created primary-beam-uncorrected images (simply the primary beam image multiplied by the primary beam), and for each image we created a mask using the same method described above. We then calculated the rms of the primary-beam-uncorrected map multiplied by the mask, and divided this by the primary beam.

These images should now be aligned to the same pixel grid, have the same resolution, and have associated noise maps. The final step is to combine them, weighting each pixel by its inverse variance. In order to remove boundary artefacts associated with combining images with very different noise levels, an apodizing mask was applied to the edges of each image in the combination. The apodizing function chosen was a Gaussian with a standard deviation of the beamsize, and the apodization was applied out to 3 times the beamsize from the edge of each map. We investigated various options for apodization (e.g.\ a cosine function or different apodization lengths), and found that this simple choice removed most of the obvious edge effects. 

Lastly, a new noise map was calculated for the combined image using the same approach used to calculate a noise map for the $uv$-combined image. For reference, the image-combined map can also be downloaded.\footnote{\url{https://doi.org/10.5683/SP3/YWBVWH}}

\subsection{Comparison between \textit{uv}-plane combination and image-plane combination}

\begin{figure}
\begin{center}
\includegraphics[width=0.5\textwidth]{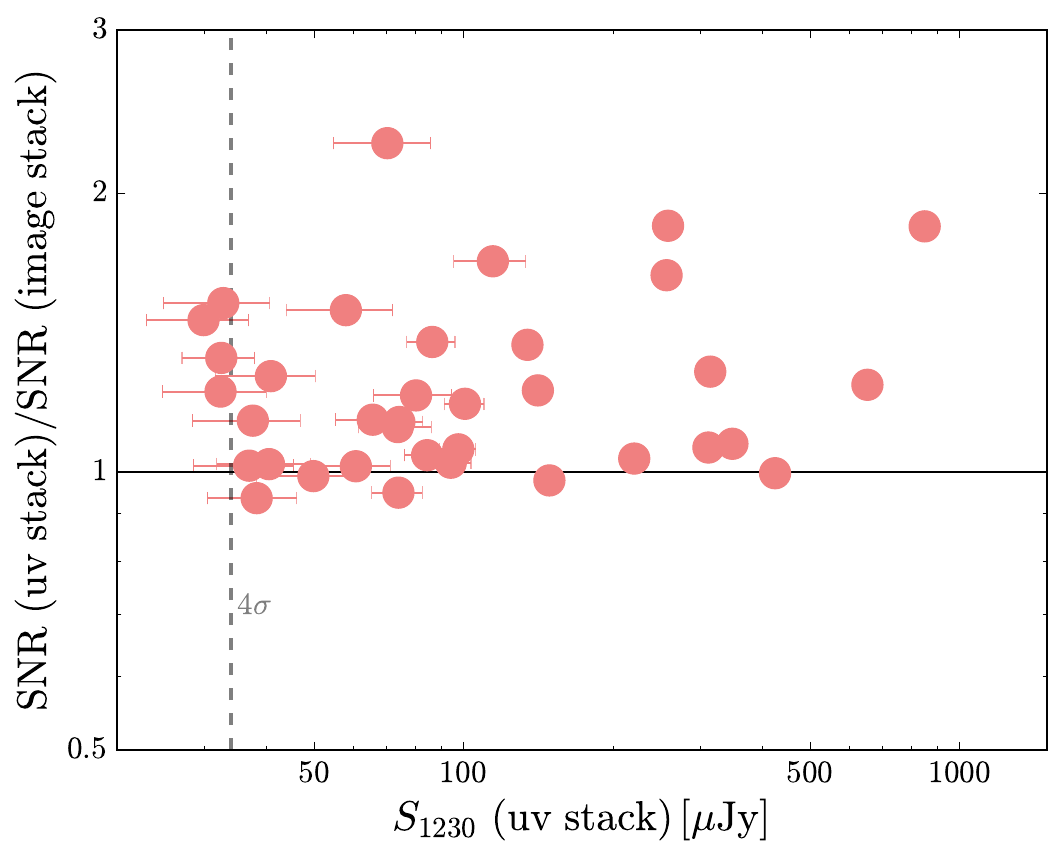}
\end{center}
\caption{Ratio of the peak S/N from the $uv$ combination to the peak S/N from the image combination, as a function of 1230-$\umu$m flux density in the image combination. The peak S/N values from the $uv$ combination are generally higher than for the image combination (the mean ratio is 1.2).}
\label{fig:test_snr}
\end{figure}

The optimal way to combine interferometric images is to add the observations in the $uv$ plane, then image the entire data set. However, the different frequencies and $uv$ sampling of the various observations may lead to unwanted behaviour. Moreover, adding the individual pointings is much more straightforward in the image plane than in $uv$ space. Therefore we would like to compare the properties of the two images ($uv$-space combination and real-space combination) to check that they are consistent with one another, and to see which one performs better.

We focus on the region defined by our deep central map with a 250\,k$\lambda$ taper, which was made going out to 0.2 times the primary beam (see Fig.~\ref{fig:snr}). We extracted all sources with a peak S/N$\,{>}\,4$ within this region from both maps using the same source-extraction procedure as described in Section~\ref{sec:results}; this threshold was simply chosen so that enough overlapping sources could be extracted for comparison. There are 36 peaks with this significance in the $uv$-space combined map and 29 peaks in the image-space combined map. In Fig.~\ref{fig:test_images} we show all of the sources found with a peak S/N$\,{>}\,4$, from which it can be seen that all bright and obvious sources are in agreement. However, we do see more sources detected in the $uv$ combination compared to the image combination, especially around the edges of our selected region, where the $uv$ map does much better.

To quantify the difference between the two maps, in Fig.~\ref{fig:test_snr} we plot the ratio of the peak S/N from the $uv$ combination to the peak S/N from the image combination as a function of 1-mm flux density in the $uv$-space combined map. For the galaxies with a peak S/N$\,{>}\,4$ in the $uv$-combined map but not the image-combined map, we simply extract the value of the image-combined map at the location of the $uv$ detections in order to include them on this plot. 
We find a relatively uniform improvement in S/N from combining the $uv$ data as a function of source brightness. The mean peak S/N of all matching sources is higher by about 1.2. Of course there is a balance between the increased complexity and computing resources required to combine the data in the $uv$ plane versus combining in the image plane, but our results indicate that the improvement from combining the data in the $uv$-plane is worthwhile.

\end{document}